\documentclass[10pt]{IEEEtran}
\AtBeginDocument{%
  \providecommand\BibTeX{{%
    \normalfont B\kern-0.5em{\scshape i\kern-0.25em b}\kern-0.8em\TeX}}}

\usepackage{ifthen}
\newboolean{treport}

\ifthenelse
{ \equal{\jobname}{\detokenize{sub}} }
{ \setboolean{treport}{false} }
{}

\ifthenelse
{ \equal{\jobname}{\detokenize{tr}} }
{ \setboolean{treport}{true} }
{}

\usepackage[english]{babel}
\usepackage{blindtext}
\usepackage{comment}
\usepackage{hyperref}
\hypersetup{
    colorlinks=true,
    linkcolor=blue,
    filecolor=magenta,      
    urlcolor=cyan,
    citecolor=black
}

\usepackage{balance}
\usepackage{placeins}
\usepackage{tabularx}
\newcolumntype{L}[1]{>{\raggedright\arraybackslash}p{#1}}
\newcolumntype{C}[1]{>{\centering\arraybackslash}p{#1}}
\newcolumntype{R}[1]{>{\raggedleft\arraybackslash}p{#1}}
\newcolumntype{B}[1]{>{\centering\let\newline\\\arraybackslash\hspace{0pt}}b{#1}}

\usepackage{algorithm}
\usepackage[noend]{algorithmic}

\usepackage{subfigure}
\usepackage{enumitem}
\setlist{nolistsep}
\usepackage{graphicx}
\usepackage{epstopdf}
\usepackage{grffile}
\usepackage{amssymb}
\usepackage{amsmath}
\usepackage{multicol,multirow}
\usepackage{rotating,url}
\usepackage{enumitem}
\usepackage{color}
\usepackage{xspace}
\usepackage{ifpdf}
\usepackage{mdwlist}
\usepackage{colortbl}
\usepackage{caption}
\usepackage{amssymb}
\usepackage{booktabs}
\usepackage{multirow}
\usepackage{enumitem}
\usepackage{xfrac}
\usepackage{afterpage}
\usepackage{pdflscape}
\usepackage{longtable}
\usepackage{hhline}
\usepackage{lscape}
\usepackage[showboxes]{textpos}

\newlength{\oldtextfloatsep}

\newcommand{\eg}{e.g., \@}

\newcommand{\eat}[1]{}

\usepackage{enumitem}
\usepackage{mdwlist}
\setlist{nolistsep}
\setlist[description]{noitemsep,topsep=0pt,parsep=0pt,partopsep=0pt,leftmargin=0pt}
\setlist[itemize]{noitemsep,topsep=0pt,parsep=0pt,partopsep=0pt,leftmargin=12pt}
\usepackage{amssymb}
\usepackage{pifont}

\newcommand{\cmark}{\ding{51}}%
\newcommand{\xmark}{\ding{55}}%

\newcommand{\new}[1]{\textcolor{black}{#1}}
\newcommand{\done}[1]{} 

\newcommand{\flowtree}{{Flowtree}\xspace}
\newcommand{\Flowtree}{{Flowtree}\xspace}

\newcommand{\Flowtrees}{{Flowtrees}\xspace}
\newcommand{\flowtrees}{{Flowtrees}\xspace}

\newcommand{\Flowstream}{{Flowyager}\xspace}
\newcommand{\FlowStream}{{Flowyager}\xspace}
\newcommand{\flowstream}{{Flowyager}\xspace}

\newcommand{\flowql}{{FlowQL}\xspace}
\newcommand{\Flowql}{{FlowQL}\xspace}
\newcommand{\flowQL}{{FlowQL}\xspace}
\newcommand{\FlowQL}{{FlowQL}\xspace}

\newcommand{\flowdb}{{FlowDB}\xspace}
\newcommand{\Flowdb}{{FlowDB}\xspace}

\newcommand{\FlowDB}{{FlowDB}\xspace}

\newcommand{\flowagg}{{FlowAGG}\xspace}
\newcommand{\Flowagg}{{FlowAGG}\xspace}

\newcommand{\FlowAGG}{{FlowAGG}\xspace}

\newcommand{\MAWI}{{MAWI}\xspace}
\newcommand{\IXP}{{IXP}\xspace}
\newcommand{\ISP}{{ISP}\xspace}

\newcommand{\iftr}[2]{\ifthenelse{\boolean{treport}} {#1} {#2}}

\long\def\comment#1{}

\begin{document}



\title{Exploring Network-Wide Flow Data with Flowyager}

 \author{
 {\rm Said Jawad Saidi$^1$ \quad Aniss Maghsoudlou$^1$ \quad Damien Foucard$^2$\\ 
Georgios Smaragdakis$^{2,1}$ \quad Ingmar Poese$^4$ \quad Anja Feldmann$^{1,3}$ }\\
                \footnotesize $^1$Max Planck Institute for Informatics  \quad $^2$TU Berlin  \quad $^3$Saarland
University \quad $^4$BENOCS GmbH\\
        }




\maketitle
 \setlength{\TPHorizModule}{\paperwidth}
\setlength{\TPVertModule}{\paperheight}
\TPMargin{5pt}
\begin{textblock}{0.7}[0.0,0.0](0.05,-0.25)
  \noindent
  \footnotesize
  If you cite this paper, please use the IEEE TNSM reference:
  
  Said Jawad Saidi, Aniss Maghsoudlou, Damien Foucard, Georgios Smaragdakis, Ingmar Poese, Anja Feldmann. 2020.
  Exploring Network-Wide Flow Data with Flowyager.
  In \textit{IEEE Transactions on Network and Service Management.}
  IEEE,https://doi.org/10.1109/TNSM.2020.3034278
\end{textblock}

\begin{abstract}
Many network operations, ranging from attack investigation and mitigation to traffic management, require answering
network-wide flow queries in seconds. Although flow records are collected at each router, using available traffic
capture utilities, querying the resulting datasets from hundreds of routers across sites and over time, remains a
significant challenge due to the sheer traffic volume and distributed nature of flow records.

In this paper, we investigate how to improve the response time for \emph{a priori unknown network-wide} queries. We
present \flowstream, a system that is built on top of existing traffic capture utilities. \flowstream generates and
analyzes tree data structures, that we call \flowtrees, which are 
succinct summaries of the raw flow data available by capture utilities. \flowtrees are self-adjusted data structures that drastically reduce space and transfer
requirements, by 75\% to 95\%, compared to raw flow records. \flowstream manages the storage and transfers of
\flowtrees, supports \flowtree operators, and provides a structured query language for answering flow queries across
sites and time periods. By deploying a \flowstream prototype at both a large Internet Exchange Point and a Tier-1
Internet Service Provider, we showcase its capabilities for networks with hundreds of router interfaces. Our results
show that the query response time can be reduced by an order of magnitude when compared with alternative data analytics
platforms. Thus, \flowstream enables \emph{interactive network-wide queries} and offers unprecedented drill-down
capabilities to, e.g., identify DDoS culprits, pinpoint the involved sites, and determine the length of the attack.
\end{abstract}


\section{Introduction}\label{sec:introduction}\label{sec:intro} 

Network operators have to continuously keep track of the activity in their
networks over both long and short time windows.
Over long time windows, e.g., days or hours, network operators are interested in
provisioning network capacity or making informed peering decisions.
Over short time windows, e.g., minutes, network operators would like to identify and rectify unusual events,
\eg attacks or network disruptions.
To that end, they typically rely on either
flow-level or packet-level captures from routers within their network \new{\cite{Flow-Monitoring-Explained}}. For a
summary of tasks and how previous work tackled them see Table~\ref{table:sys-problems}.

Flow captures include 5-features: source (src) and destination (dst) IP addresses, port
numbers, protocol ID--to summarize traffic information per flow--Packet and
byte count~\cite{TMatrices:2004,DiagnosingAnomalies:2004}.  
Packet captures gather packet
headers~\cite{NewDirections:2002,Building-Netflow:2004,AutomaticallyInferring:2003,libpcap}.
Unfortunately, gathering data for every packet is often too expensive at high-speed links.
Thus, flow-level and packet-level capture tools rely on sampling packets, e.g., 1 of every 10k
packets~\cite{Estimating-Flow-Distributions:2003}.

Among the most popular 
capture tools are NetFlow~\cite{rfc3954}, IPFIX~\cite{rfc7011},
sFlow~\cite{sFlow}, and libpcap~\cite{libpcap}.
All major router and high-end switch vendors (Cisco, Juniper, Alcatel-Lucent,
and Huawei) offer flow capture
capabilities~\cite{rfc3954,rfc7011,sFlow}\footnote{NetFlow is a Cisco trademark, so other vendors market
  the NetFlow support with other names, e.g., Juniper Networks use the
  trademark Jflow or cflowd.} in their commodity as well as high-end
products.\footnote{
NetFlow and IPFIX capabilities are available in router series, e.g.,
Cisco IOS-XR, IOS and Catalyst router~\cite{Cisco-Netflow}, Juniper M-, T-, and MX-series
routers~\cite{Juniper-routers}, Alcatel-Lucent 7750SR~\cite{Alatel-lucent-routers}, Huawei NE-series
routers~\cite{Huawei}, and switches, e.g., Cisco (5600, 7000, 7700), Enterasysthese
(S- and N-series), and servers, e.g., VMware (vSphere 5.x).}

Recently, query-driven solutions, e.g., Sonata~\cite{gupta2018sonata}, Stroboscope~\cite{tilmans2018stroboscope}, and
Marple~\cite{marple2017}, made it possible to compile specific queries into telemetry programs and collect data from all
queried network nodes. These solutions provide exceptional flexibility, but they require the network
operator to know \emph{a priori} (i) the nature of the network problem, (ii) the network-related query that has to be compiled
into telemetry programs, (iii) the network node where the telemetry capability is available, and (iv) the node where the query
has to be executed. Unfortunately, in large networks with hundreds of interfaces, operational issues arise at different
parts of the network and the queries that are required are not known in advance. In many cases, network engineers have to
try different queries to locate the source and type of problem interactively. Thus, it takes a prohibitively large time
to compile such queries into telemetry programs. Another obstacle toward adopting such solutions is that this requires hardware
investments by the network operator. For example, Marple relies on P4-programmable software switches that are not yet
widely adopted by Internet Exchange Points (IXP) operators and Internet Service Providers (ISP).

To the best of our knowledge, there is at this point in time no system
that offers answers to \emph{a priori unknown network-wide} queries in
a scalable \emph{interactive} manner, even though the necessary raw network
data, e.g., via NetFlow~\cite{Cisco-Netflow,Building-Netflow:2004},
sFlow~\cite{sFlow}, IPFIX~\cite{rfc7011}, or libpcap~\cite{libpcap} is
collected by most operators.

From an operational point of view, fast exploration of large volumes of network
flows over time and across sites is useful to answer a range of operational
queries (see Table~\ref{table:sys-problems}). 
Yet, network operators need to be able to tackle such tasks in a unified and systematic way with reliable and scalable
tools. Existing data analytics systems, e.g., Spark~\cite{Spark}, are not tailored to analyze network data 
when it comes to scalability, interactivity, handling of geo-distributed data, 
or answering a priori unknown network-wide queries.

In this paper, we design, implement and evaluate a system, \flowstream, that is able
to answer \emph{a priori unknown network-wide queries} with fast response, and, thus, enables \emph{interactive} exploration
of network data \emph{across network sites} and \emph{over time}. The
architecture of our system is built around the following requirements:
\begin{description*}
\item [(1) Scalability:] The system should grow with the network size, the number
  of data sources, and the analysis requirements. Hereby, it should enable
  distributed deployment and not require all data to be transferred to a
  central location.

\item [(2) Reuse of \emph{existing} flow captures:] As it takes significant effort to
  deploy novel network capture utilities, the system should work on top of
  existing, widely deployed, and supported flow capture capabilities,
  such as NetFlow, sFlow, IPFIX, or libpcap. In high-speed links, these tools typically sample
packets~\cite{Estimating-Flow-Distributions:2003} to provide summaries of flow activity.

\item [(3) Support of \emph{interactive} and \emph{ad-hoc} que\-ries:] To
  ea\-si\-ly ex\-pl\-ore network
  data, the system needs to offer an interface that is flexible and interactive
  \new{(meaning response times in the order of seconds)} 
  so as to improve user productivity and enable drill-down capabilities. Possible
  queries vary and a system should not only focus on batch-style known queries but also
  enable quick \emph{ad-hoc} exploration of the data, i.e., answer queries that
  \emph{are not known in advance}, and allow for follow-up queries. 
  Answering network-wide queries should not require custom code or
  scripting as network operators usually neither have the required time
  nor the resources (e.g., storage or computing).
  The goal is to reduce the response time of queries from hours or dozens of minutes to seconds
  and, thus, enable interactive and drill-down queries.

\item [(4) Support of queries \emph{across network sites} and \emph{over} \emph{time}:] Most que\-ries are not just
  for some specific time period or network site. Rather, they correlate data
  spanning multiple periods, across network sites, and at different granularities,
  e.g., per site, region, time of day, and event. The system should be able
  to collect, index, and store summary data across multiple sites and over time.

\end{description*}

\begin{table}[t]
\scriptsize
\begin{center}
  \begin{tabularx}{\linewidth}{L{.52\linewidth}|L{.46\linewidth}}
  \toprule

{}Application & Related Work \\
\midrule
{Aggregated flow statistics (range queries over IP/ports/time/location)}
&
\cite{Gigascope,sarlis2015datix,cormode2003finding,ben2017constant,curtis2011devoflow,gupta2018sonata,marple2017}
\\
{Counting traffic}
& \cite{Gigascope,huang2017sketchvisor, basat2017elephant, Yu2019dShark,
  cormode2004improved,cormode2008finding,Cormode2005multigraph,
  curtis2011devoflow,gupta2018sonata,ben2017constant}
  \cite{narayana2016compiling, li2016flowradar,
    huang2018sketchlearn,tilmans2018stroboscope,Elastic-Sketch-SIGCOMM2018, yu2013software,bajpai2016network,prieto2007gap}
\\
{Traffic matrix}
& \cite{Yu2019dShark,narayana2016compiling}
\\
{DDoS diagnosis}
& \cite{yu2013software, huang2017sketchvisor, Yu2019dShark, covarianceDDoS, sekar2006lads, earlyDDoS,gupta2018sonata,narayana2016compiling}
\\
{Super-spreaders Detection}
& \cite{yu2013software, huang2017sketchvisor,gupta2018sonata}
\\
{top-K number of flows}
& \cite{basat2017elephant, bajpai2016network, prieto2007gap, Efficient-frequent-topk:2005-metwally}
\\
{Flows above threshold T (Heavy Hitters)}
& \cite{Elastic-Sketch-SIGCOMM2018, huang2018sketchlearn,yu2013software,huang2017sketchvisor,basat2017elephant, cormode2003finding,ben2017constant, schweller2007reversible,curtis2011devoflow, gupta2018sonata, tang2019mv}
\\
{Heavy Changers Detection }
&\cite{Elastic-Sketch-SIGCOMM2018, yu2013software,huang2017sketchvisor, schweller2007reversible, Cormode2004deltoid, tang2019mv}
\\
{Blackhole Detection} 
&\cite{tammana2016simplifying, tammana2018distributed, arzani2018007,Yu2019dShark}
\\
{Port-based / 4/5-tuple queries}
& \cite{Gigascope,Elastic-Sketch-SIGCOMM2018,huang2018sketchlearn,li2016flowradar,huang2017sketchvisor,basat2017elephant,Yu2019dShark,gupta2018sonata}
\\
\bottomrule
  \end{tabularx}
  \end{center}
\caption{Typical network queries and systems to tackle them. Currently, 
  no system addresses all of them.}
\vspace{-2em}
\label{table:sys-problems}
\end{table}

Although most networks gather raw flow data, answering network-wide
queries is difficult due to:
(a) the distributed nature of data collection (per interface and router)
    at different locations, i.e., at multiple border and/or backbone routers,
(b) the massive and ever-increasing size of the flow data (despite sampling)
    incurring an excessive cost to store, transfer, and analyze flow data--indeed,
    it often has to be deleted after some time to be able to store more recent data,
and
(c) the international footprint with the requirement to comply with local
legislation which may prohibit the transfer of raw data.

To achieve the above, we need data structures that generate
\emph{succinct} and \emph{space-efficient summaries}, as well as
\emph{indexing} of network flow captures that are light (easy to transfer), can be
analyzed locally, and enable answering \emph{interactive a priori unknown network-wide queries}.
These data structures should be used to \emph{accurately} and \emph{quickly}
answer queries and tackle network management tasks that involve \emph{multiple
sites} and/or span \emph{multiple periods} in a \emph{user-friendly} and
\emph{unified} way.\\

\noindent The contributions of our paper are:

\begin{itemize}
\item We design, deploy and evaluate \flowstream, a system built on top of
  existing voluminous network captures, that enables interactive data
  exploration. We show that with \Flowstream the query response time for
  network-wide queries can be reduced from hours or minutes to seconds.

\item We propose a lightweight self-adjusting data structure,
  \flowtree, that inherits the performance of previously proposed hierarchical
  heavy hitter structures for computing flow summaries. \Flowtree summarizes
  elephants as well as mice flows and supports multiple
  operators, such as merge, compress, and diff, to
  summarize information across multiple sites and time periods.

\item We propose an SQL-inspired language, \flowql, which
  provides a unified interface to ask arbitrary ad-hoc queries about flow
  captures, including drill-down queries.

\item We show that when answering a wide range of queries, \flowstream
significantly outperforms the state of the art data analytics systems, namely, ClickHouse, and Spark. 
 
\item We share our experience of rolling out \flowstream at
  different operational environments, namely a large IXP and a tier-1 ISP,
  and showcase how to tackle various network management
  tasks.  We will make \Flowstream and its code available for non-commercial
  use under the following link ~\cite{flowyagergit}.

\end{itemize}


\section{State of the art}\label{sec:related}

Existing network analytics systems, such
as~\cite{Arbor:SIGCOMM2010,caceres2000measurement},
typically transfer the raw traces to
a centralized data warehouse for archiving and processing. However,
transferring the raw traces is increasingly expensive due to the data volume
---e.g., Terabytes of flow data generated in a single day can be out of sync, and
all need to be transferred.
Moreover, additional constraints are posed by national regulations when
networks operate at regions under different jurisdictions:
for example, transferring data that includes user identifiers,
e.g., IP addresses allocated to EU citizens, without their consent,
violates the EU General Data Protection 
Regulation (GDPR)~\cite{EU-GDPR}. Fines are steep, namely up to 4\% of
worldwide turnover or 20 million Euros, whichever is higher.

\noindent{\bf Network monitoring systems:}
Alternative proposals suggest to enable powerful custom data collection per query and
realize this by combining traffic mirroring and deterministic packet sampling.
These include
query-based monitoring such as
Stroboscope~\cite{tilmans2018stroboscope},
network troubleshooting using mirroring~\cite{OFRewind,wundsam2010network},
analysis of in-network packet traces~\cite{Yu2019dShark,teixeira2020packetscope},
as well as monitoring links on-demand as shown
by Gigascope~\cite{Gigascope},
pruning-based solutions such as Cheetah~\cite{tirmazi2020cheetah}
or other SDN-based monitoring, such as \cite{ding2020incrementally} or
\textsc{precision}%
~\cite{basat2020designing}.
The main 
disadvantage of these systems is that the target flows, sites, and periods of
interest need to be known in advance, which is often not the case in practice.

Streaming network telemetry systems, from more classic approaches such as
A-GAP~\cite{prieto2007gap} to the numerous modern solutions, such as
Sonata~\cite{gupta2018sonata}, FlowBlaze~\cite{pontarelli2019flowblaze} or
Poseidon~\cite{zhang2020poseidon},
build on the same ideas but require programmability
from network devices, e.g., P4 switches or FPGA.
These systems assume that users can predefine what is relevant
and optimize the monitoring accordingly, often following a top-down
approach~\cite{yu2019network}. As a consequence, if, potentially, all flows
are of interest, these systems can degrade to ``standard'' flow monitoring which 
for large networks is challenging.
Marple~\cite{marple2017} adds flexibility to network-wide monitoring but requires P4-programmable capabilities that
have not been yet widely adopted in wide-area networks by ISP and IXP operators.

\noindent{\bf Big data analytics systems:} Some operators directly feed their
flow captures into state-of-the-art analytics systems, often based on the
map-reduce principle, e.g., Spark~\cite{Spark} and Hadoop~\cite{CCRLee2013}, or
column-based databases, e.g., ClickHouse~\cite{ClickHouse}. This has scalability issues. Thus, recently proposed big data analytic
systems%
---see \cite{Global-Analytics:2015,WANalytics:2015,viswanathan2016clarinet,hsieh2017gaia,huang2019yugong} as well as
\cite{d2019survey} and references within%
--suggest to use a distributed setup whereby data is locally
preprocessed, e.g., by aggregation or sampling, and then centrally
analyzed. This reduces the need to transfer the raw data.
Note that none of the
above focuses on network management tasks.
Thus, their programming interface follows the map and
reduce paradigm which differs from network operation tasks.
Even though such
systems can provide significant speedup for tasks that can be parallelized, not
all network management tasks may benefit.
Like \Flowstream, such big data analytics systems are \textit{flexible} w.r.t. the
queries supported.
Yet, unlike \Flowstream, they typically are not
\textit{compatible} with existing network monitoring software, do not fully
support principled \textit{aggregation} (over time, space and flows), do not
offer any \textit{history}, and do not give any performance (accuracy or runtime)
\textit{guarantees}.

\noindent{\bf Data summaries--Heavy Hitters:} Previous work on computing
network summaries has focused on how to efficiently compute
heavy hitters (HH)~\cite{Building-Netflow:2004,basat2019randomized,NewDirections:2002,harrison2020carpe}
and
hierarchical heavy hitters (HHH)~\cite{DiamondHHH:2004,ben2017constant,MitzenmacherHHH:2012}
using minimal resources to be able to compute them on the router itself.
These solutions provide an online summary of the (hierarchical) heavy hitters
for a fixed observation window, at one location, and only on
a given subset of the data.
In contrast, to answer interactive
network management queries (see Table~\ref{table:sys-problems}), we need summaries
over different subsets of the data,
per site/router and across sites/routers,
and at many different time granularities, from minutes to days --- or
even months.

Heavy hitters change as data is aggregated: as more data comes in, popularities increase
overall. Consequently, the threshold to be considered a heavy hitter should be raised.
In contrast, some HHH data structures, e.g.,~\cite{ben2017constant} use a single manually
defined absolute threshold (e.g., frequency above $ 1000 $) to characterize heavy hitters,
resulting in a data structure unable to adapt its definition of heavy hitter as the underlying data changes.
\Flowstream builds upon heavy hitter data structures by adding support for
\textit{aggregation} (over time, location, and flows) and adding 
\textit{flexibility} w.r.t.\ the supported queries.

\noindent{\bf Data summaries--Sketches:} Another approach for computing network
summaries are sketches, e.g.,~\cite{cormode2005s,Elastic-Sketch-SIGCOMM2018,ivkin2019know} as well as systems that utilize sketches for network monitoring and debugging~\cite{liu2016one,
huang2017sketchvisor,yu2013software,wellem2019flexible,wang2020fast}.
The capabilities of sketches include counting, top-K, HH, as well as HHH. They are
highly space-efficient data structures that support many types of queries.
Yet, most do not support range queries, e.g., queries that involve a range of
sites and/or time periods. Moreover, extracting an estimate from sketches
is often not time-efficient.
We note that the focus of sketches is
similar to that of HHH, i.e., computing online summaries for a fixed
observation window with minimal resources.
\Flowstream could be built upon sketches but we decided to build upon a HHH
data structure.

\section{\Flowstream Architecture}\label{sec:architecture}

To address the challenges outlined in the introduction, we build a scalable
distributed network data analysis architecture, \flowstream.
Its \emph{input} is existing per-interface network \emph{flow captures}, either
flow summaries---reporting on packet, byte, or flow counts per 5-tuple (src/dst IP address,
src/dst port, protocol)---or packet-level summaries (e.g., trace
sample). We emphasize that we do \emph{not} propose yet another
NetFlow.
Its \emph{output} is network reports including packet, byte, or flow counts
across network sites and time periods. Prime users, i.e., network operators,
can access the data via \flowql, an SQL-inspired query language that returns
results in seconds and, thus, enables interactive ad-hoc queries with drill-down
capabilities. For a comparison between \flowstream and other approaches, we refer to Table~\ref{table:sys-comparison}.

\begin{table}[t]
  {\scriptsize
  \begin{center}
    \begin{tabular}{lB{2em}B{4em}B{2em}B{2em}B{4em}}
\toprule
{}                              & Net. Mon. & Analytics & HHH   & Sketch   & \Flowstream\\
\midrule
Input: Packets                  & \cmark & \xmark & \xmark & \xmark &  \cmark \\
\hline
Input: Flows                    & \cmark & \xmark & \xmark & \xmark &  \cmark \\
\hline
Distributed Queries         & \cmark & \xmark & \xmark & \xmark &  \cmark \\
\hline
Online                          & \xmark & \cmark & \xmark & \xmark &  \cmark \\
\hline
Arbitrary Queries           & \xmark & \cmark & \xmark & \xmark &  \cmark \\
\hline
Query language                  & \xmark & \cmark & \xmark & \xmark &  \cmark \\
\hline
Summarization                   & \cmark & \cmark & \cmark & \cmark &  \cmark \\
\hline
Low Installation Cost       & \cmark & \xmark & \cmark & \cmark &  \cmark \\
\hline
Low Maintenance Cost        & \cmark & \xmark & \cmark & \cmark &  \cmark \\
\hline
Adaptivity to Data              & \cmark & \xmark & \cmark & \cmark &  \cmark \\
\bottomrule
\end{tabular}
\end{center}
}
\caption{Comparison of systems w.r.t. functionality offered. \cmark: full support, \xmark: no support.}
\label{table:sys-comparison}
\end{table}

\begin{figure}[t]
\centering
  \includegraphics[width=1.1\linewidth]{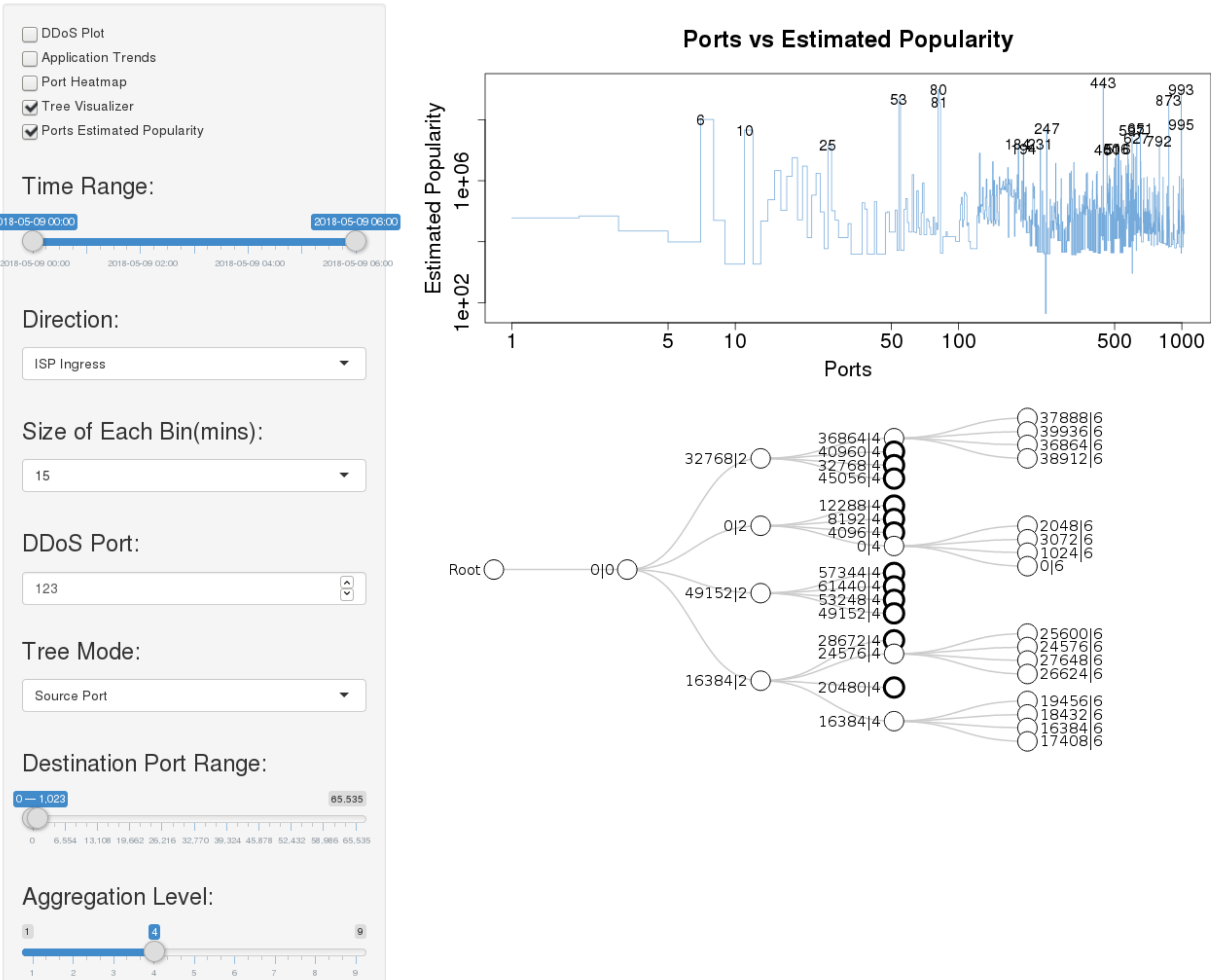}
\caption{\Flowstream: Interacting with 1-feature \flowtrees.}
\label{fig:1d_demo}
\centering
  \includegraphics[width=1.1\linewidth]{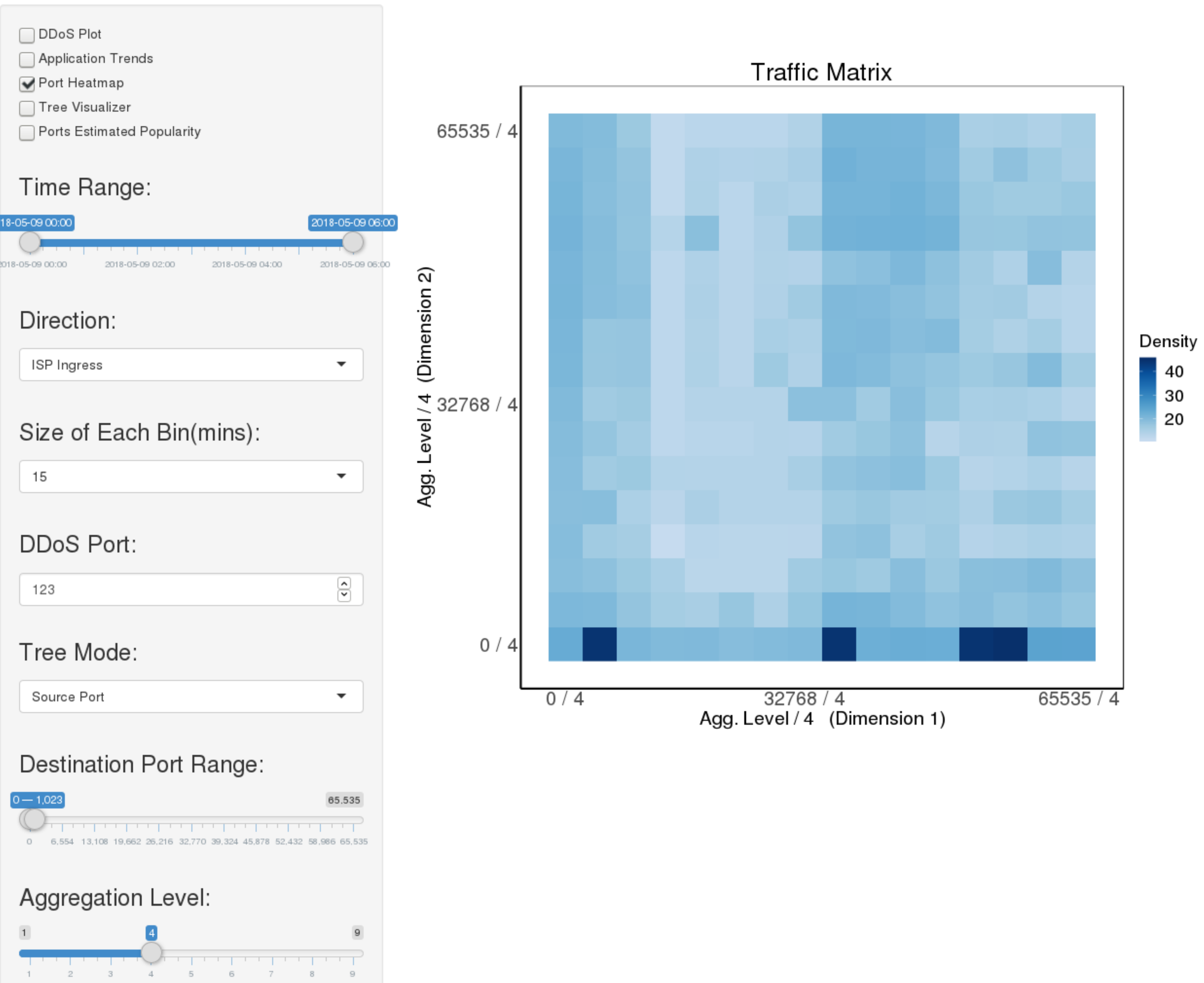}
\caption{\Flowstream: Interacting with 2-feature \flowtrees.}
\label{fig:2d_demo}
\end{figure}

 To underline \flowstream's capabilities for exploring network data, we show in
  Fig~\ref{fig:1d_demo} and Fig.~\ref{fig:2d_demo} screenshots of \flowstream's
  Web interface. The Web interface highlights that searches are possible across
  time ranges, site sets, and feature sets. Moreover, it showcases \flowstream's
  drill-down capabilities that are also visually supported.

\Flowstream is a modular system that consists of three main components:

\begin{enumerate} 

\item \emph{\FlowAGG}, which takes existing flow (or packet) captures as input and
computes flow summaries, using \emph{\flowtrees} (see below), which it stores and
exports.
Besides, \emph{\FlowAGG} may, if it has enough storage, keep a
local copy of the flow captures themselves.

\item \emph{\FlowDB}, which takes flow summaries as
input, stores, and indexes them, while using them to answer \flowql queries.
It can use \FlowAGG internally to compute further flow summaries.

\item \emph{\flowql}, which uses the flow summaries kept within \FlowDB to answer
interactive or batch-style queries including Hierarchical Heavy Hitter/top-K
queries, Above-Thresh queries, or top-K heavy changer queries across time and
sites.

\end{enumerate}

\done{As a module I would also add the Flowyager Operators. (b).. \Flowstream Operators, which are applied to \flowtrees in order to execute queries using these succinct data
structures or integrate.}

\begin{figure} [t]
  \captionsetup{skip=.25em}
  \centering
  \includegraphics[width=1\linewidth]{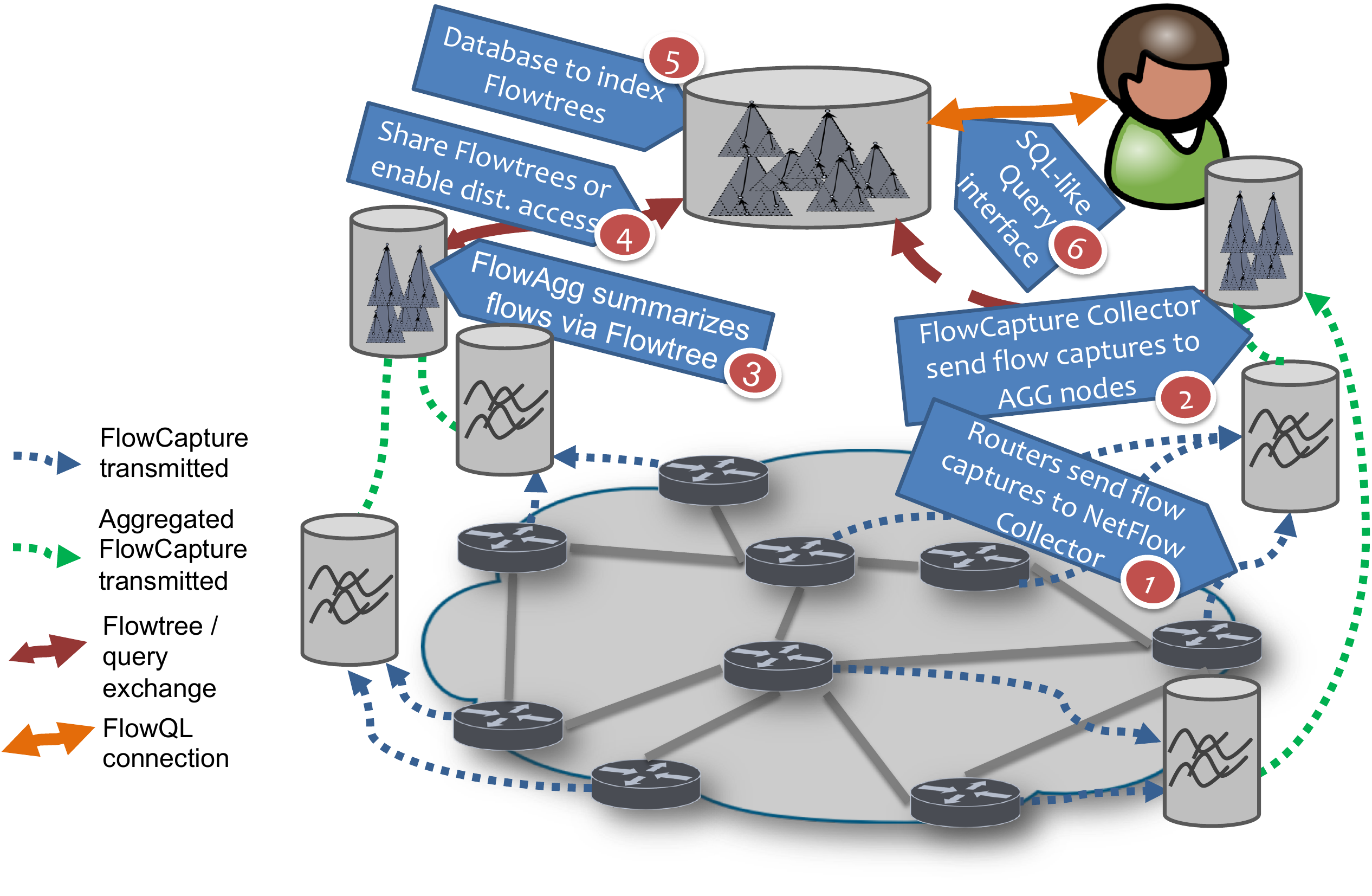}
  \caption{\FlowStream architecture.}
  \label{fig:ft-system-overview}
\end{figure}

\begin{figure} [t]
  \captionsetup{skip=.25em}
  \centering
  \includegraphics[width=1\linewidth]{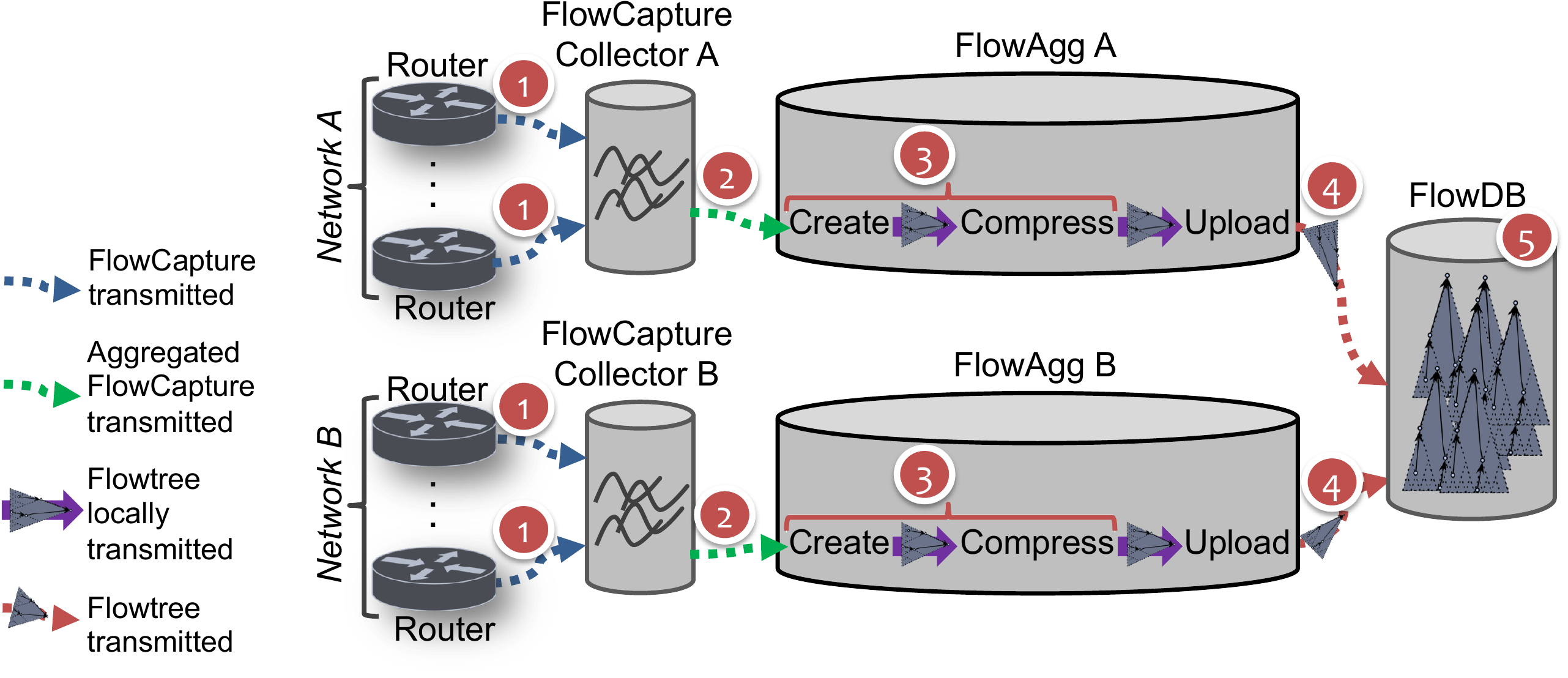}
  \caption{\FlowStream Processing Pipeline.}
  \label{fig:ft-pipeline}
\end{figure}

To better understand the system architecture, Figure~\ref{fig:ft-system-overview}
gives an overview of the overall system, while Figure~\ref{fig:ft-pipeline} presents
\flowstream's processing pipeline.
Each router sends its data to a NetFlow
collector~\textcircled{1}, which forwards it to one of potentially many
distributed FlowAGG instances~\textcircled{2}. Each FlowAGG instance computes
summaries~\textcircled{3} and then uploads these either to another FlowAGG
instance or directly to \FlowDB~\textcircled{4}\footnote{For simplicity we
  restrict our discussion to a centralized
  instance of \FlowDB. However, it is possible to use a hierarchical
  design similar to what has been proposed for
  logs of distributed
  servers~\cite{Akamai-Infrastructure-Monitor,cohen2010keeping}}.
\FlowDB then
processes the summaries~\textcircled{5} and uses them to answer user
queries~\textcircled{6}.

\textbf{\flowtree} is a data summary of a stream of raw flow data that supports efficient 1-d HHH extraction and other
operators. \Flowtrees are the data primitives of \flowstream. Details on the design and implementation of \flowtree data
structure and \flowtree operators are presented in Section~\ref{sec:module-flowtree}.  

\textbf{\Flowagg} uses a separate plug-in, written in C, for each data
source, including IPFIX, NetFlow, sFlow, and libpcap.
  
\textbf{\Flowdb} 
is responsible for collecting and storing the \flowtrees.  It
also provides an interface that the user of the \flowstream can use to answer network-wide queries based on the stored
\flowtrees, \textbf{\Flowql}, whose design is largely inspired by GSQL~\cite{Gigascope} which uses an SQL-like query
language. Using GSQL directly does not suffice due to the unique capabilities of \flowstream. Details on the design and
implementation of \Flowdb are presented in Section~\ref{sec:module-flowdb}.

In total, it took approximately 21k lines of code (LoC) in C and C++ to realize
\flowstream. About 16k LoCs are
for \Flowdb, 1.5k for \FlowAGG, 2.5k for \Flowtree library, and 1k for 
shared components.

\section{Flowtree}\label{sec:module-flowtree}

\Flowtree is the data structure that is used as a data primitive in
\flowstream. Before we dive into the details of \Flowtree and its operators, we
provide background on Hierarchical Heavy Hitter (HHH) data structures.

\subsection{Hierarchical Heavy Hitters}

To enable \flowstream we need succinct summaries from flow captures that are
light to transfer, yet, allow for real-time, interactive queries using different
flow feature sets. A \emph{flow feature} refers to any of the components of a
flow's 5-tuples, namely protocol, src and dst IP, src and dst port. A
\emph{feature set} includes a subset of the possible 5 flow features.

We take advantage of the fact that most of the data on the Internet is skewed
in the sense that Zipf's law~\cite{Zipf,zhang2002characteristics,breslau1999web}
typically applies.  However, flat summaries, i.e., histograms, do not
suffice. Rather, we need hierarchical heavy hitters (HHH)
\footnote{The set of
  HHH for a single hierarchical attribute with popularity counts and a
  threshold $\theta$ corresponds to finding all nodes in the hierarchy such
  that their HHH count exceeds $\theta * N$, whereby the HHH count is the sum
  of all descendant nodes which have no HHH ancestors.}.
  HHH utilize attribute
hierarchies and identify the most popular elements across a hierarchy. For IPv4
prefixes, we use the network prefix length as an obvious feature hierarchy. As
such, 10.1.2.0/23 is the parent of 10.1.2.0/24 and 10.1.3.0/24. For ports, we can
use port ranges, e.g., 80/15 is the parent of 80/16 and 81/16. Each feature
hierarchy, by default, uses a mask. An IP a.b.c.d is part of the prefix
a.b.c.d|$n_1$ and a.b.c.d|$n_1$ is a more specific prefix and, thus, a child of
a.b.c.d|$n_2$ if $n_1 > n_2$. The same applies to ports, whereby, e.g., 0|8
refers to the ports from $[0,63]$. It is possible to define
custom hierarchies, e.g., all Web ports, all DNS ports, or all well-known ports.

Ideally, one would use 5-dimensional hierarchical heavy hitters (5-d HHH),
across all flow features. Unfortunately, this is infeasible due to its
computational complexity~\cite{DiamondHHH:2004,FHH-Cormode:2008}.
Rather, we use 1-d HHH which can be updated in amortized $O(1)$ time per entry
while maintaining the accuracy for HHH and space efficiency of $O(H/\epsilon
\text{log}(\epsilon N))$, whereby $N$ is the number of items processed, $H$ is
the number of hierarchy levels, and $\epsilon$ bounds the
precision~\cite{DiamondHHH:2004,FHH-Cormode:2008}.

\begin{figure}[t]
  \centering
  \captionsetup{skip=.45em}
  \includegraphics[height=3cm]{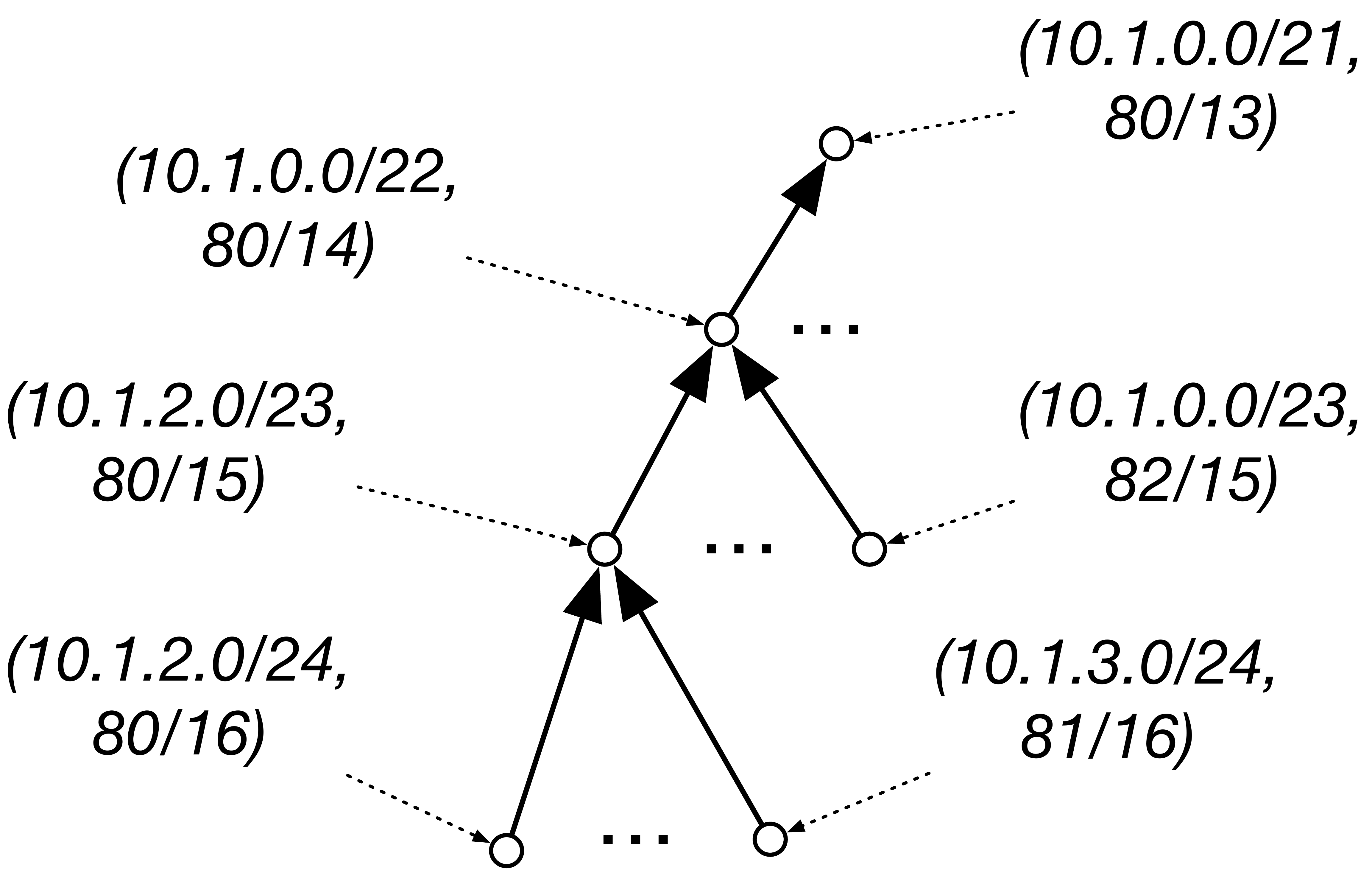}
 \caption{Example: 2-Feature flow hierarchy.}
\label{fig:2f-example}
\vspace{-0.5cm}
\end{figure}

Contrary to previous work, we do not restrict the 1-d HHH to a single flow
feature. Our first key functionality is that we can generalize 1-d HHH by defining a
\emph{joined hierarchy} for a given feature set, e.g., a joined hierarchy for
both dst IP and dst port, whereby, the parent of 10.1.2.0/24|80/16, as well as
10.1.3.0/24|81/16 (IP range|port range) is 10.1.2.0/23|80/15. The parent of
10.1.2.0/23|80/15 is 10.1.0.0/22|80/14 and its great-grandparent is
10.1.0.0/21|80/13. For visualization of a sample 2-f
hierarchy see Figure~\ref{fig:2f-example}.
In effect, we rely on \emph{generalized flows}: Flows summarize related packets
over time at a specific aggregation level.  Possible feature sets include
``4-feature'' flows (i.e., (src IP, dst IP, src port, dst port)), 
``2-feature'' flows, e.g., (dst IP, dst port) (DIDP). 

The joined hierarchy can capture
the correlation of more than one dimension, e.g., the correlation between IP activity and port activity. It allows
identifying heavy hitters on sets of features, and thus, investigating more complex use cases. For example, in an
attack, both the target IP and port are important to investigate the type of attack. In general, any query that involves
multiple features can be potentially benefited by this joined hierarchy.

Our second key functionality is that
if the 1-d HHH data structure supports the 
operators \emph{merge ($\cup$)} and \emph{compress},
we can compute summaries across time and/or space.
In effect, these two operators allow us to add the features
\emph{time} and \emph{location}. Given two data structures, $A_1$ for time
period $t_1$ (location $l_1$) and $A_2$ for $t_{2}$ ($l_2$), we get the joined data
structure by $A_{12} = (A_1 \cup A_2)$. The \emph{compress} operator is especially useful in reducing the memory footprint of the structure. This operator prunes the tree leaves, and if needed the internal nodes, whose contributions are less than some configurable thresholds, and summarizes their contribution to their parents.

Other operators are \emph{diff}, \emph{query}, \emph{drill-down},
\emph{HHH} resp.\ \emph{TOP-k}, \emph{Above-x}
The diff operator is useful to identify changes,
the drill-down operator to explore sub-regions.
The HHH and Above-x operators allow us to find popular feature sets.
The operators are used for interactive queries via \flowql.

\begin{figure*}[t]
\begin{minipage}[t]{0.3\linewidth}
\input{alg/flowtree_build}
\end{minipage}\hfill
\begin{minipage}[t]{0.3\linewidth}
\input{alg/flowtree_stat}
\end{minipage}\hfill
\begin{minipage}[t]{0.31\linewidth}
\input{alg/flowtree_operators}
\vspace*{-.4cm}
\subfigure[Queries]{\label{fig:query-1}
  \includegraphics[width=.34\linewidth]{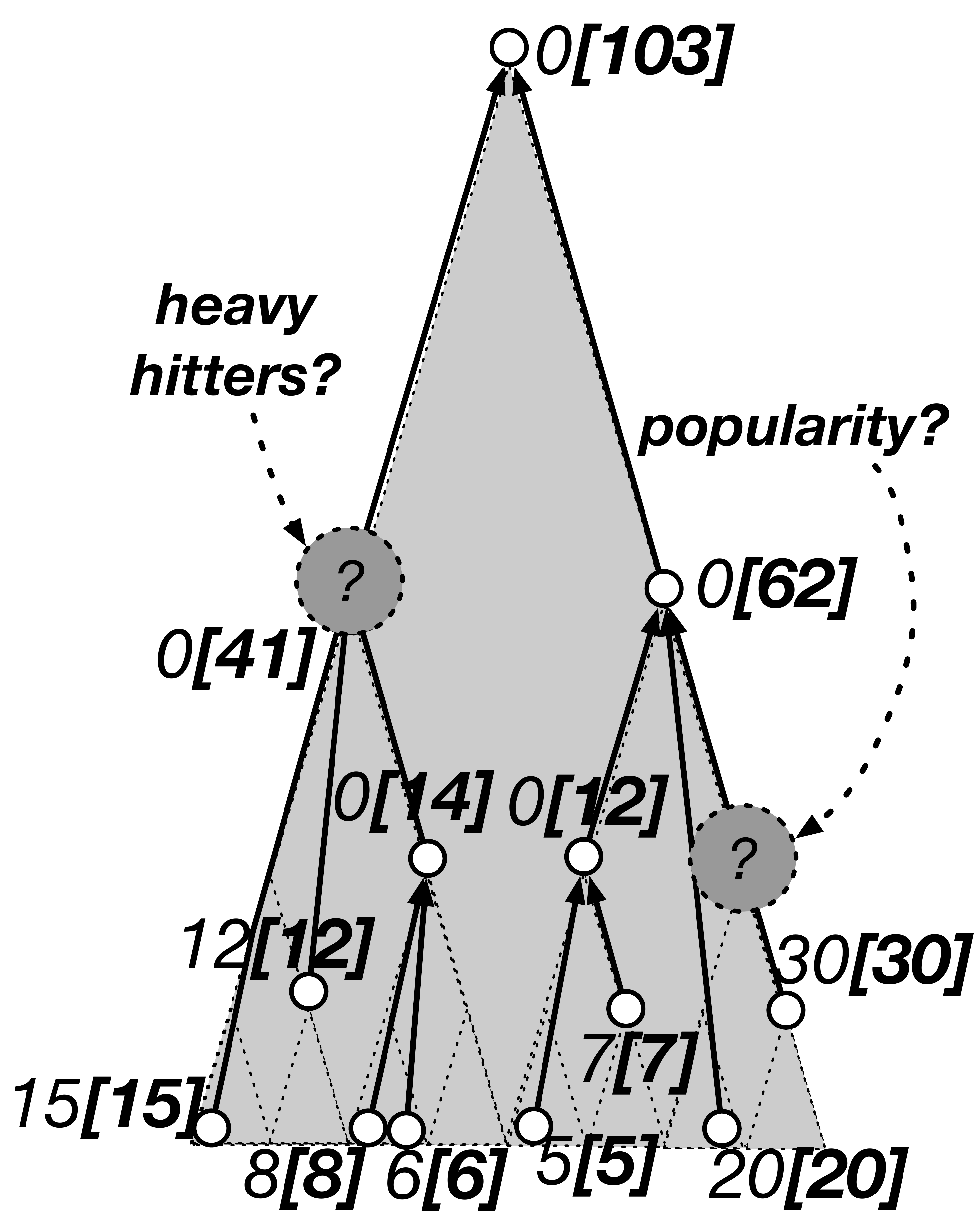}
}
\subfigure[Above-k]{\label{fig:query-3}
  \includegraphics[width=.23\linewidth]{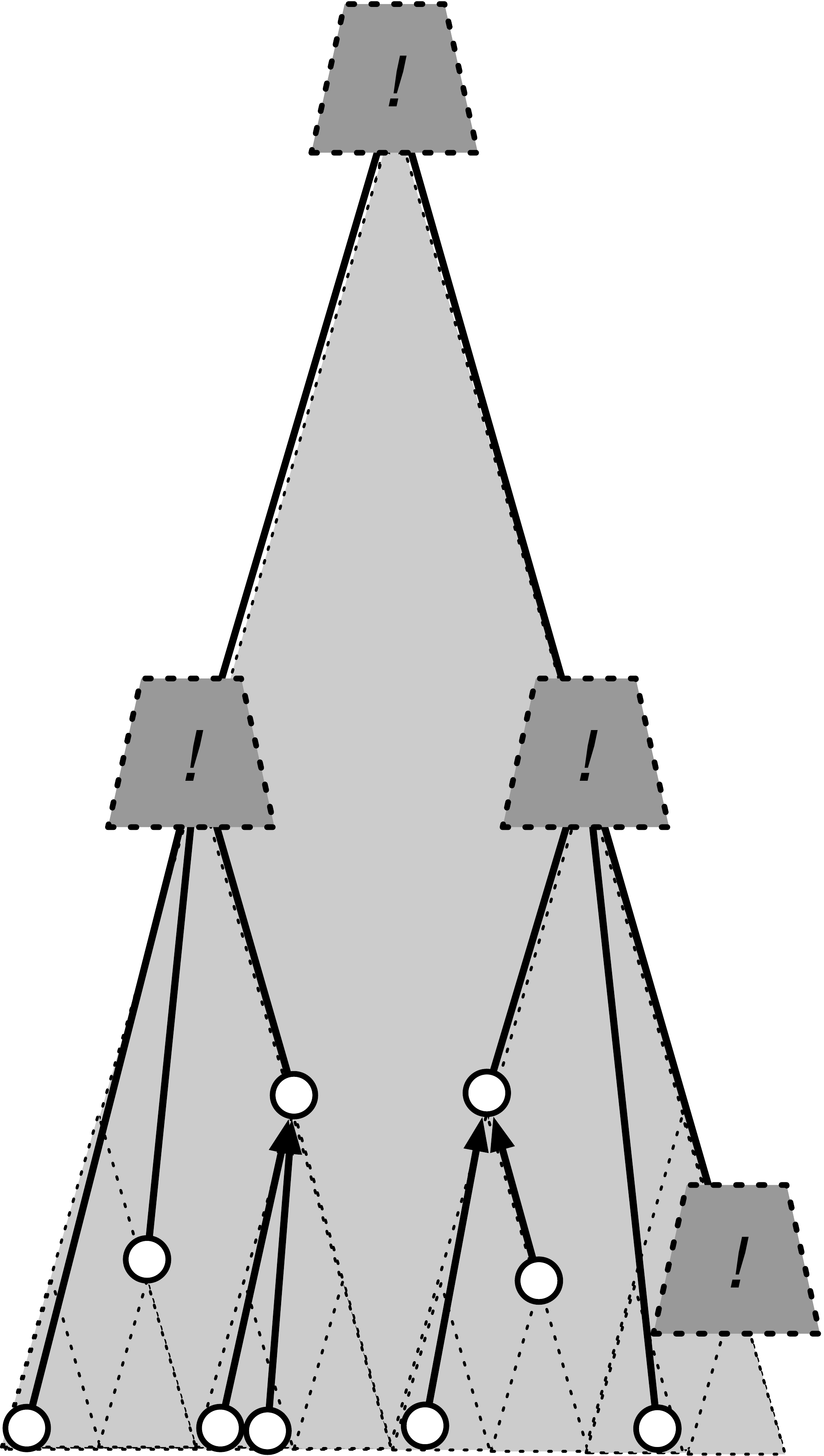}
}
\parbox[b]{1.4cm}{\scriptsize (a) Node in \\\flowtree: 10.\\
  (b) Node not \\in \flowtree:\\ 50\\
  (c) Query: \\Above-29. \\Result see\\ Boxes.}
\caption{\Flowtree queries.}
\label{fig:query}
\end{minipage}
\end{figure*}

\vspace{1cm}
\subsection{\Flowtree Data Structure}
 
After evaluating different 1-d HHH data structures, including those of
Cormode et al.~\cite{DiamondHHH:2004,FHH-Cormode:2008},
Basat et al.~\cite{ben2017constant}, and
Mitzenmacher et al.~\cite{MitzenmacherHHH:2012},
 we decided to augment the structure by Cormode et al.:
 this data structure is self-adjusting and
 its entries can be easily
 extracted via enumeration; thus, it provides natively drill-down capabilities. 
 \Flowstream does \emph{not} intrinsically depend on this data
 structure; rather, it can be built on top of any data structure that supports
 abstract hierarchies and the basic operators.

 \noindent{\bf \Flowtree data structure:}\label{sec:concept}
 Generalized flows form a tree via its hierarchy where each node
 corresponds to a flow. An edge exists between any two nodes $a$, $b$ if
 $a$ is a sub\-node of $b$ in the feature hierarchy, i.e., if $a \subset b$
 ---see Figures \ref{fig:flow-examples-wo-pops-2} and \ref{fig:flow-examples-w-pops-2-app}.
 We annotate each node with its popularities, including packet count, flow
 count, and byte count for UDP and TCP.  The popularity of a node is the sum of
 its own popularity and the popularity of the children---see Figure~\ref{fig:def-pops-2}.


 However, during the construction of the trees, we only keep the nodes'
 ``complementary popularity'', namely the popularity (pop) that is not covered
 by any of the children. Thus, it is possible to prune such a tree by pushing
 the contribution of the pruned nodes to their parent. This is a \emph{key
   functionality} for efficiently updating our self-adjusting data
 structure. \flowtree keeps ``popular'' nodes and prunes ``unpopular'' ones by
 summarizing them at their parent. \flowtree inherits the insertion and
 self-adjusting strategy from Cormode et al.\ but rather than allowing the
 number of nodes to grow unlimited, we limit the maximum number of nodes that a
 tree can contain by repeatedly pruning (compressing) the tree when
 necessary. Still, \flowtree closely matches the excellent performance and
 accuracy bounds for 1-d HHH in terms of space efficiency and precision.

\subsection{Flowtree: Visualizing the Concepts}\label{sec:flowtree-concepts}

\begin{figure}[t]
\begin{minipage}[t]{1\linewidth}
  \centering
  \subfigure[1-feature \flowtree: IP.]{
    \label{fig:flow-examples-wo-pops-1}
    \label{fig:flow-examples-wo-pops}
    \includegraphics[width=.45\linewidth]{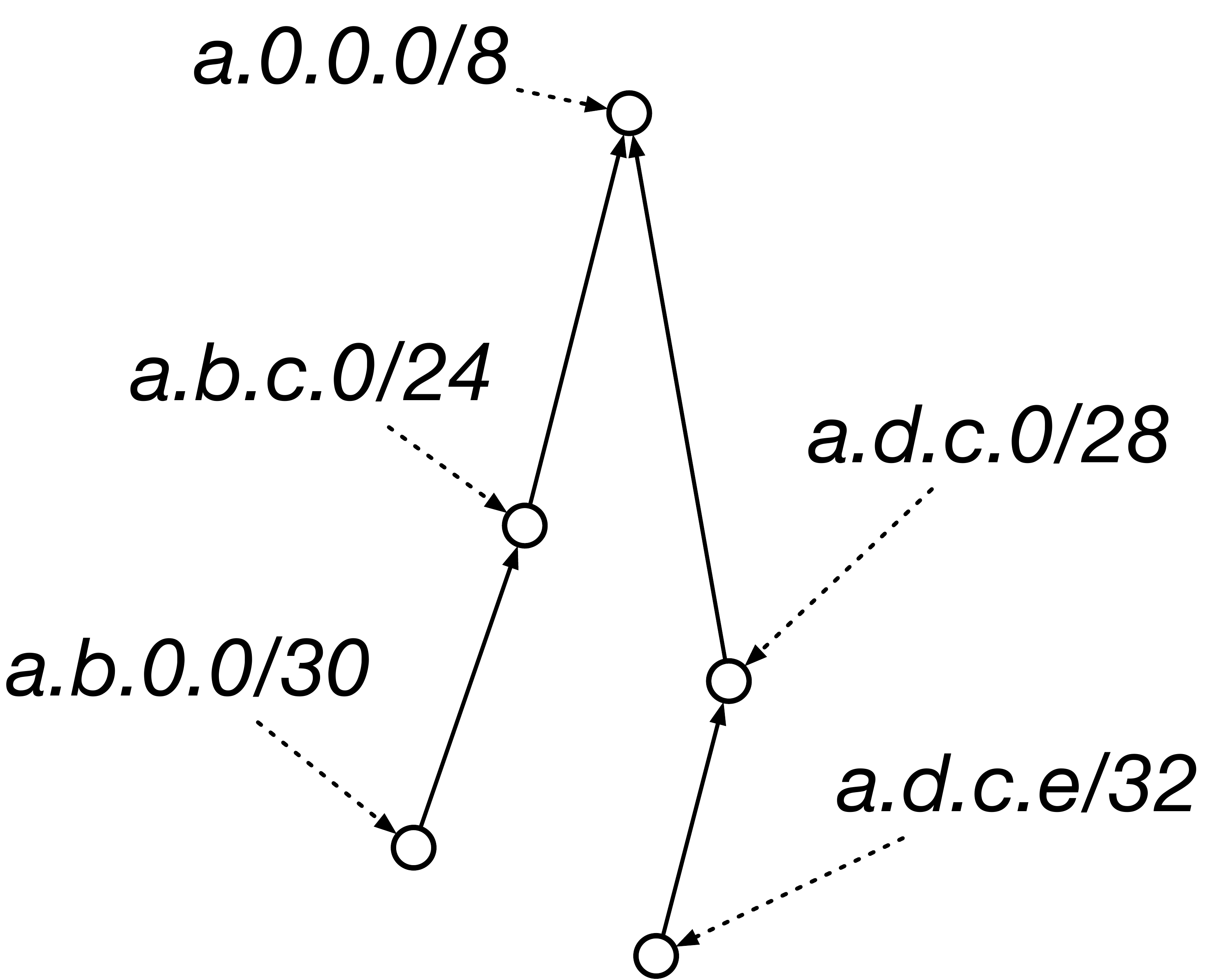}
  }
  \subfigure[Comp\_pop]{\label{fig:def-pops-1}
    \includegraphics[width=.2\linewidth]{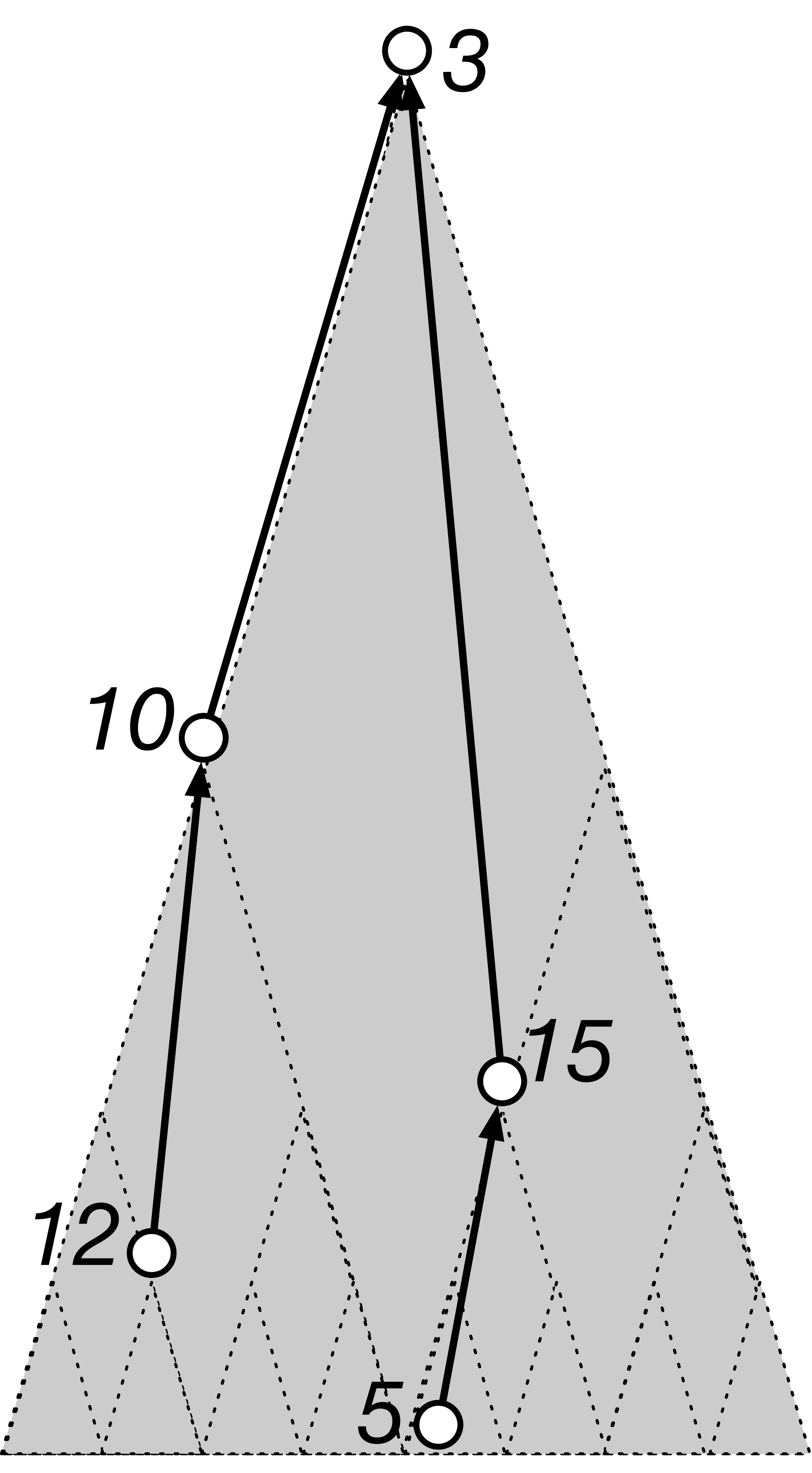}
  }
  \subfigure[Popularity]{\label{fig:def-pops-2}
    \includegraphics[width=.2\linewidth]{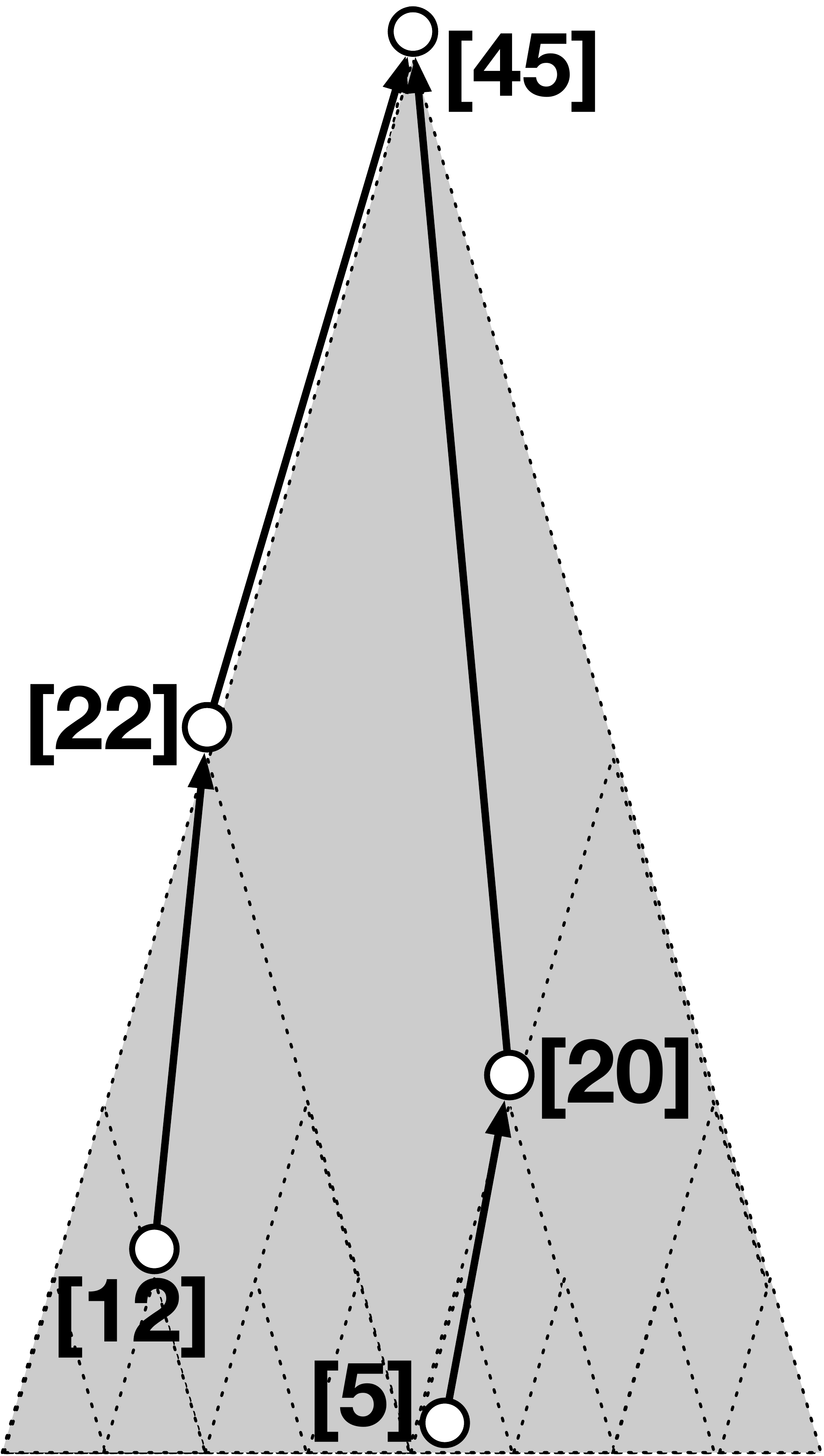}
  }
  \vspace*{-.5em}
  \caption{\Flowtree concept.}
  \label{fig:flow-examples-flowtree-appw-pops-app}
\end{minipage}
\end{figure}

\begin{figure}[t]
\begin{minipage}[t]{1\linewidth}
  \centering
  \subfigure[4-feature \flowtree: \newline src/dst IP and port.]{
    \label{fig:flow-examples-wo-pops-2}
    \includegraphics[width=.45\linewidth]{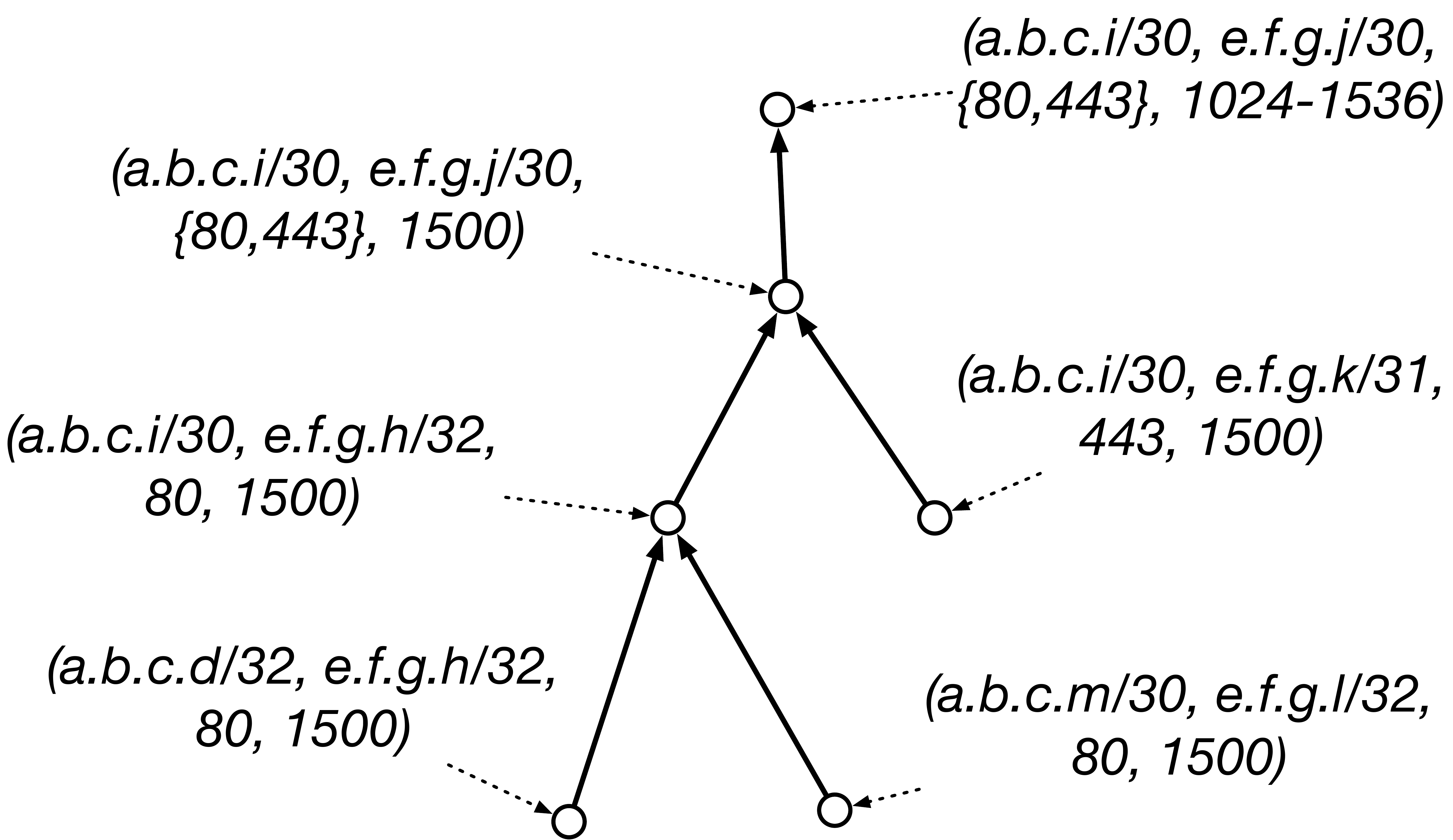}
  }
  \subfigure[4-feature \flowtree with \newline complementary\_pop.\ and pop.]{
    \label{fig:flow-examples-w-pops-2-app}
    \includegraphics[width=.45\linewidth]{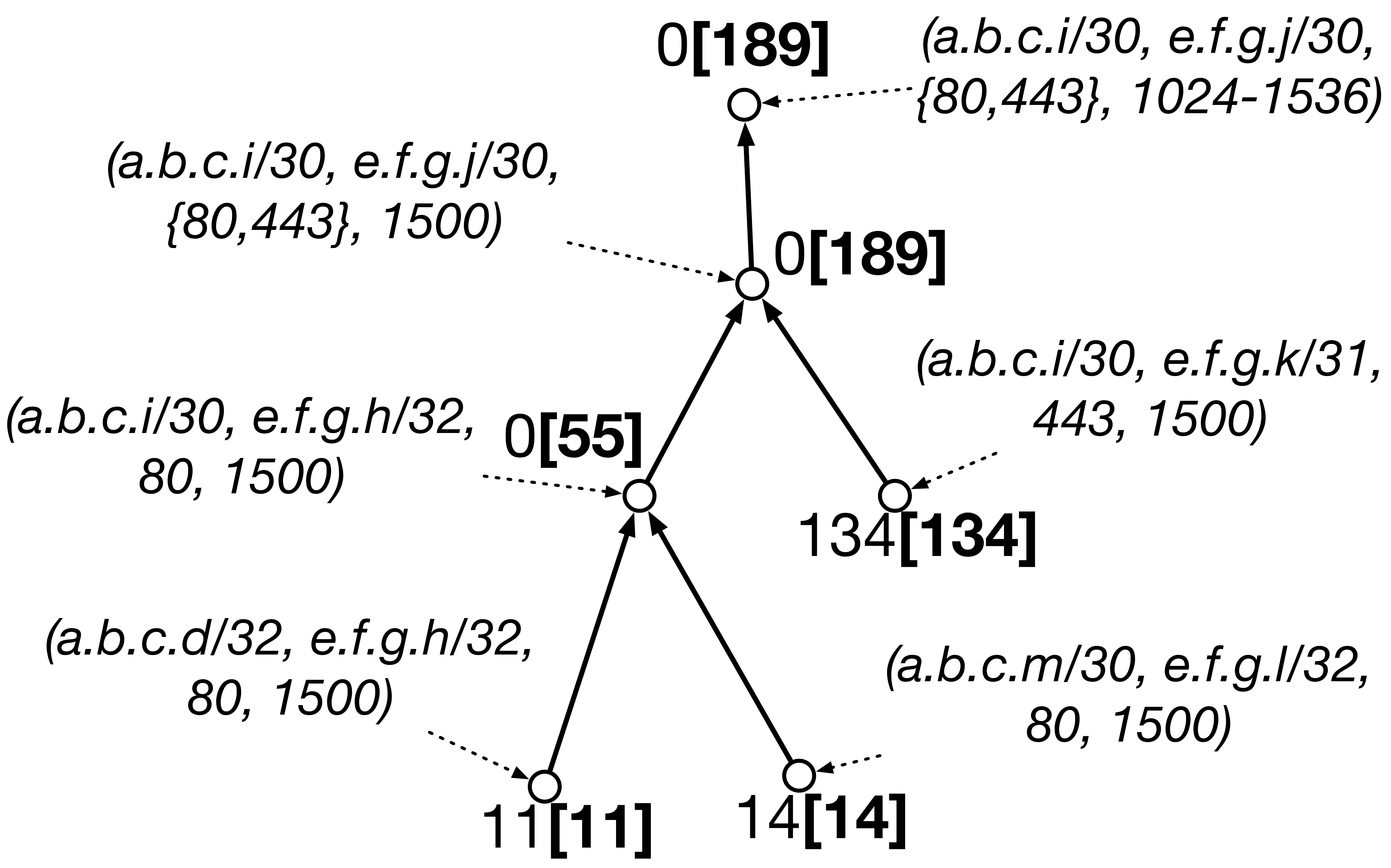}
  }
  \vspace*{-.5em}
  \caption{4-feature \flowtree.}
  \label{fig:def-pops}
\end{minipage}
\end{figure}

We start with the visualization of the differences between popularities and
complementary popularities in Figure~\ref{fig:def-pops}.  Next, we show the two
different feature hierarchies, namely a 1-feature hierarchy on IP addresses, and
a 4-feature hierarchy on src/dst IP addresses and src/dst ports with and without
popularities, see Figures~\ref{fig:flow-examples-wo-pops}
and~\ref{fig:def-pops}.


Initially, a \flowtree has exactly one entry---the root. When adding a node, we
add a new leaf node if necessary and a subset of the nodes on the path to the
first existing parent,  (in the worst case the root) and update the statistics of
the leaf node. We call these intermediate nodes as internal nodes.  Thus, each node maintains the complementary popularity
(comp\_pop), the popularity (pop) that is not covered by any of the children, see Alg.~\ref{alg:build-flowtree}.  Popu\-la\-ri\-ties are
computed from the complementary popularities by summing the complementary
popularities of all nodes in its subtree including its own. This can be done
via a depth first search in O(\# nodes) time, see Alg.~\ref{alg:compute_stats}.
This uses two functions for finding parents of a node. parent(node) 
refers to the direct parent in the feature hierarchy while find\_parent(node) refers to the
parent in the \flowtree. 

Updating an existing node corresponds to finding it, which takes time O(1) using
an appropriate hash-map. Adding a new node may take up to O(\# hierarchy level)
time (using an appropriate hash-map). Yet, the expected number of new nodes is
small if the distribution of the data is skewed.

To limit \flowtree memory footprint, we periodically or on demand, delete nodes with low
popularity. We first compute the popularities by using the stats function in Alg.~\ref{alg:compute_stats} and
then prune nodes whose complementary resp.\ absolute popularity are below an adjustable
threshold. This ensures that at any time the number of nodes in a \flowtree is
proportional to the number of processed flows resp.\ less than a predefined
maximum.  The complementary popularity of a deleted node as well as its children
are pushed to its parent. The overall cost of such a compression step is O(\#
nodes).
Note that since only nodes with small popularity are deleted, the
complementary popularity of an interior node is a good estimate of the
cardinality of the contributing flow set. 
Finally, to control the rate of the growth of the tree and preventing the frequent addition and deletion of internal nodes, we insert the internal nodes with a probability of $p$. The default value of $p$ is 0.3.

 \vspace{-0.4cm}
\subsection{\Flowstream Operators}\label{sec:flowyager-oper}

\noindent{\bf Query and drill-down:}
The base operators are \emph{query} (see Fig.~\ref{fig:query}) and
\emph{drill-down}.
If the feature $f$ is 
a node in the \flowtree, the answer is computed from the node statistics. Otherwise,
we find the potential node, $q$, that corresponds to $f$ and
estimate its popularity based on the popularity of the predecessor of $q$,
$p$, and its children, $C$. We split the children into two subsets:
$C_f$ and $C_o = C - C_f$, whereby $C_f$ includes those that are a subset of
$f$ in the hierarchy.
Now, $\sum_{c \in C_f} \text{pop}(c)$ is a lower
bound for the popularity of $f$ 
and two estimates of $f$'s popularity are
$\text{pop}(p) - \sum_{c \in C_o} \text{pop(c)}$ or $\text{comp\_pop}(p) +
\sum_{c \in C_f} \text{pop}(c)$, see Fig.~\ref{fig:query}. If the feature set
does not correspond to a node $p$, the query is expanded to a tree-walk
starting at the smallest possible parent of $p$.
The output 
of the query are then all nodes and their popularities that match the input
feature set.
For example, src\_ip = a.b.0.0|16 and src\_port = 80|16 start at node
(a.b.0.0|16,80|8) and outputs only the nodes where src\_port is 80 and
src\_ip is a subprefix of a.b/16. Drill-down queries retrieve the children of
a node. 
Note that we can derive estimates for all flows, from mice to elephants:
even for low-popularity nodes, the number of flows remains a good
estimate for the number of contributing flows.

\noindent{\bf Above-t:} Results in a tree-walk and all nodes whose popularity are
above the threshold value are returned.

\noindent{\bf Top-k :}

To compute the top-k, we identify
the \flowtree entry with the largest popularity, delete its contribution, and
then iterate. Hereby, we use a priority queue.

\begin{figure}[t!]
\subfigure[Merge]{\label{fig:merge}
  \includegraphics[height=1.7cm]{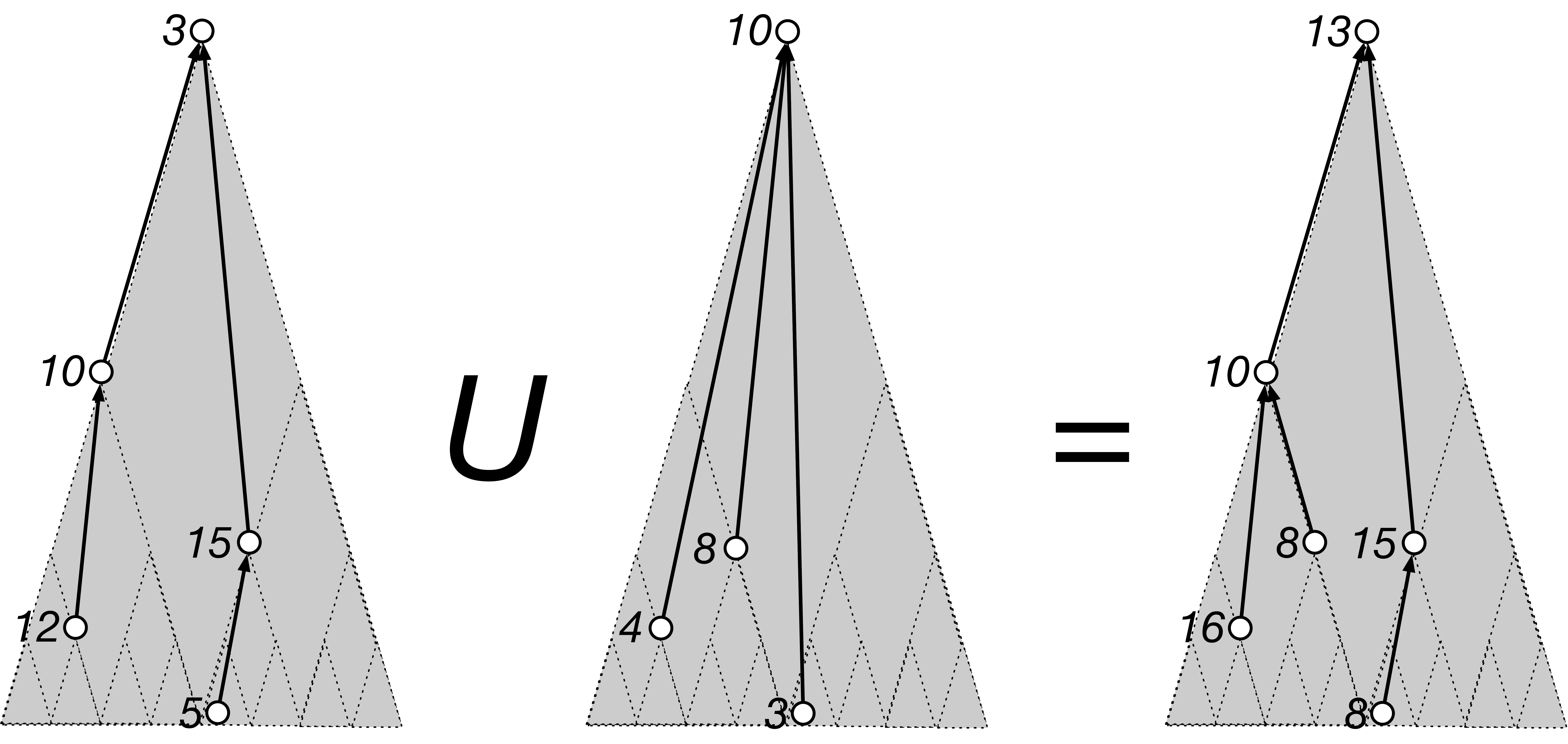}
  }\hfill
\subfigure[Diff]{\label{fig:diff}
  \includegraphics[height=1.7cm]{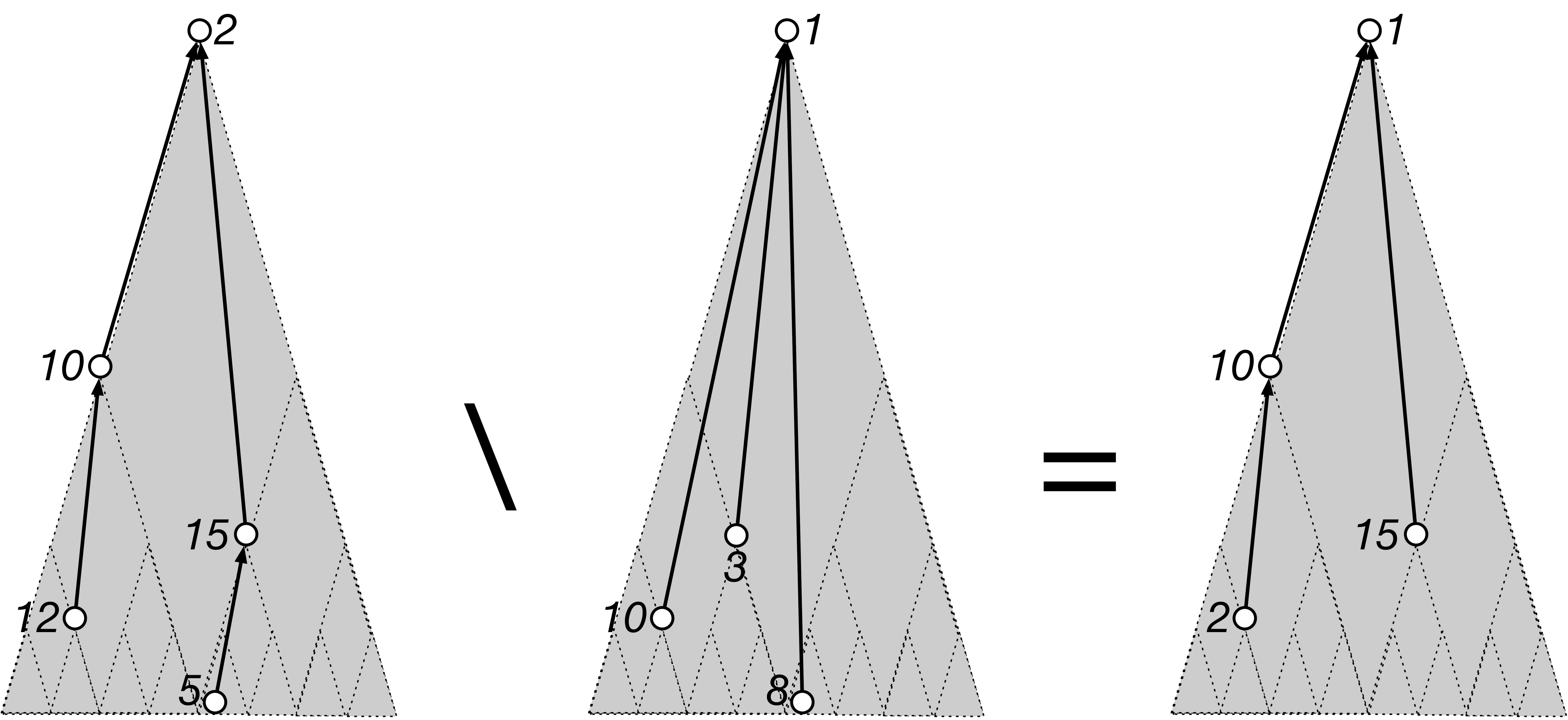}
}
\caption{\Flowtree Operators: Merge and Diff}
\label{fig:operators:merge-diff}
\end{figure}

\noindent{\bf Merge:}
We merge two \flowtrees by adding the nodes of one to the other. Note that
the update will only be done for the complementary popularities---
see Alg.~\ref{alg:merge} and Fig.~\ref{fig:merge}---, with missing nodes
being assigned a popularity of zero.
The statistics have to be recomputed and, to reduce the memory footprint, we
compress the joined tree.
If the total absolute contributions of the two trees differ significantly,
one should rescale the complementary popularities of the trees before
merging.

\noindent{\bf Diff and HeavyChanger:} Just as one can merge \flowtrees, one
can also compute the difference between two trees.  This is a merge
operation with subtraction instead of addition---see Alg.~\ref{alg:diff} and Fig.~\ref{fig:diff}.
Heavy changers are detected by using Top-k on diff of the two trees.

\flowtrees maintain counters for various features of the flows. In the current implementation, we use counters
for packet, byte and flow counts. This structure supports cardinality-based queries but is limited to the elements (features) already in the
tree (nodes). It is possible to maintain additional counters and support additional cardinality-based queries, e.g.,
using counters for ports, but at the cost of requiring additional space. In some cases, this is necessary. For example,
such cardinality-based queries will enable the detection of non-volumetric attacks, e.g., semantic attacks. By
allocating more space and maintaining more counters, it is possible to detect different types of attacks, e.g., ``slow''
DDoS attacks (Slowloris). We plan to explore the accuracy of cardinality based queries and the effect of allocating more
space and maintaining more counters in \flowtrees as part of our future in future work.

\section{FlowDB}\label{sec:module-flowdb}

\begin{figure} [t]
  \captionsetup{skip=.25em}
	\centering
	\includegraphics[width=.9\linewidth]{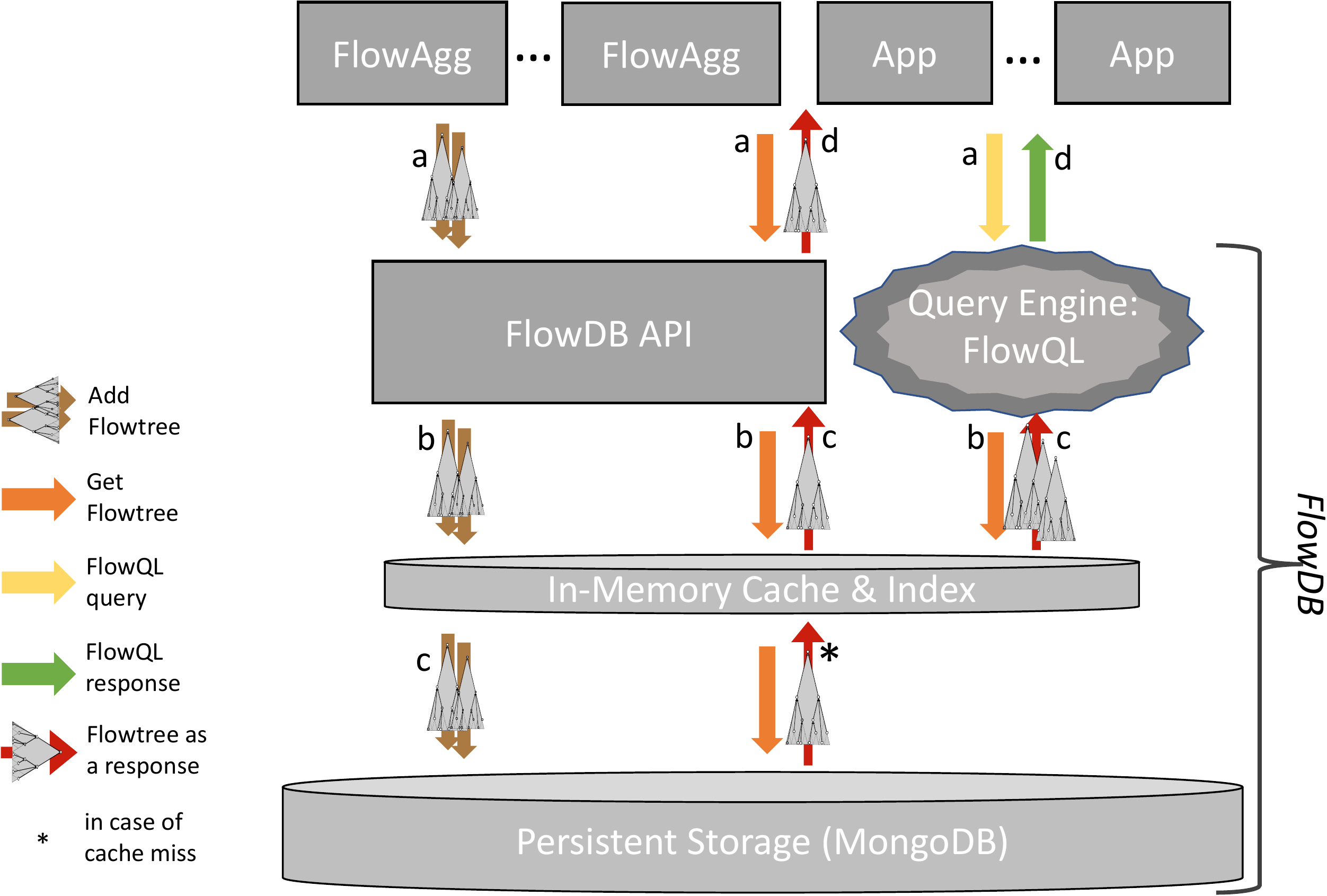}
	\caption{\flowdb overview.}
	\label{fig:fdb-architecture}
\end{figure}

\Flowdb collects and stores \flowtree summaries computed by \flowagg in
persistent storage.  Each \flowtree has a unique key that is made from its timestamp which along with its granularity reflects a time interval, the id of the site/location, and its feature-set.
The values are the \flowtrees, which are stored as byte buffers.
Figure~\ref{fig:fdb-architecture} visualizes \flowdb's architecture.

\subsection{\Flowdb Implementation}\label{subsec:flowdb-impl}

Currently, our database of choice is MongoDB~\cite{mongodb} because it is
lightweight, although any other key-value datastore can be used. To accelerate
query processing, we use an in-memory index and an in-memory cache. The
in-memory index is a collection of T*-trees that track \flowtrees and enable
range queries over different time periods. The in-memory cache uses a least
recently used (LRU) policy to keep recently added or queried trees in
memory. \Flowdb is designed with parallelization in mind: it is capable of
receiving multiple streams of \Flowtrees from multiple \FlowAGG daemons while
answering queries to multiple users at the same time. Parallelization is employed in performing major tasks such as  
	handling requests from \FlowAGG daemons and remote API calls, storing \flowtrees in persistent storage, and query processing. 
Upon receiving a query, the system first checks whether the queried trees are
in memory. In case of cache misses, it retrieves trees from storage.

The system is highly configurable in terms of memory usage, by setting a
maximum number of \flowtrees in memory, cache eviction interval, degree of
parallelization, etc. The maximum number of \flowtrees in memory controls the memory footprint of \flowdb.
To access the database, \flowdb offers both an API with
the services Add \Flowtree and Get \Flowtree and an interface for \flowql. FlowAGG
and other components of \Flowstream use the Apache Thrift Remote Procedure
Call (RPC) framework~\cite{Apache-thrift} for communication.

To enable \emph{Geo-Distributed Query Execution}, the in-memory index 
keeps track of whether a \flowtree is stored locally or at a remote
\Flowdb. Thus, if necessary, all remote \flowtrees can be fetched via the
\Flowdb API to answer a \flowql query.  In our planned geo-distributed query execution, we partition site-IDs and map a site-ID to a \FlowDB instance. Once a \FlowDB instance receives a query, it will check whether the given site-ID is stored locally. If the required \Flowtree is not stored locally, it can issue a request to the target \FlowDB instance and retrieve the \Flowtree. Once the \Flowtree is retrieved, it will be merged with the \Flowtrees that are already present and the intended query is fulfilled.  The evaluation of this feature is beyond
the scope of the current manuscript.

\subsection{\Flowql Query Language}

To realize \flowql, we took inspiration from SQL keywords, yet we developed our
own grammar. We used ANTLR~\cite{ANTLR4} to generate the parser for the grammar.
We offer an interactive command-line shell as well as a graphical user interface
using R~shiny~\cite{shiny}--cf. the screenshots from Fig~\ref{fig:1d_demo} and Fig.~\ref{fig:2d_demo}.
More specifically, with \flowql the user chooses their operator via a
\texttt{SELECT} clause, one or multiple time periods via a \texttt{FROM}
clause, and the feature set via a \texttt{WHERE} clause.

\begin{description}
\item [SELECT:] specifies the answer type. Allowed values include
  \textit{`pop'} for popularity or flow/byte/packet count,
  \textit{`top-K'} for the top-k most popular flows,
  \textit{`HHH-P'} for the 1-d hierarchical heavy hitters with flow counts above P\% of total traffic,
  \textit{`hc-K'} for the top-k heavy changers,
  \textit{`above-T'} for all flows with popularity above $t$, and \textit{`*'}
  for all flows satisfying the WHERE clause.
\item [FROM:] specifies one or multiple time periods.
\item [WHERE:] selects the feature sets and one or multiple conditions.
	Possible feature elements are \textit{site\_id}, \textit{src\_ip}, \textit{dst\_ip},
	\textit{src\_port}, \textit{dst\_port}, \textit{proto}.
	Possible values are ANY or any region, IP prefix, or port range (using the IP|mask resp.\ the
	port|portmask syntax). Combinations are feasible via (AND, OR, and ()).
\end{description}
\noindent Thus, \flowql queries have the following syntax:\\
{\scriptsize
\noindent{\hspace*{.25cm}\texttt{SELECT [pop, top-k, hc-k, above-t, hhh-k, *] \\
		\hspace*{.25cm}FROM (time YYYY-MM-DD hh:mm to YYYY-MM-DD hh:mm)+\\
		\hspace*{.25cm}WHERE ([Conditions via AND, OR, ), (, feature = value])+}}
}

Using \flowql, we found that we often wanted to repeat the same query across
multiple time bins or sites. Thus, we added two iterators:
\texttt{answer-bin-x} that iterates across time bins of size x minutes
and
\texttt{site\_id=ITR-x|n} that iterates across all sites
within a site set, specified with an interval, e.g., $[x, x+2^n-1]$, or
using a pattern.

To be able to drill-down and inspect a specific time-range in more detail, we additionally provide drill-down queries. In a drill-down query, a particular granularity in which one desires to inspect the traffic should be specified. For instance, to see the result of a query in 15-minute time bins, one should specify \textit{bin15} in the query.

\subsection {Query Execution}\label{subsec:query-execution}

Upon receiving a FlowQL query, first, the WHERE clause is converted into a Disjunctive Normal Form. This results in breaking down the current query into smaller queries, which we call \textit{mini-queries}.\\
Each mini-query is then processed independently. For each mini-query, the corresponding trees are fetched considering the time-range, granularity, and feature sets.
For instance, for a query requiring \textit{src\_port=X}, 1-feature trees, \textit{SP} in this case, are fetched.
In a non-drill-down query, trees with the highest granularity existing in FlowDB are fetched. For a drill-down query, trees with the granularity specified in the query are fetched. If the specified granularity does not exist in FlowDB, multiple lower-granularity trees are merged using the MERGE operator to build trees with the specified granularity.
Consider the following query which asks for bin-30:\\
{\small
\noindent{\hspace*{.25cm}\texttt{SELECT pop(any,byte,bin30) FROM (time 2018-05-09 00:00 to 2018-05-09 23:59) WHERE site\_id=ANY and src\_port=X}}
}\\
This is a drill-down query to zoom into a full-day time-range in half-an-hour bins. 
Now assume that there are no 30-min granularity trees in FlowDB for the specified time-range, but there are 15-minute granularity trees. Then for each time-bin, two 15-minute trees will be merged to build the required granularity.\\
If the number of trees to be merged is large, the merge operation is performed in parallel to speed up the merge process.  In a heavy changer query, two time-ranges should be provided and the trees fetched for each of these two time-ranges are diff'ed using the DIFF operator.\\
Then, the final trees are processed using different Flowtree operators to fulfill the query conditions, e.g. \textit{src\_port=X}. If the query is \textit{pop}, knowing the popularity is as easy as finding the corresponding node in the tree and returning the popularity value. If the node is not in the tree, an estimation using the parent's popularity is returned as previously described in ~\ref{sec:flowyager-oper}. \\
If the query is \textit{above-T}, ABOVE-T operator with threshold T is used. For the \textit{top-K} and \textit{hhh-P}, the TOP-K operator will be used. In top-K, it should return the top K flows with any non-zero popularity. In hhh-P, P is the threshold for the fraction of total contributions.\\

\section{Experimental Deployments}
\label{sec:data}

We rolled out and tested \flowstream in three different types of networks,
namely a large European IXP (\IXP), a tier-1 ISP (\ISP), and our testbed using
a sample dataset (\MAWI)---see Table~\ref{tab:deployment-summary} for an overview.
In this paper, we report on experiments on stored data that we use for
reproducibility.  At two locations, the \IXP and the \ISP, we are
in the process of moving towards live data import after extensive testing on site.

\begin{table}[t]
\centering
  \captionsetup{skip=.25em}
{\footnotesize
  \begin{tabular}{|C{0.8cm} |R{1.6cm} R{1.0cm} R{.8cm} R{.8cm} R{.7cm}|}
\hline
\bf{Dataset} & \bf{Time range} & \bf{\#Interface} & \bf{Input Size} & \bf{Type} & \bf{Time bin} \\
\hline
IXP & Sep'19 1--7 & $\approx$ 1,250 & $\approx$~10TB & Flow & 15m \\
\hline
ISP & Apr'19 1--2 & $\approx$ 1,300 & $\approx$~25TB & Flow & 15m \\
\hline
MAWI & May'18 9--10 & 2 & $\approx$~1TB & Packet & 1m \\
\hline
  \end{tabular}%
  }
\caption{Deployment overview: \IXP, \ISP, and \MAWI.}
\label{tab:deployment-summary}
\end{table}

\noindent{\bf Ethical considerations:} We are fully aware of the sensitivity of
network data and, therefore, only work with a subset of the packet header
information, namely src IP, dst IP, src port, dst port, protocol, whereby all
IPs have been consistently anonymized per octet (bijective substitution using a hash function), even though this may negatively
affect prefix aggregation.  Note that the live operational deployment of
\flowstream will not require such anonymization.

	\begin{table}
	\begin{center}
		\scriptsize
		\begin{tabular}{|l|c|c|c|c|c|}
			\hline
			Short Form & Meaning\\
			\hline
			\hline
			SIDI & src IP and dst IP  \\
			\hline
			SPDP & src port and dst port \\
			\hline
			SISP & src IP and src port \\
			\hline
			SIDP & src IP  and dst port \\
			\hline
			DISP & dst IP and src port \\
			\hline
			DIDP & dst IP and dst port \\
			\hline
			SI & src IP  \\
			\hline
			DI & dst IP  \\
			\hline
			SP & src port \\
			\hline
			DP & dst port \\
			\hline
			FULL & src IP,  dst IP,
			 src port, and dst port \\
			\hline
		\end{tabular}
	\end{center}
	\caption{Overview of the feature sets of Flowtree.}
	\label{table:trees:feature_sets}
\end{table}

\noindent{\bf \IXP Dataset:}
This dataset consists of IPFIX flow captures at one of the largest Internet
Exchange Points (IXPs) in the world with more than 800 members and more than
8~Tbps peak traffic. The IPFIX flow captures are based on random sampling of 
1~out of 10k packets that cross the IXP switching fabric. The anonymized
capture includes information about the IP and transport layer headers, as well as packet
and byte counts. To evaluate the system at real-world scales, we included all
sites during the first week of 
September 2019. Each site corresponds to the router interface of an IXP member
connected to the IXP's switching fabric.

We deployed \Flowstream within a virtual machine (VM) on
a server at the IXP's premises.  The VM is assigned 400 GB of memory and 40 threads
on a machine with two Intel-Xeon-gold 6148 CPUs each with 40 threads. 

\noindent{\bf \ISP Dataset:} This dataset consists of approx.\ 1,300 NetFlow
streams (one per interface) from a major tier-1 ISP.  We receive 
NetFlow data from 40 routers located in 30 cities
in 4 European countries, as well as the US. The ISP's internal systems preprocess
the raw NetFlow streams into 26 separate ASCII data streams.  
The NetFlow packet sampling is identical across all the routers. We
include all data from Apr.\ 01, 2019 (00:01:00 UTC) to Apr.\ 03, 2019 (02:01:00 UTC).
We deployed \flowstream as a Docker container with 94 GB memory and 32 threads
on a machine with two Intel Xeon E5-2650 CPUs.

\noindent{\bf \MAWI Dataset:} This dataset consists of packet-level capture
collected at the transit 1~Gbps link of the WIDE academic network to its
upstream ISP on May~9-10, 2018. Each packet capture lasts for 15~mins and
contains around 120~M packets. The anonymized trace is publicly
available~\cite{Mawi} and we use it to be able to release sample queries and
results.  We interpret each direction as a site. For this dataset, we deployed
\Flowstream on a testbed machine, with 128 AMD-EPYC 7601 CPUs and 1.5TB memory.

\noindent{\bf \flowstream setup:} In terms of the basic setup for the \flowstream
evaluation, we choose fixed time periods rather than a fixed number of
flows. The advantage of the former is that we can easily summarize across time
and that we can even look at coarser time granularities. The advantage of the
latter is a constant number of entries 
to summarize. We choose the former rather than the latter as summarizing and
investigating across time are typical network operator tasks. We
keep \flowtrees for every 15~minutes for every site for the \IXP and
\ISP datasets and 1~minute for the \MAWI dataset. We generate 11~different
feature trees, namely all four 1-feature trees, all six 2-feature trees, and a
4-feature tree, see Table ~\ref{table:trees:feature_sets} for the details. By default, we limit each \flowtree to 40k nodes. 
1-feature port \flowtrees are limited to 10k nodes. In addition, we generate
aggregated trees for 15~minutes, 1~hour, 1~day, and 1-week time granularities,
each with at most 40k nodes. This results in one tree per site for each time
granularity and a single tree for all sites for each time granularity.

\noindent{\bf Big data analytics setup:} We compare \flowstream's performance
with \emph{task-specific data-parallel Python scripts}, as well as installations
of a prominent big data analytics platform, namely \emph{Spark}~\cite{Spark},
and a column-based state of the art database, namely
\emph{ClickHouse}~\cite{ClickHouse}. Each installation was done on the same VM
as \flowstream. Note that this implies that Spark was not deployed on a physical
cluster of machines but in a multi-threaded environment.

\section{\Flowstream Prototype Evaluation} \label{sec:flowstream_prototype_evaluation}

Next, we describe our experience with deploying \flowstream, which
we will make publicly available for non-commercial use.  Our evaluation
highlights the four main strengths of \Flowstream: reduced storage
footprint, low transfer cost, rapid response to a wide range of queries,
and high accuracy.
Since these characteristics are related to our choice of underlying data
structure and its resp.\ parameters, we start by evaluating \Flowtree---
the current basis of \flowstream.

\begin{figure*}[t!]
\centering

\begin{minipage}[t]{0.45\linewidth}
\subfigure{
	\includegraphics[width=1\linewidth]{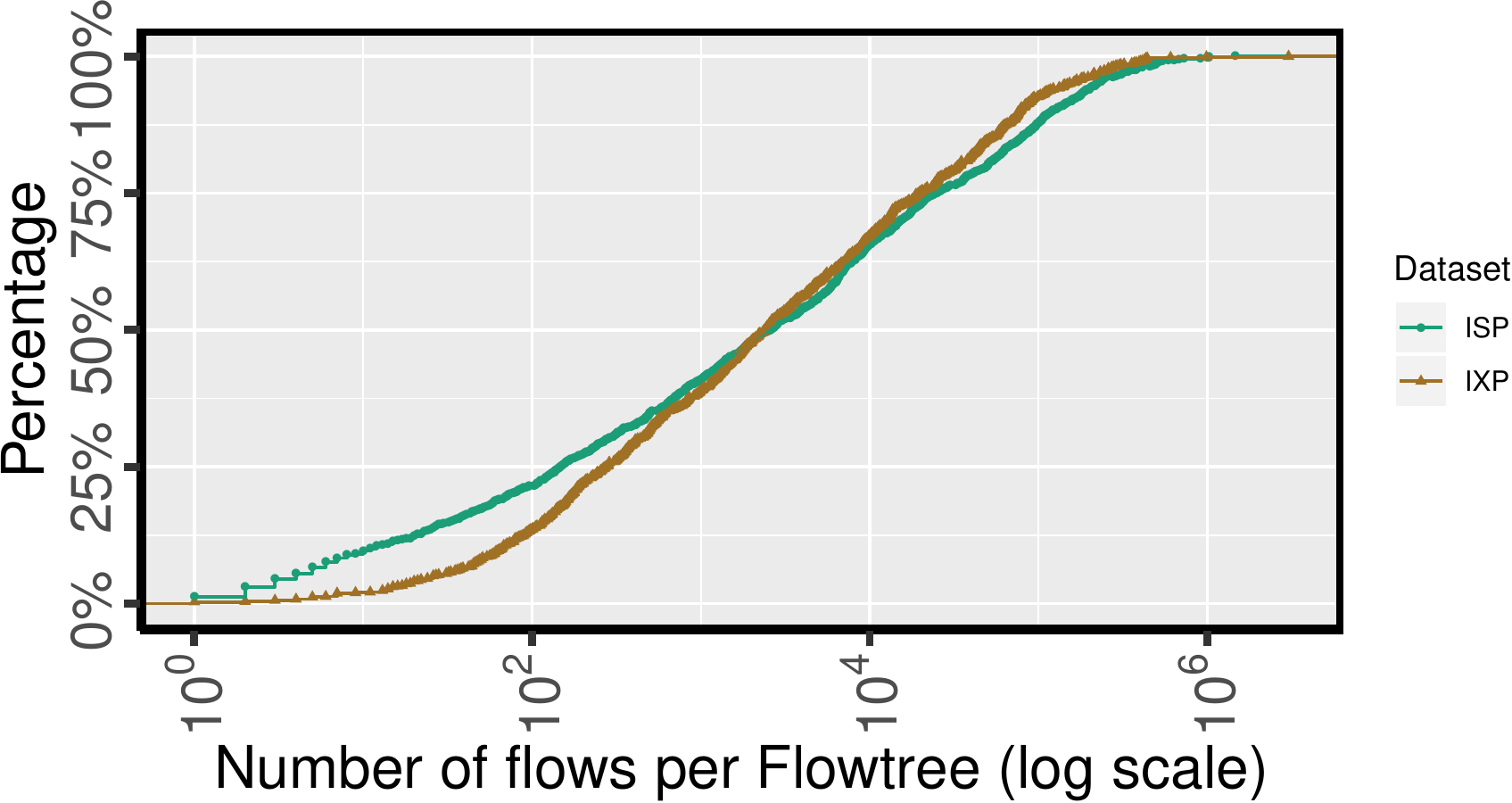}
}
\caption{ECDF of \# of entries--all sites (IXP and ISP).}
\label{fig:input_file_entries}
\end{minipage}
\hfill
\begin{minipage}[t]{0.51\linewidth}
\subfigure{
	\includegraphics[width=0.85\linewidth]{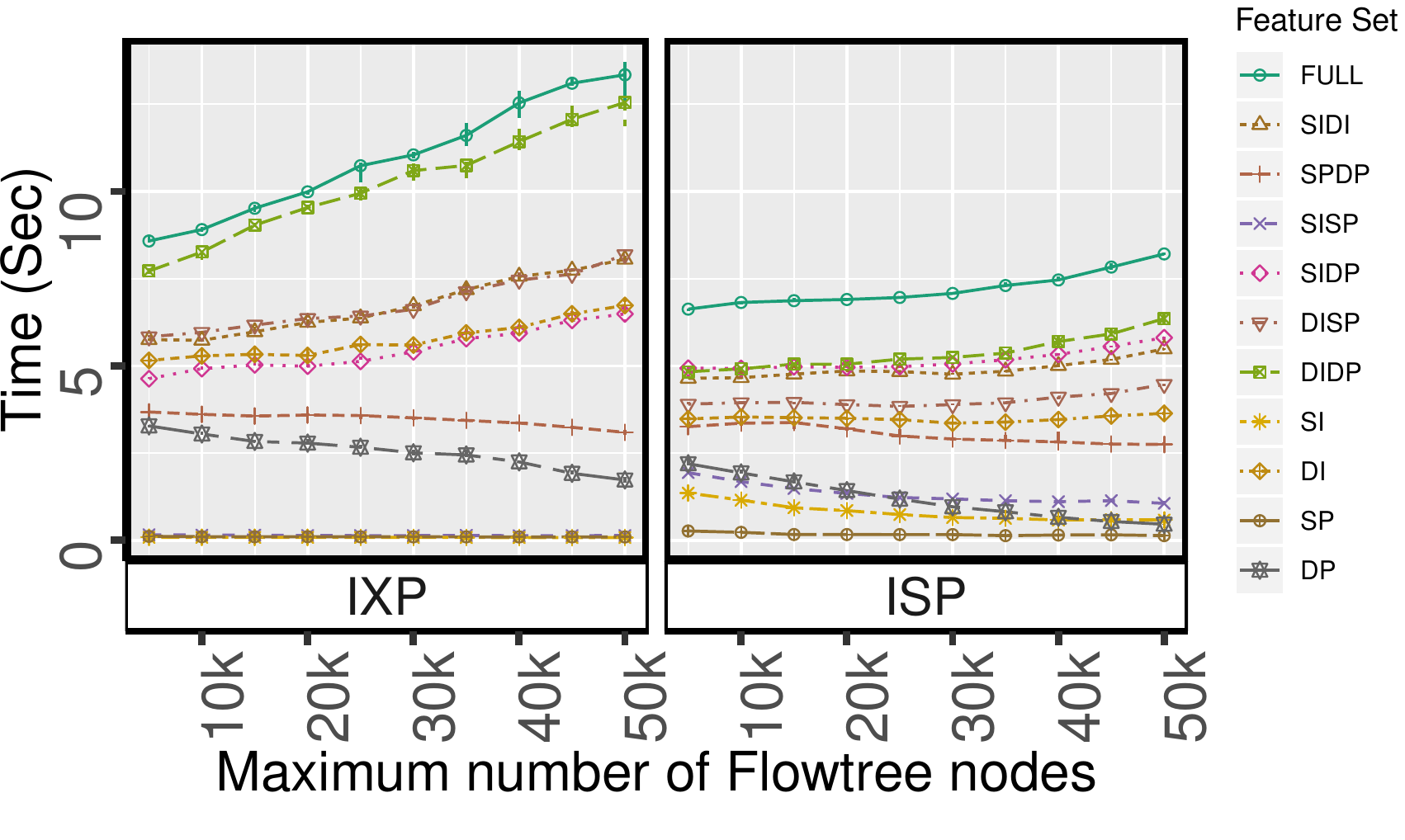}
}
\vspace{-10pt}
\caption{\Flowtree build time (IXP/ISP: four/one 15-min.\ trees) 
	  vs.\ max.\ \# of nodes per feature set.}
\label{fig:ft-runtime-all}
\end{minipage}
\end{figure*}

\subsection{\Flowtree Evaluation}
\label{sec:flowtree-eval}

\done{We have to address a comment from the shadow PC: we should make clear that the reduction of transfer and space
  with flowtree is between 75-95\% (we claim this in the abstract).\\
this claim has already been addressed in the part  Flowtree space-saving}

\noindent{\bf Input data skewness:} One motivation for using HHHs
is to take advantage of the skewed input data. We indeed confirm that
the flow captures are skewed in the sense that for all feature sets, all time
periods, and all sites with enough traffic, the traffic volume follows a skewed
distribution.

\begin{figure*} [t!]

	\captionsetup{skip=.25em}
	\centering
	\subfigure[$ARE$ vs.\ \# of \flowtree nodes (all feature sets at IXP).]{
	\includegraphics[width=0.48\linewidth]{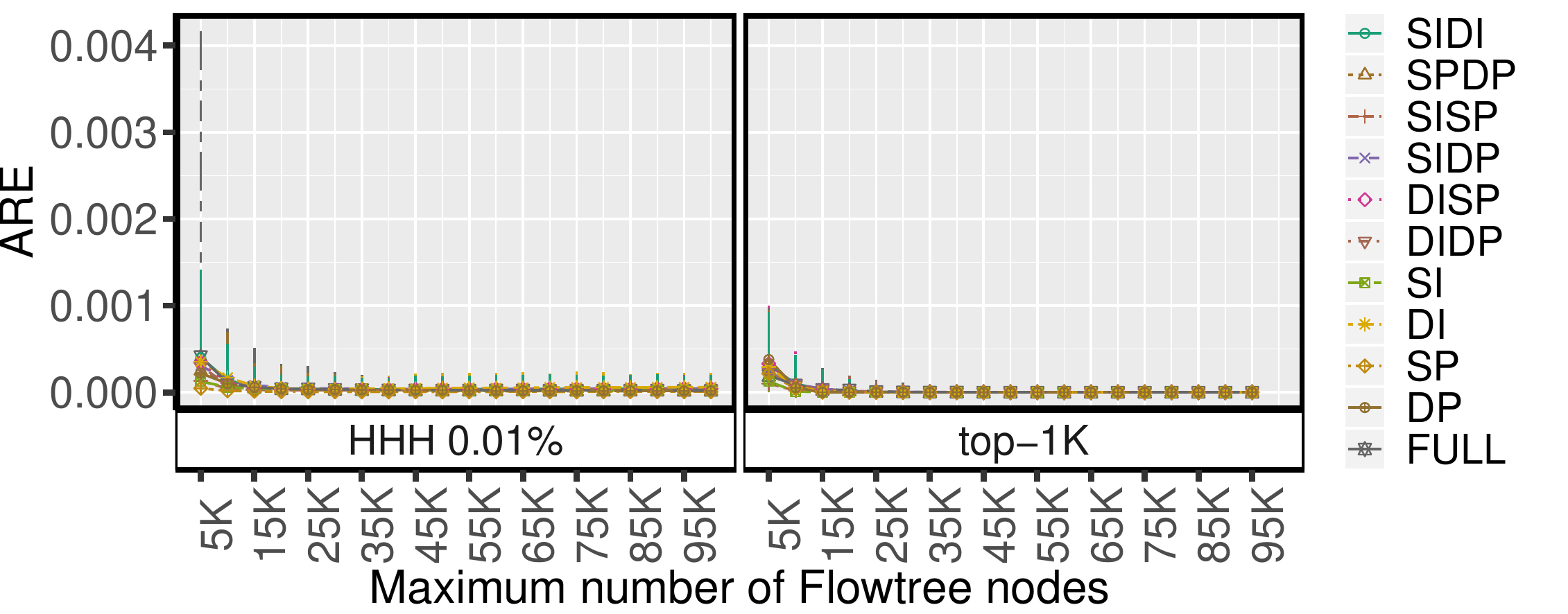}
	\label{fig:are-vs-ntrees}
}
\subfigure[$F1$-score vs.\ \# of \flowtree nodes (all feature sets at IXP).]{
\includegraphics[width=0.48\linewidth]{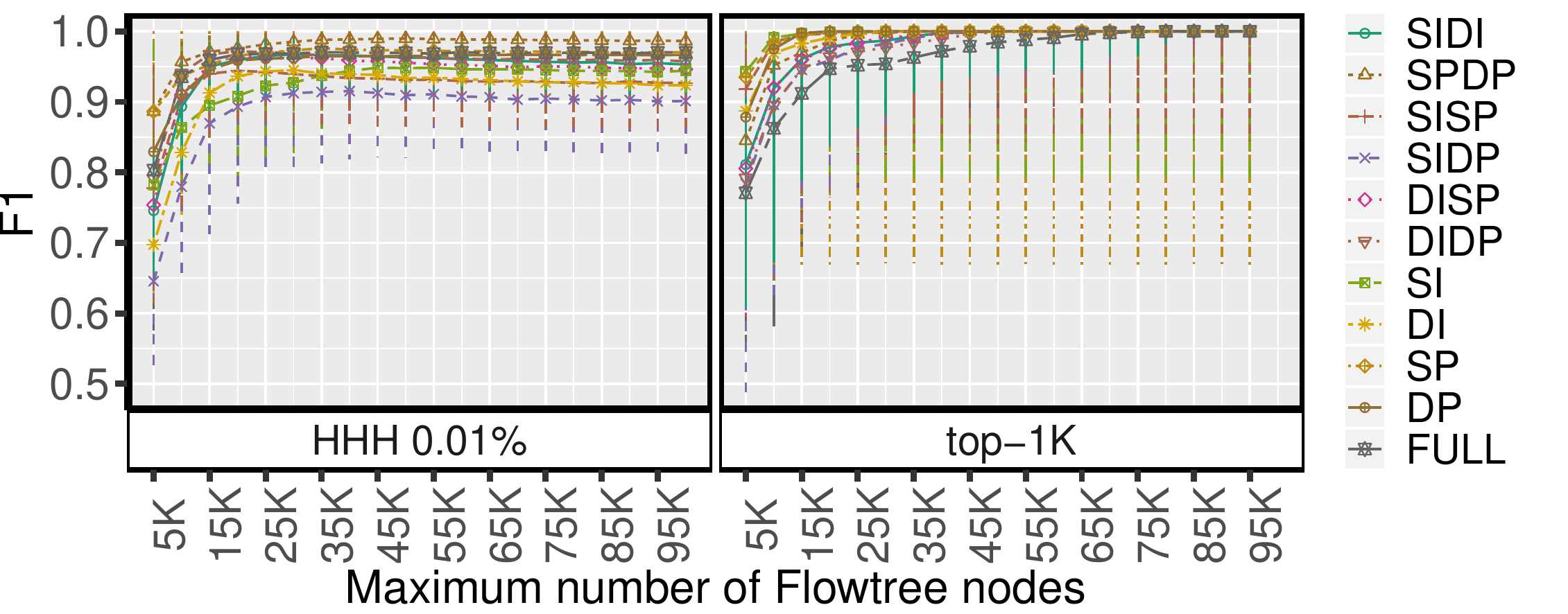}
\label{fig:f1-vs-ntrees}
}
\caption{Accuracy of \flowtree for commonly-used queries (all feature sets at IXP).}

\end{figure*}

Next, the data structure should be able to summarize time periods with small as
well as large numbers of flows as underlined by
Figure~\ref{fig:input_file_entries}, which shows the empirical cumulative
distribution (ECDF) of the number of flow entries per 15-minute \flowtree for
the IXP and the ISP datasets using a logarithmic x-axis. We find a huge
skew. More than 37.5\% of the time periods (per site) have less than 1,000
entries, yet more than 12.5\% have more than 50k entries. This underlines that
the data structure has to be very flexible to efficiently summarize time
periods with many as well as few flows.

\noindent{\bf \flowtree creation time:}
Next, we focus on the worst-case runtime to generate \flowtrees,
which, in part, depends on the deployed hardware\footnote{At the IXP we have
Intel Xeon Gold 6148 CPUs; at the ISP we only have Xeon-E5-2650 CPUs}. We focus
on one hour of data, the busy hour, for the largest site at the IXP and 15 minutes
of data--again busy hour and largest site, for the ISP to get an upper bound on
the runtime.  Note that the data includes more than 6.5M flows that have to be
processed. We compute \flowtrees for each 11 feature set while varying the
maximum number of \flowtree nodes from 5k to 50k. We repeat the experiments 10
times and measure the runtime, in terms of wall time, for generating trees as
reported by the C++ chrono library\footnote{We choose setup to similar to
  \cite{ben2017constant,Elastic-Sketch-SIGCOMM2018} which also use wall time and
  preload the input data into memory.}.

Figure~\ref{fig:ft-runtime-all} shows the 10th and 90th percentile of
the tree creation times vs.\ the maximum
number of \flowtree nodes. All runtimes are well below 15
seconds for 1-hour resp.\ 15 minutes input files; thus, even if we have to process
flows from 1,000+ sites, the deployed hardware, with moderate parallelization, is
sufficient for generating all 11-feature \flowtrees in real time.
In the worst case we needed 20 min to process traces from all 1,000+ sites over one hour; that is, Flowtree would only not become a bottleneck if the throughput tripled and input from 1,000+ sites were to be processed. In that case, aggregating firs over different subsets of the flow space would be necessary.
We notice different behavior for different features: The (destination IP, port)
feature trees are very fast to compute, which can be explained by the fact
that they exhibit the most skewed input distribution. The full (4-feature) trees
take the longest---not surprising given that this feature combination potentially
has the largest number of tree nodes.

We also notice that from one feature set to the next, the runtime sometimes
decreases and sometimes increases as we increase the maximum
number of tree nodes.  The reasoning behind this surprising behavior is as follows.
When the number of \flowtree nodes increases, while compressions happen less
frequently, they take more time to run, given that they have to process a larger
input. If the data is skewed, the increase of the compression runtime with the
number of nodes is limited while the reduction in the average delay
between two compressions is significant. Reversely, if the data is not less
skewed, the increase in compression runtime outbalances the reduction in
inter-compression delay.


\noindent{\bf \flowtree accuracy:}\label{sec:flowtree accuracy}
Next, we look at the accuracy of the query results with focus on advanced
queries, namely the 1-d HHH and top-K queries for \flowtrees with different featureset. Our metrics are
the Average Relative Error, $ARE$, and the $F_1$ score.
The $ARE$ is the average of the ratios between the errors and the ground-truth
values; that is, in our case,
$\frac{1}{n}\sum_{i=1}^{n}\frac{\rvert f_i - \hat{f_i}\rvert}{f_i}$ with $n$ 
the number of flows, $f_i$ the flow popularity and $\hat{f_i}$ 
the estimated flow popularity.
The $F_1$ score is the harmonic mean of precision and recall; accordingly,
it accounts for both false positives and false negatives and ranges from 0 to
1---1 being the best value (perfect precision and recall) and 0 the worst.
We calculate the $ARE$ and the $F1$ score for the 1-d HHH and top-K queries, with
thresholds of 0.01\% and K=1000 respectively, for each 15-minute \flowtree and all sites over the IXP's
busy hour, letting the maximum number of nodes in the \flowtrees vary
from 5k to 100k.  Note that we only evaluate the queries if a \flowtree
summarizes at least 10k flows within the 15-minute time period since
otherwise, the results would be a fraction of a flow, which does not exist.  To
generate the ground truth, we use a \flowtree with an unrestricted number of
nodes.  Finally, we only accept exact matches: in case of HHH, if a generalized flow $f$ is
in the actual heavy hitters it has to be returned by the HHH query; if the
HHH query returns instead, a parent or child of $f$ in the tree, this is a miss.

Figure~\ref{fig:are-vs-ntrees} plots the median $ARE$ values vs.\ the maximal
number of nodes in the \flowtree and includes 10th and 90th percentiles as
error bars in top-K and HHH queries. Our experiment shows that even for 10k \flowtrees the median ARE
values are less than $0.0002$ for all feature sets. Moreover, the main reason
for ARE variations are flows with relatively small popularity.

The results for the $F1$ scores---see Figure \ref{fig:f1-vs-ntrees} which shows
the median together with the 10th and 90th percentile vs.\ the number of nodes
per tree---confirm the excellent performance of \flowtree. Even for small
trees, the median numbers are well above 0.9 for most feature sets. Moreover,
the number of outliers is small.


\noindent{\bf \flowtree vs.\ RHHH:}
Next, we compare \flowtree to a state-of-the-art data structure, the constant
time updates in hierarchical heavy hitters (RHHH)~\cite{ben2017constant}.
More precisely,  RHHH is a randomized version of the deterministic HHH algorithm
(dHHH) proposed by Mitzenmacher et al.~\cite{MitzenmacherHHH:2012}.
RHHH has $O(1)$ update complexity, improving the $\Omega(H)$ update complexity
of its deterministic counterpart, where $H$ is the number of hierarchy levels.

While both \flowtree and RHHH take in the maximum node count as input, RHHH
(and dHHH) have an additional input parameter: the HHH-threshold.  The
HHH-threshold determines if a frequent item is a heavy hitter, and, thus, if
a node should be maintained in the tree. This complicates the usage of RHHH
since neither the number of flows nor their popularity distribution is known
in advance. Setting the threshold too high creates a very shallow tree with
high aggregation, e.g., /16s and /8s, which does not keep enough detail.
Setting the threshold too low may result in a tree with more nodes than the
maximum node count. Indeed, we run into these limitations when executing 
the publicly available code~\cite{RHHH-code} on the corresponding input.
Hence, we evaluated the two systems under similar conditions, i.e.,
with the dataset that was used to evaluate RHHH~\cite{ben2017constant} (CAIDA).
The evaluation dataset comes from Equinix-Chicago trace of
CAIDA~\cite{Equinix-dataset}---this contains 20 Million packets (no sampling)
from a 1Gbps link in the colocation facility in Chicago.
In contrast, note that \flowtree is self-adjusting.

We used a number of metrics:
(1) system runtime,
(2) F1, on top 1k sources or destination of the input
trace are present in the trees of 1k, 10k, 20k, and 40k nodes,
(3) accuracy (ARE), i.e., how well the \flowtree or RHHH estimate the counters of the
heavy hitters, either single IPs or aggregations.

\begin{table}
\begin{center}
{\scriptsize
\begin{tabular}{|l|c|c|c|c|c|}
\hline
Data structure & 1k & 5k & 10k & 20k & 40k \\
\hline
\hline
\flowtree & .19 (.31) & .77 (.92) & .92 (.99) & .98 (.99) & .99 (.99)\\
\hline
RHHH & .42 (.11) & .50 (.57) & .91 (.92) & . 92 (.94) & .93 (.95) \\ 
\hline
\end{tabular}
}
\end{center}
\caption{
F1 score on top 1k src (dst) IPs for 1k, 5k, 10k, 20k, and 40k
node \flowtree and RHHH trees.
}
\label{table:trees:F1-comparison}
\end{table}

The system runtime for creating RHHH trees is, as expected, quite constant:
around 26 seconds.  For \flowtree the time is higher, around 50 seconds, even
as the number of nodes increases.

With regard to F1 score--identifying the correct set of heavy hitters--we find
that if RHHH is not tuned, its performance is poor: very few of the IP heavy
hitters are present and the trees are very small; the F1 of \flowtree
is significantly better.
Table~\ref{table:trees:F1-comparison} reports on the top-1k heavy hitter
IPs indeed in the tree for \flowtree vs. RHHH with different total numbers of nodes%
(each time the threshold in RHHH is adjusted to produce 1k heavy hitters, i.e., the same output as Flowtree).
For trees with up to 10k nodes, \flowtree includes a significantly larger
number of heavy hitters than RHHH but beyond 10K nodes the differences get
smaller.

\begin{figure}[t!]
	\centering
	\includegraphics[height=2.8cm]{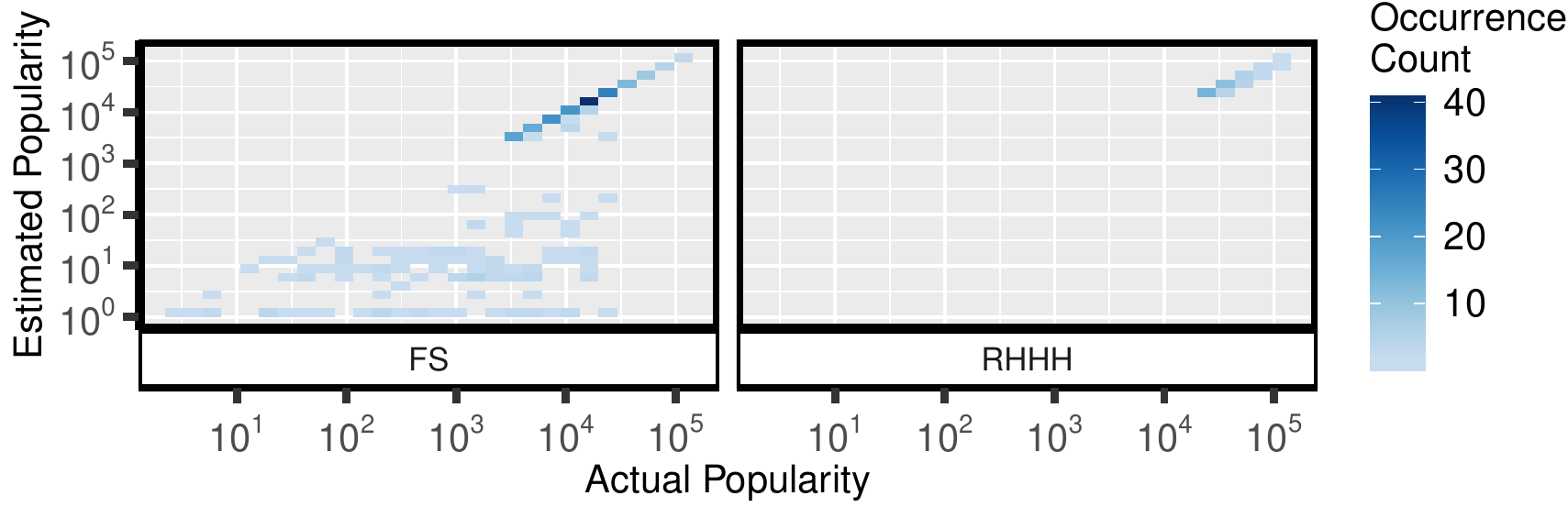}
	\caption{Comparison of estimated vs actual popularities using \flowtree (left) and RHHH (right).}
	\label{fig:trees:accuracy-comparison}
\end{figure}

Next, we turn our attention to the accuracy of the estimated values for each
heavy hitter. We plot in Fig.~\ref{fig:trees:accuracy-comparison} the estimated
value using \flowtree (left) and RHHH (right) compared to the actual value for
the 1k node trees---ARE on the top .1\% IPs of 0.71 for Flowtree vs. 0.92 for RHHH.

The closer a point is to the diagonal the higher its
accuracy. At first glance, RHHH might look better. However, it only contains a
small subset of the relevant HHs as many top-1k entries are aggregated by
RHHH. Thus, \flowtree again significantly outperforms RHHH.

\begin{figure*}[t!]
  \captionsetup{skip=0em}

	\begin{minipage}[t]{0.5\linewidth}
	\subfigure{
		\centering
		\includegraphics[height=4cm]{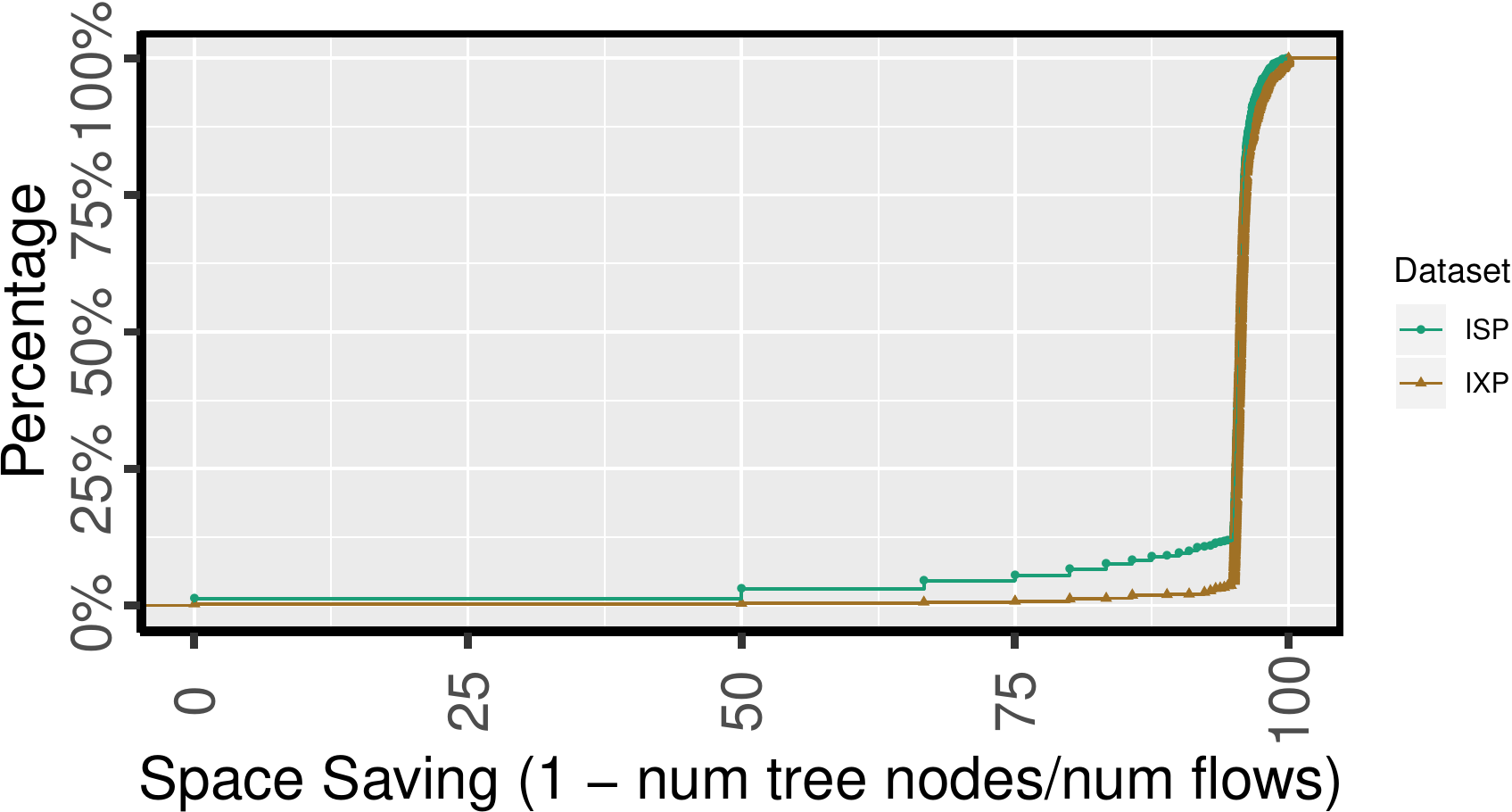}
		\label{fig:fs-tree_count}
	}
	\caption{ECDF of space-saving for all \flowtrees (all time intervals/IXP sites)}
	\end{minipage}\hfill
	\begin{minipage}[t]{0.5\linewidth}
	\subfigure{
		\centering
		\includegraphics[height=4cm]{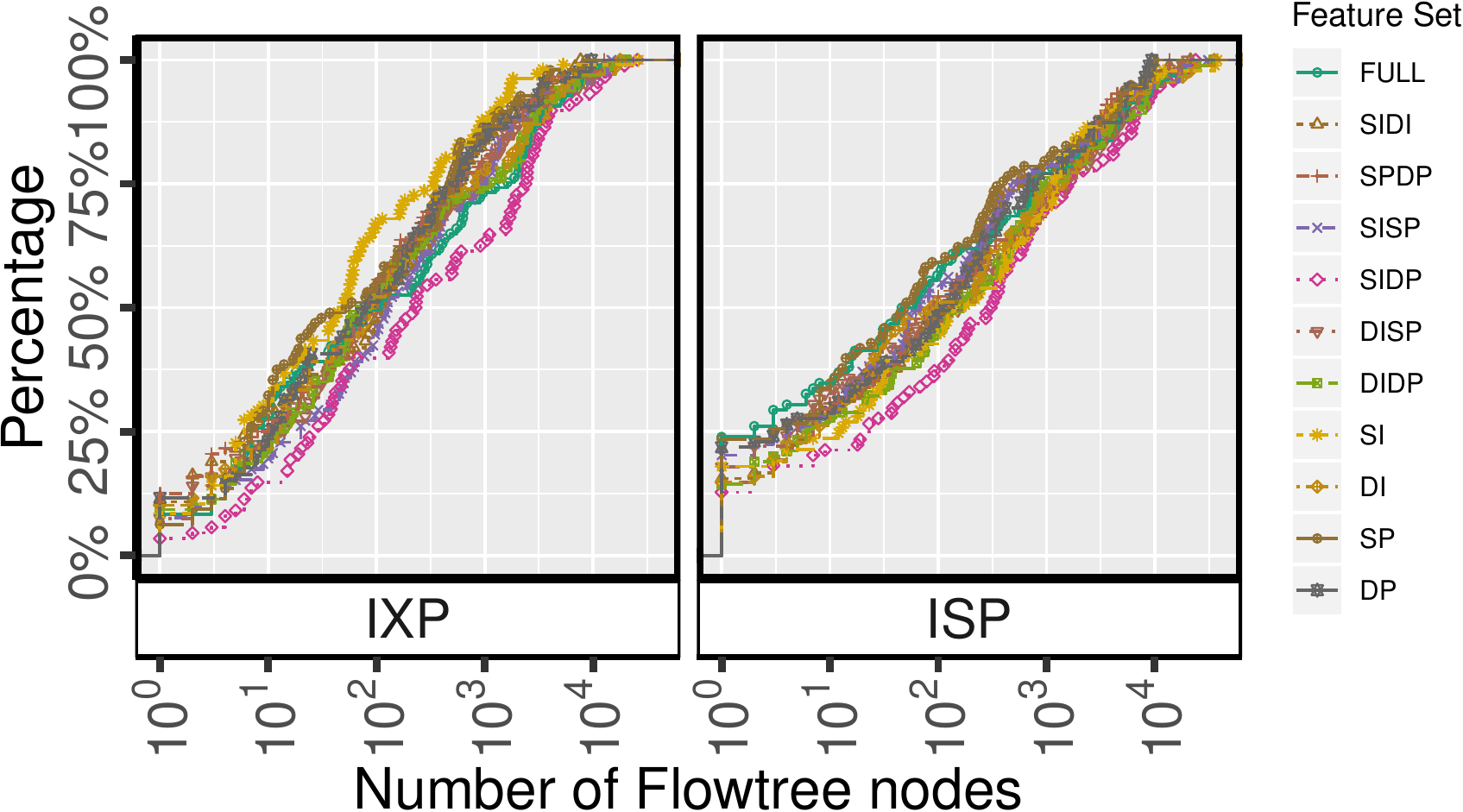}

		\label{fig:ft-tree_sizes_bin_lIXP_rISP_tmp}
	}
	\caption{ECDF: \# of \flowtree nodes (IXP and ISP).}
	\end{minipage}
	%
	%
	\begin{minipage}[t]{0.5\linewidth}
	\subfigure{
		\centering
		\includegraphics[height=4cm]{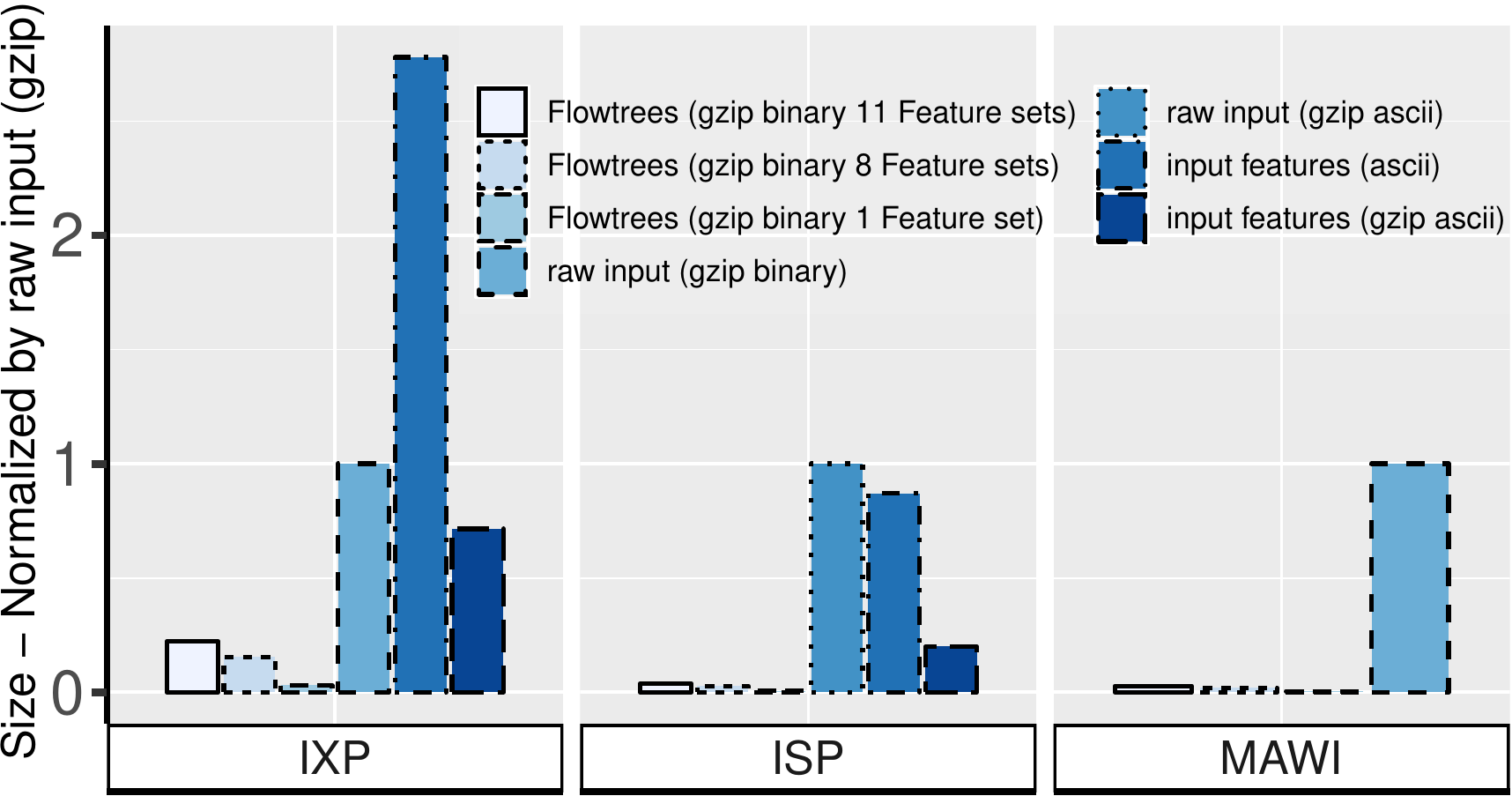}
		\label{fig:fs-compression-15}
	}
	\caption{Space usage vs.\ raw compressed (gzip) input data.}
	\end{minipage}
	\begin{minipage}[t]{0.5\linewidth}
		\centering
			\subfigure[ IXP]{
				\includegraphics[height=0.38\linewidth]{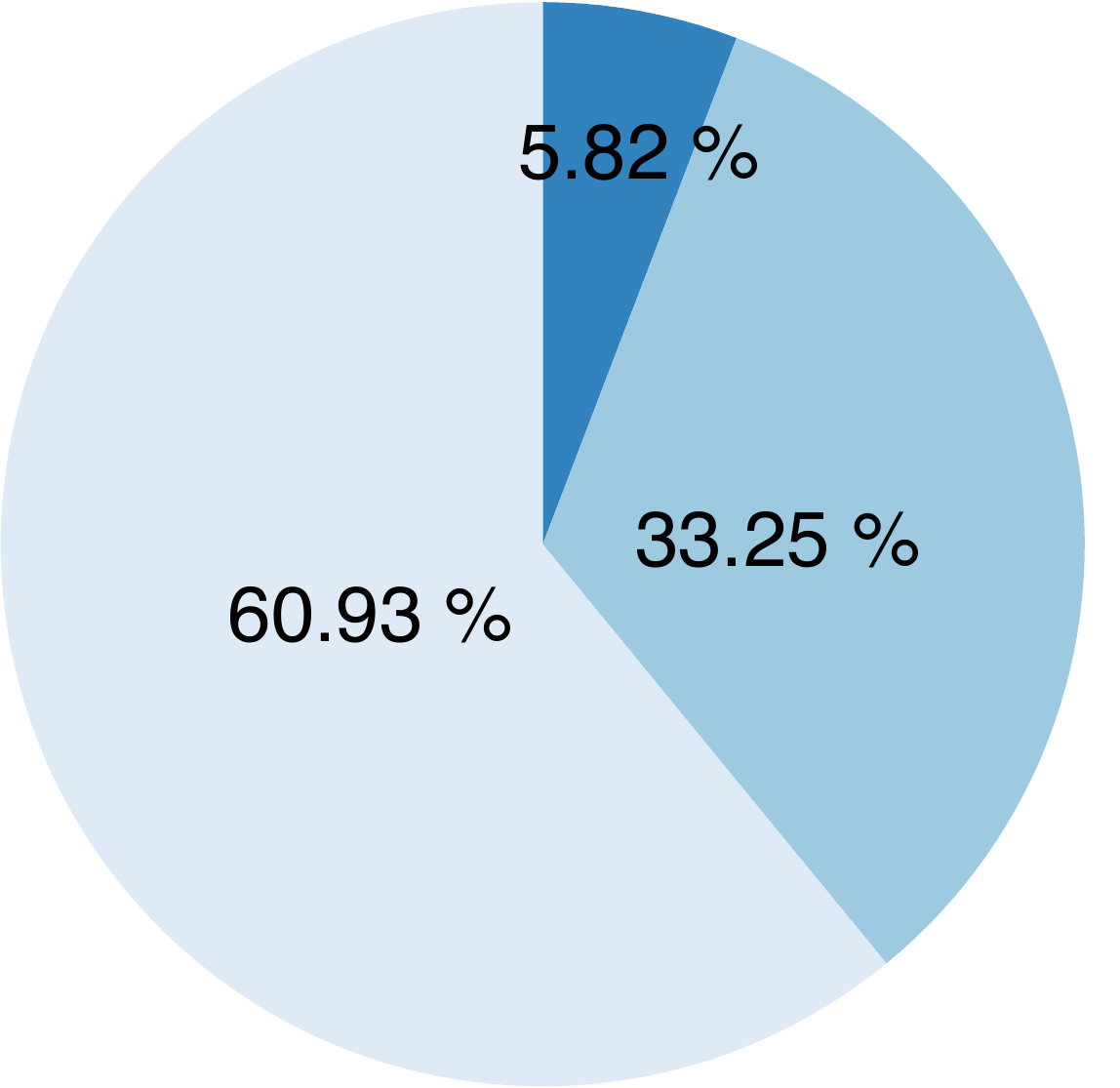}
				\label{fig:additional-tree-pie-ixp}
			}
			\subfigure[ ISP]{
				\includegraphics[height=0.38\linewidth]{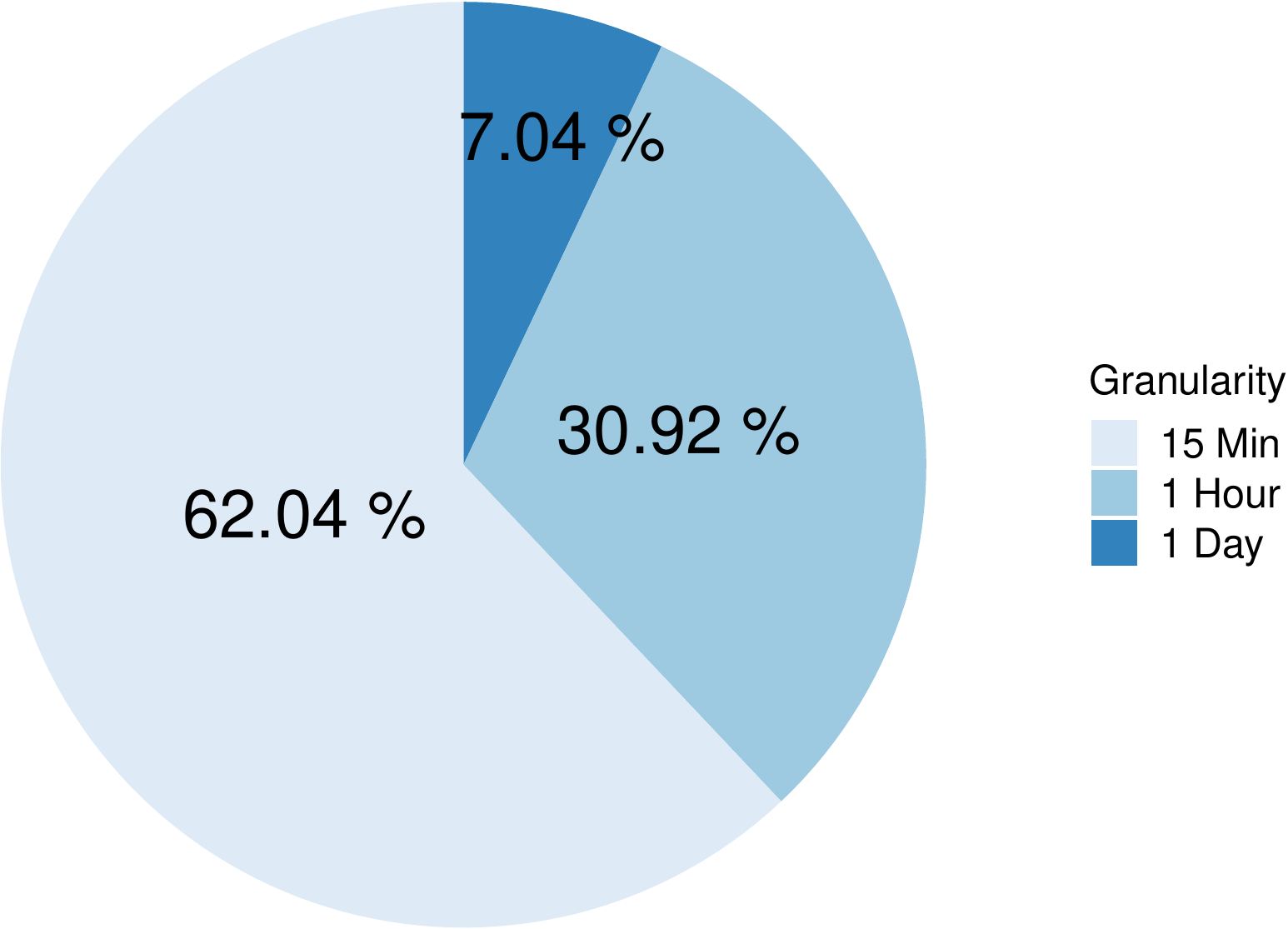}
				\label{fig:additional-tree-pie-isp}
			} 
		\caption{Pie Chart: MongoDB footprint.}
		\label{fig:additional-tree-pie}
	\end{minipage}

\end{figure*}

\noindent{\bf \flowtree space saving:}
Given that we can compute \flowtrees efficiently and that they accurately
answer 1-d HHH queries, we move on to study their space efficiency. Given the
$F1$ scores and $ARE$ values we, for the rest of this paper, choose 10k nodes
for the 1-feature \flowtrees for src and dst ports and 40k nodes for all other
feature combinations. (While 20k may be sufficient, using 40k does not increase
the storage resp.\ communication overhead significantly, as we apply a final
compress operation before using any \flowtree.)

To highlight the ability of \flowtree to compress its input,
Figure~\ref{fig:fs-tree_count} plots the ECDF of
\flowtrees space saving ($ 1 - \frac{ \texttt{\#nodes in tree} }{
\texttt{\#input flows} } $) for all sites and all 15-minute time intervals.
For almost all \flowtrees the space savings are well above 95\%.  This is also
underlined by Figure~\ref{fig:ft-tree_sizes_bin_lIXP_rISP_tmp} which shows
the ECDF of the number of actual \flowtrees nodes. Note that a \flowtree will
always contain less than 40k/10k nodes because we always run a final compression.
Alternatively, it might simply happen that the data did not contain enough
different feature combinations in the first place.

\subsection{\Flowstream Evaluation} \label{sec:flowstream_evaluation}

\noindent{\bf \flowstream space efficiency:} Given the above results regarding
the capabilities of \flowtree, it is not surprising that \flowstream achieves
excellent compression ratios. For the IXP (ISP), we see that compared to the
original compressed IPFIX data (original compressed ASCII flow summaries), the
single full-feature \flowtree in compressed binary format has a space saving of 97\% resp.\ 99.5\%.
With additional feature sets, e.g., all
1-feature \flowtrees and three 2-feature \flowtrees, we still reach space saving of 92\%
resp.\ 97.5\%. If we include all 11 possible feature combinations, the space saving
is 89\% resp.\ 96\%. Even if we normalize not by the raw input data but only
against the necessary features for the \flowtrees, the space savings are
still excellent, e.g., more than 97\% for the 1-feature \flowtree at the ISP.
For a visualization of the space efficiency relative to the size of the raw
compressed (gzip) input data, see Figure~\ref{fig:fs-compression-15}.

While 15-minute time granularity is excellent for ans\-wering detailed queries,
many queries involve coarser time granularities. Thus, it can be useful to
add time as another feature and add 1-hour as well as 1-day aggregated \flowtrees
by merging (and then compressing) the smaller-time-granularity \flowtrees.
\flowstream does so automatically.  While this needs some extra memory, it adds
less than 40\% overhead---see Figure~\ref{fig:additional-tree-pie}---while
offering the potential to significantly reduce query response time. Moreover,
should space become an issue, \flowstream may decide to permanently delete
smaller-time aggregates while keeping higher-time aggregation summaries.
This is one of the design features that enable resource management
with \flowstream. It is always possible to still keep coarse grain summaries of
previous time periods or site sets even if disk space is running out.

  \begin{table}[tb]
  \tiny
    \caption{Benchmark queries for \flowstream evaluation. Note that these queries correspond to those identified in Table~\ref{table:sys-problems}.}
    \vspace*{-1.5em}
    \scriptsize
    \begin{center}
      \begin{tabularx}{\linewidth}{L{0.08cm}L{0.8cm}L{2.5cm}L{4.15cm}}
      \toprule
            & {}Benchmark & Goal & Query \\
            \midrule 
            1 &
            {Agg\-re\-ga\-ted flow statistics}
            & Computing total traffic with specific features from IP/ports/time/location
            & SELECT pop(PROTO,COUNTMODE[,BIN])  FROM (time YYYY-MM-DD hh:mm to YYYY-MM-DD hh:mm) WHERE ( site\_id = ANY and dst\_ip = IP/mask and dst\_port = port/portmask )
            \\
            \hline
            2 &
            {Counting traffic}
            & Computing Traffic volume between given IP/Port subnet/addresses, for a specific site n
            & SELECT pop(PROTO,COUNTMODE[,BIN]) FROM (time YYYY-MM-DD hh:mm to YYYY-MM-DD hh:mm) WHERE (site\_id = n and src\_ip = IP/mask)
            \\
            \hline
            3 &
            {Traffic flows}
            & Displaying flows belonging to given subnets / IP addresses, passing through a specific site
            & SELECT *(PROTO,COUNTMODE[,BIN]) FROM (time YYYY-MM-DD hh:mm to YYYY-MM-DD hh:mm) WHERE (site\_id = n and src\_ip = IP/mask)
            \\
            \hline
            4 &
            {Traffic matrix}
            & Finding popular flows from a subnet to subnets for all sites
            & SELECT above(K,PROTO,COUNTMODE[,BIN]) FROM (time YYYY-MM-DD hh:mm to YYYY-MM-DD hh:mm) WHERE (site\_id = ANY and src\_ip = ANY and dst\_ip = ANY)
            \\
            \hline
            5 &
            {DDoS diagnosis}
            & Finding the src IPs from which a dst IP (victim) has received abnormal traffic.
            & SELECT top(K,PROTO,COUNTMODE[,BIN])  FROM (time YYYY-MM-DD hh:mm to YYYY-MM-DD hh:mm) WHERE site\_id = ANY and dst\_ip = {[victim\_ip]}
            \\
            \hline
            6 &
            {Super-spreader Detection}
            & Finding hosts that send packets to more than k unique dst during a time interval (requires multiple queries)
            & SELECT above(K,PROTO,COUNTMODE[,BIN]) FROM (time YYYY-MM-DD hh:mm to YYYY-MM-DD hh:mm) where (site\_id = ANY and src\_ip = ANY)
            \\
            &
            & 
            & SELECT * FROM (time YYYY-MM-DD hh:mm to YYYY-MM-DD hh:mm) where (site\_id = ANY and dst\_ip = {[pop\_ip]})
            \\
            \hline
            7 &
            {Top-k flows}
            &Detect Top K flows in one or more sites , going to / coming from a specific subnet or IP address
            & SELECT top(K,PROTO,COUNTMODE[,BIN]) FROM (time YYYY-MM-DD hh:mm to YYYY-MM-DD hh:mm) WHERE site\_id = n and (src\_ip = IP/mask or dst\_ip = IP/mask)
            \\
            \hline
            8 &
            {Heavy Hitters}
            &Detect all flows with popularity over threshold T, in one or more sites, going to / coming from a specific subnet or IP address
            & SELECT hhh(T,PROTO,COUNTMODE[,BIN]) FROM (time YYYY-MM-DD hh:mm to YYYY-MM-DD hh:mm) WHERE (site\_id = n and src\_ip = IP/mask)
            \\
            \hline
            9 &
            {Heavy Changers Detection }
            & Detect Top K heavily changed flows in one (or more) site(s).
            & SELECT hc(K,PROTO,COUNTMODE[,BIN]) FROM (time YYYY-MM-DD hh:mm to YYYY-MM-DD hh:mm)(time YYYY-MM-DD hh:mm to YYYY-MM-DD hh:mm) WHERE site\_id = n
            \\
            \hline
            10 &
            {Full/4/5 tuple queries}
            & Counting / Detecting flows belonging to a specific protocol/application
            & SELECT *(PROTO,COUNTMODE[,BIN]) FROM (time YYYY-MM-DD hh:mm to YYYY-MM-DD hh:mm) WHERE site\_id = n
            \\
            \bottomrule
      \end{tabularx}
    \end{center}
    \label{table:app-query}
  \end{table}

\subsection{\FlowQL Evaluation}
\label{sec:flowql}
Next, we focus on the performance (query response time) of the query capabilities and the query
engine using a set of benchmark queries.  In particular, we go back to the main
tasks of a network manager---recall Table~\ref{table:sys-problems}---and pick a
benchmark query for each of the identified tasks---note that the detection of one super-spreader
requires two queries. These chosen queries are shown in
Table~\ref{table:app-query}, the table which thus contains queries for
every single important network management task tackled by related work.

To challenge \flowstream, we task it to execute these queries for a full day
for all sites in the IXP dataset.  We evaluate three different ways
of answering the queries using \flowtree, namely using \flowQL with \flowtrees and 15-minute,
1-hour, and 1-day aggregation. On the IXP machine, we execute each benchmark 10 times and measure,
just as before, the wall time as reported by the C++ chrono library.

\begin{figure} [t]
  \captionsetup{skip=.5em}
  \subfigcapskip=-10pt
  \subfigbottomskip=-2.5pt
  \centering
  \subfigure[1-day \flowtrees (cold/hot cache)]{
    \label{fig:queryresponse-1-day}
    \includegraphics[width=1\linewidth]{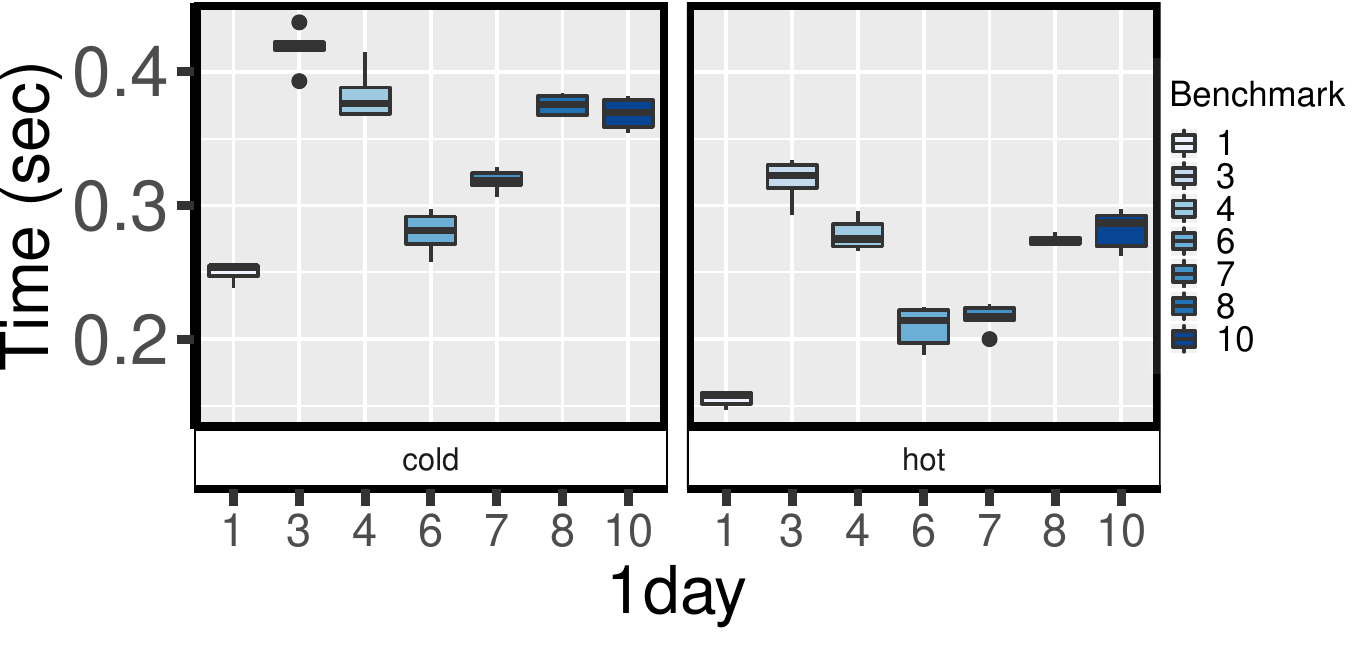}
  }
  \subfigure[1-hour \flowtrees (cold/hot cache)]{
    \label{fig:queryresponse-1-hour}
    \includegraphics[width=1\linewidth]{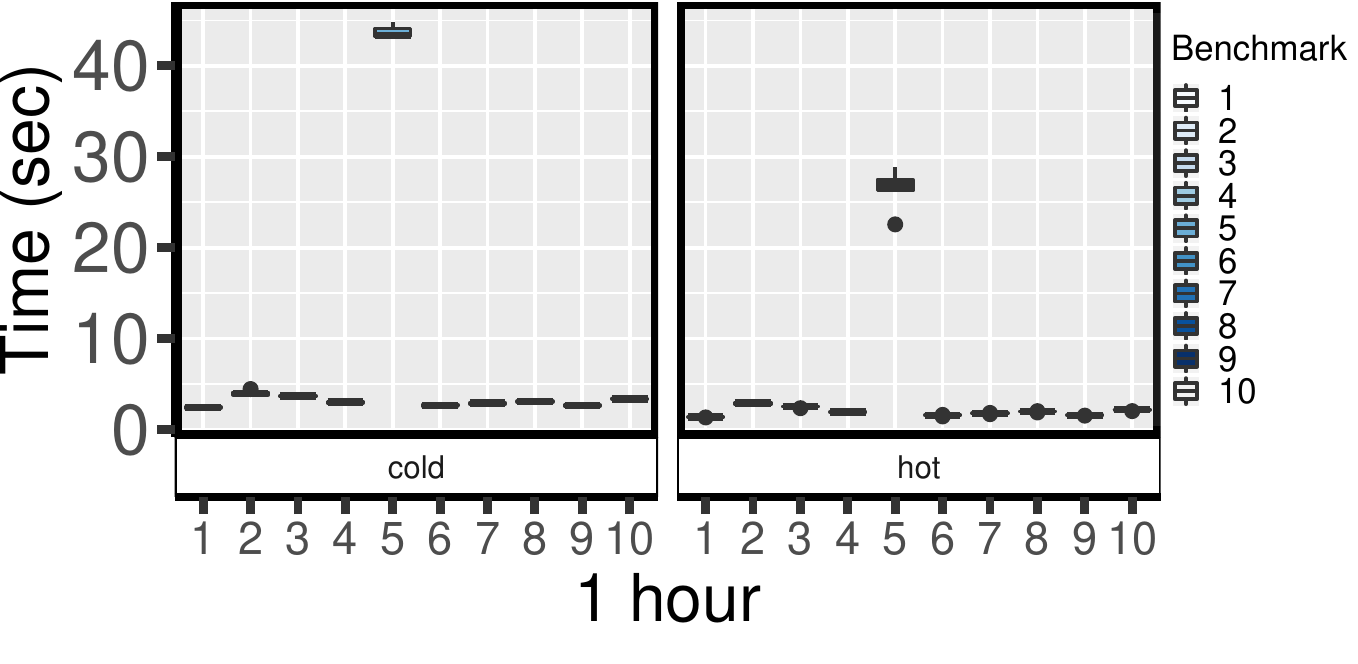}
  }
  \subfigure[15-minute \flowtrees (cold/hot cache)]{
    \label{fig:queryresponse-15-min}
    \includegraphics[width=1\linewidth]{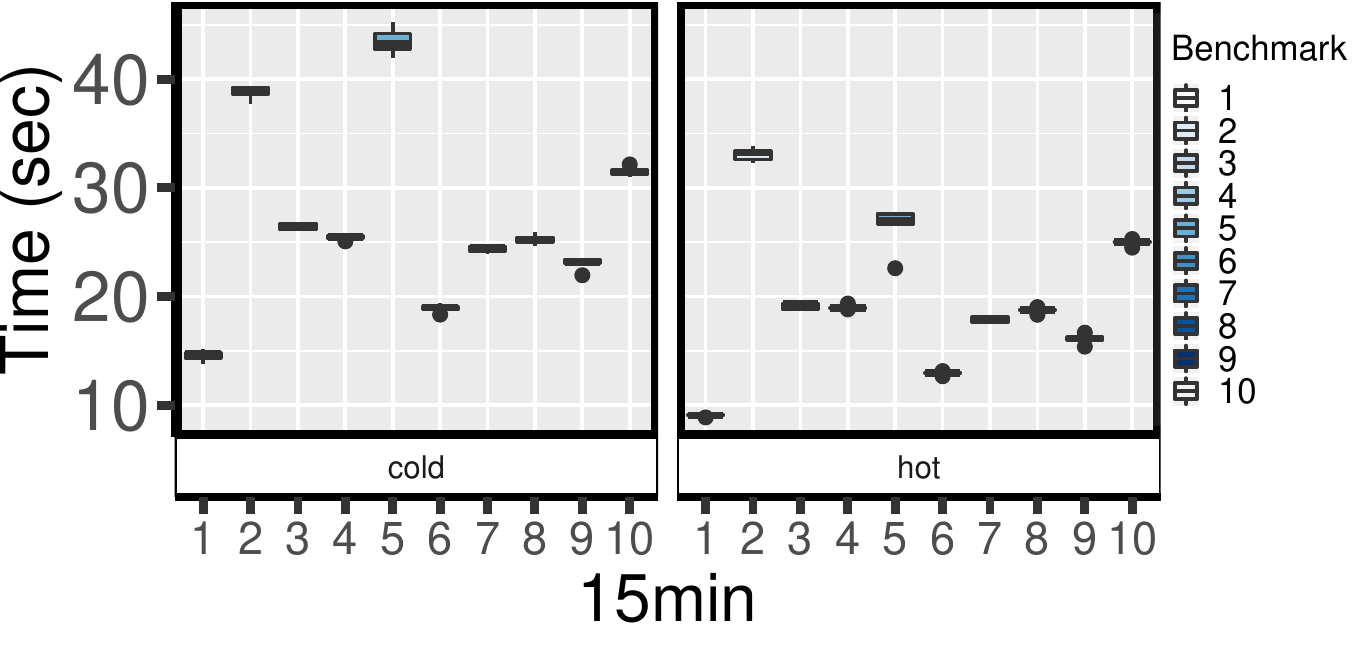}
  }

  \caption{IXP: \flowstream times (Table~\ref{table:app-query} benchmarks).}
  \label{fig:queryresponse}
\end{figure}

\begin{figure*} [t]
  \captionsetup{skip=.5em}
  \centering
  \includegraphics[width=0.7\linewidth]{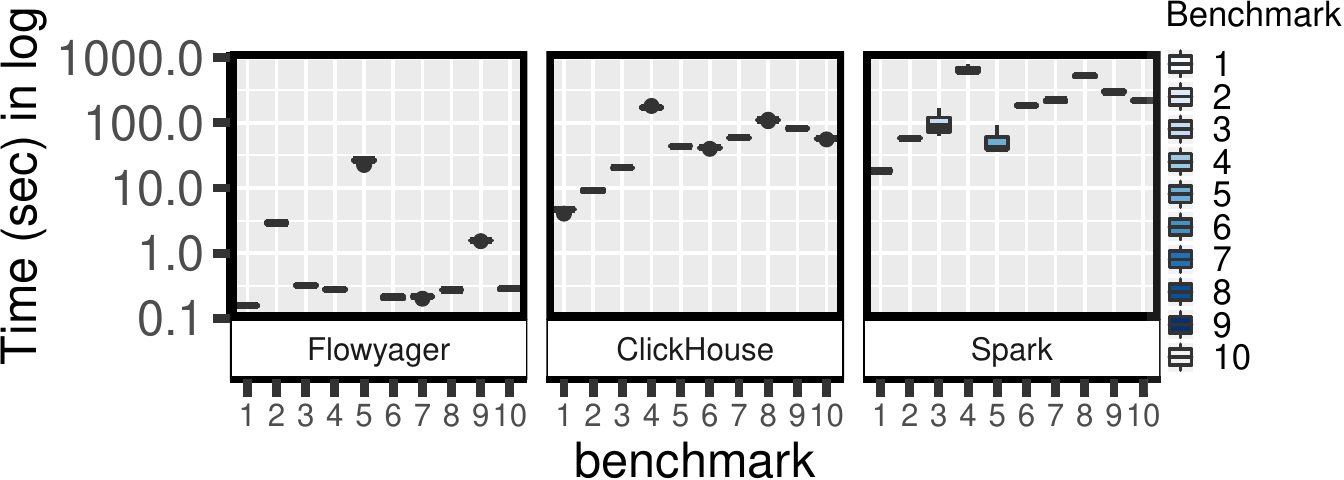}
  \caption{IXP: Query response time comparison: \flowstream vs.\ ClickHouse vs.\ Spark (Table~\ref{table:app-query} benchmarks).}
  \label{fig:queryresponse-comparison}
\end{figure*}

Figure~\ref{fig:queryresponse} shows the resulting \flowQL query response
times for each benchmark as boxplots. 
Hereby, we distinguish between cold and hot query response times. In the
hot case,  relevant \flowtrees may be retrieved from the in-memory cache. In
the cold case, we restart the in-memory cache process for each benchmark.  If we
use the 1-day \flowtrees, see 
Figure~\ref{fig:queryresponse-1-day}, the answers are readily available and the
response arrives in the blink of an eye (less than 1 second).
By using the in-memory cache we speed up query response time by
about 10 to 50\%.  We also check the accuracy of the results and find that the
results are accurate\footnote{We exclude Benchmarks 2, 5, and
9 as these benchmarks concern 60 min
time-intervals and, thus, cannot be answered using data at 1-day granularity.}.

With 1-hour trees, see
Figure~\ref{fig:queryresponse-1-hour}, the query response times typically increase by roughly a
factor of $7$, even though the number of \flowtrees that have to be
processed increases by a factor of $24$. This is possible as \flowstream takes
advantage of parallelization. For Benchmark~5 the query response time is the worst as
we have to execute an iterator across all 24 hours. Note, this is no principle
limitation of the design of \flowstream but a limitation of the implementation
which does not yet parallelize the iterators.
If we move to 15-minute trees, see Figure~\ref{fig:queryresponse-15-min},
the query response time increases further up to a factor of eight. This highlights the
efficiency obtained by using higher granularity trees in the design of \flowstream.
Note, all benchmarks are executed
using a research prototype rather than a production system.

Using an appropriate \flowtree granularity, we can answer all except one benchmark
query in less than 5 seconds, underlining that \flowstream is indeed able to
answer apriori unknown queries. This query response time enables interactive
exploration of the data.

\subsection{\Flowstream vs.\ Possible Alternatives}

Finally, we explore how well \flowstream performs compared to other systems. We
picked three alternatives, namely, using (a) task-specific data-parallel Python
scripts, (b) Spark~\cite{Spark}--a state of the art data analytics platform, and (c)
ClickHouse~\cite{ClickHouse}--a state of the art column database.  We evaluated all these systems on the same machine and dataset in IXP as previously described in \ref{sec:data}. 

First, we find that coding a custom python script for each benchmark takes a
reasonably experienced programmer at least 2-3 hours for programming and
debugging even if they can build upon a template from another
benchmark. After all, it takes time to validate that the script is indeed doing
what it is supposed to do. For some of the advanced tasks, e.g., the HHH, we
did not start from scratch but rather included existing code. Nevertheless,
this again did take additional time.  Running the Python code on a day of data
did take a mean of 39 minutes using a parallelization across 24 cores. Using 24
cores enables the script to parallelize the tasks by processing each hour of
data in a separate process.  Across all benchmarks, the Python code needed a
minimum of 19 minutes and a maximum of 54 minutes.

\begin{figure*} [t]
  \captionsetup{skip=.5em}
  \centering
  \includegraphics[width=.9\linewidth]{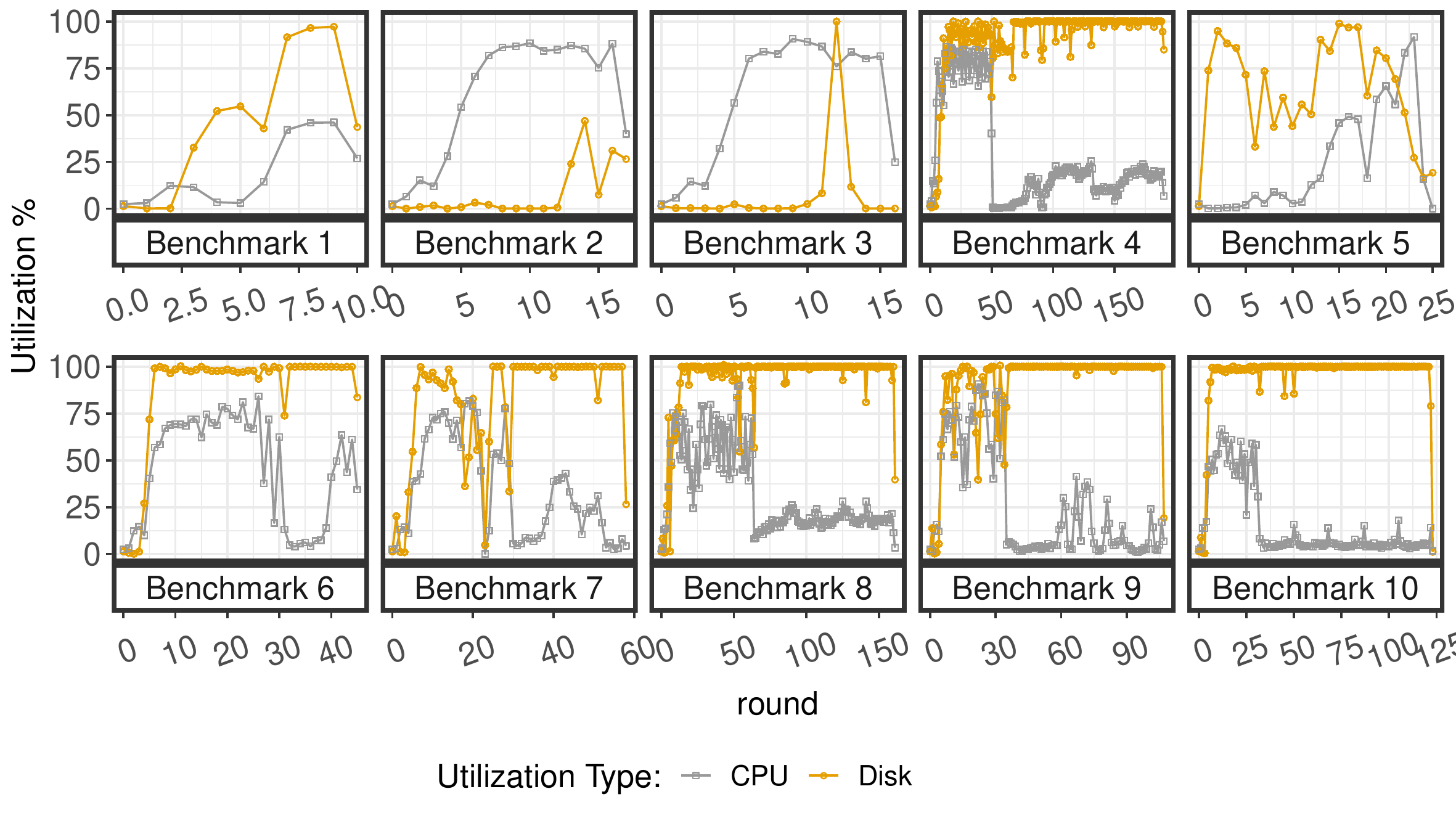}
  \caption{CPU and disk I/O usage in Spark experiments}
  \label{fig:spark-disk-cpu}
\end{figure*}

Second, we find that setting up Spark and coding the queries  
require significant time. Indeed, it is necessary to first convert the data
into a Spark-compatible format to get any reasonable performance (query response times
less than 1 hour). This takes roughly 15.5 minutes per day of data for the IXP
site. The resulting benchmark query response times are shown in
Figure~\ref{fig:queryresponse-comparison}.  Using this preprocessed data as
input, the benchmark queries take a minimum of 20 seconds and up to 800
seconds. Note that for Benchmark~8  Spark only computes heavy hitters rather
than HHH as implementing HHH on top of Spark is non-trivial. To measure the CPU usage and disk I/O usage of each Spark benchmark, we used the \emph{iostat} command sampling every 5 seconds. In Figure \ref{fig:spark-disk-cpu}, the x-axis shows the round, i.e. the 5-second period in which we sample, and y-axis shows the utilization in percentage. CPU utilization is shown in square points, while the round points show the disk I/O. We observe that in the majority of the cases, Spark is bound by disk I/O rather than CPU. This holds for benchmarks 1, 4-10. However, benchmark 2 is a drill-down query and requires multiple GROUPBY statements. Also, benchmark 3 works with only two features. Hence, the intermediate results are not too big to require frequent disk access. Therefore, unlike other benchmarks, benchmark 2 and 3 are limited more by CPU capacity than disk I/O. Indeed, this figure highlights the significant overhead of query processing using only the raw data.

Third, we set up an instance of ClickHouse. Here, it is necessary to first load
the data into the database. This takes roughly 45 minutes per day of IXP data.
On the other hand, the resulting benchmark query response times are significantly smaller
than those of Spark, see Figure~\ref{fig:queryresponse-comparison}. Again,
ClickHouse only supports a limited version of the HHH query for Benchmark~8.
Figure~\ref{fig:queryresponse-comparison} also includes the \flowstream
benchmark results from Section~\ref{sec:flowql}.  \flowstream's benchmark
performance supersedes all comparison systems.

\subsection{Summary and \flowstream Limitations}

Overall, \flowstream by far outperforms all three alternatives. Moreover,
\flowstream is adaptive and supports HHH and physically distributed execution.
We acknowledge that creating all \flowtrees does add some overhead--one day
does take roughly 4 hours. However, this is a one-time operation, and overhead only matters if we
consider archived data, but the \flowtrees can well be generated as the
flow captures arrive, recall Section~\ref{sec:flowtree-eval}. Moreover, it is
easy to do memory management within \flowstream; e.g., rather than purging older
data, we can summarize it.

The limitation of \flowstream is that its answers are only estimates. However,
these are accurate both for elephants and mice flows alike. Hereby, we want to
point out that most network-wide systems anyhow rely on highly sampled flow
captures. As such the fact that we ``only'' provide estimates does not increase
the uncertainties dramatically.  If higher accuracy is necessary, we recommend
combining \flowstream for data exploration with ClickHouse for focused
in-depth analysis. Moreover, the insights from \flowstream can be used to
instantiate online non-sampled queries using streaming network telemetry
systems, such as Sonata~\cite{gupta2018sonata}.

\section{Use-Cases}\label{sec:use-cases}

In this section, we showcase how to use \flowstream for tackling typical
network operator tasks.

\begin{figure*}[t]
  \captionsetup{skip=.75em}
  \subfigcapskip=-5pt
  \subfigbottomskip=-2.5pt
\centering
\subfigure[ISP: 123 port activity across 60 minute time bins ]{
\includegraphics[clip,width=.8\columnwidth]{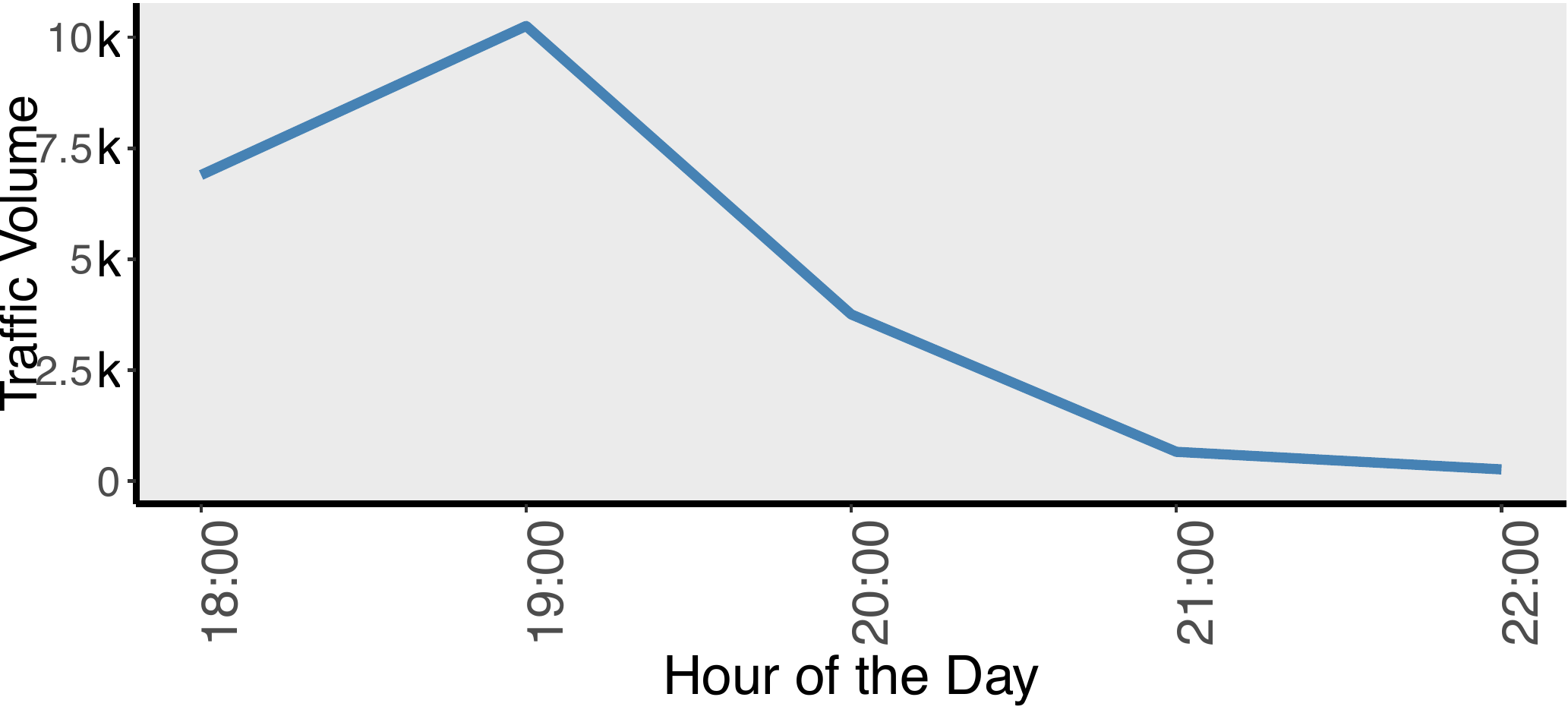}
\label{fig:ddos-ISP1}
}
\hfill
\centering
\subfigure[ISP: 123 port activity across 15 minute time bins]{
\includegraphics[clip,width=.8\columnwidth]{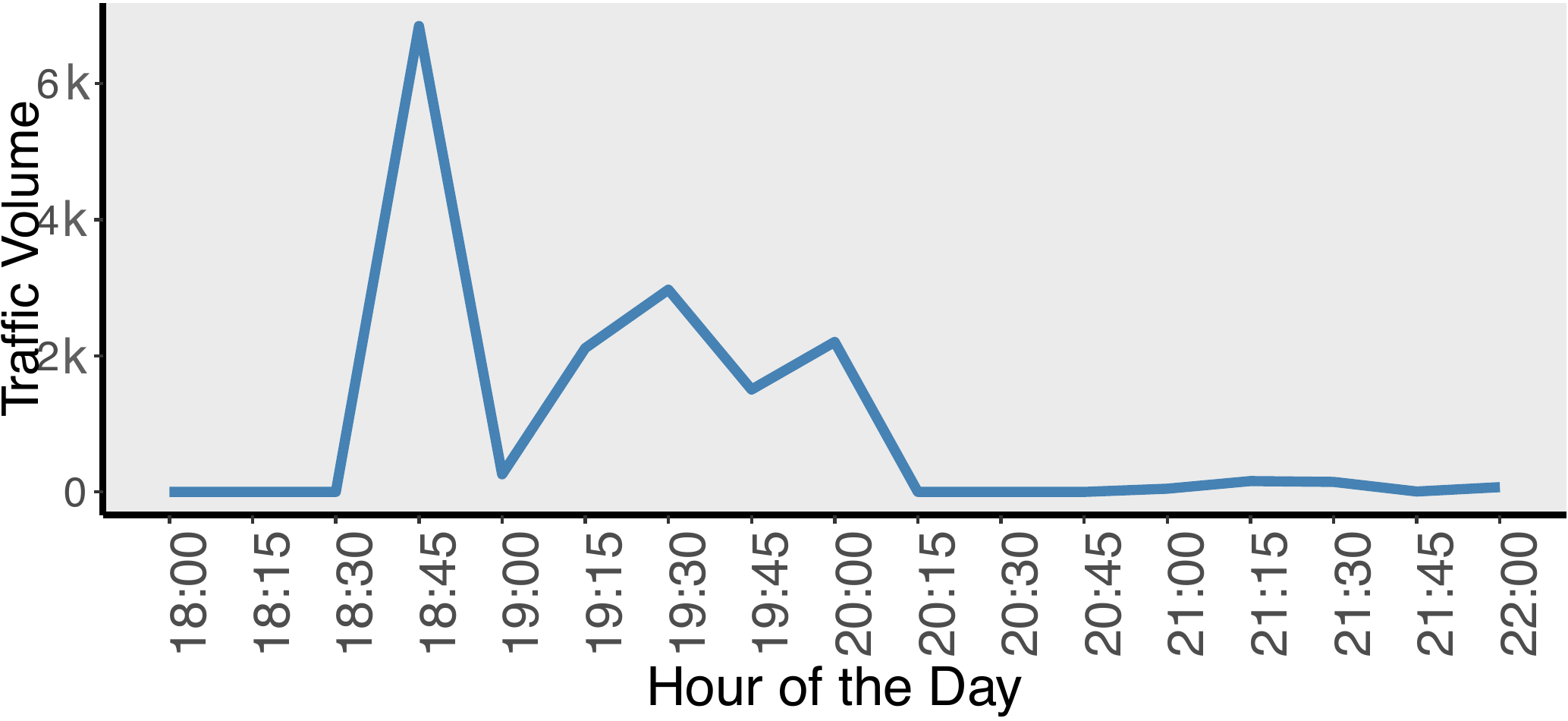}
\label{fig:ddos-ISP2}
}
\caption{ISP: DDoS NTP attack investigation.}
\label{fig:ddos-ISP}

  \captionsetup{skip=.75em}
  \subfigcapskip=-2.5pt
  \subfigbottomskip=-2.5pt
\centering
  \subfigure[Application trends]{
\includegraphics[width=.4\linewidth]{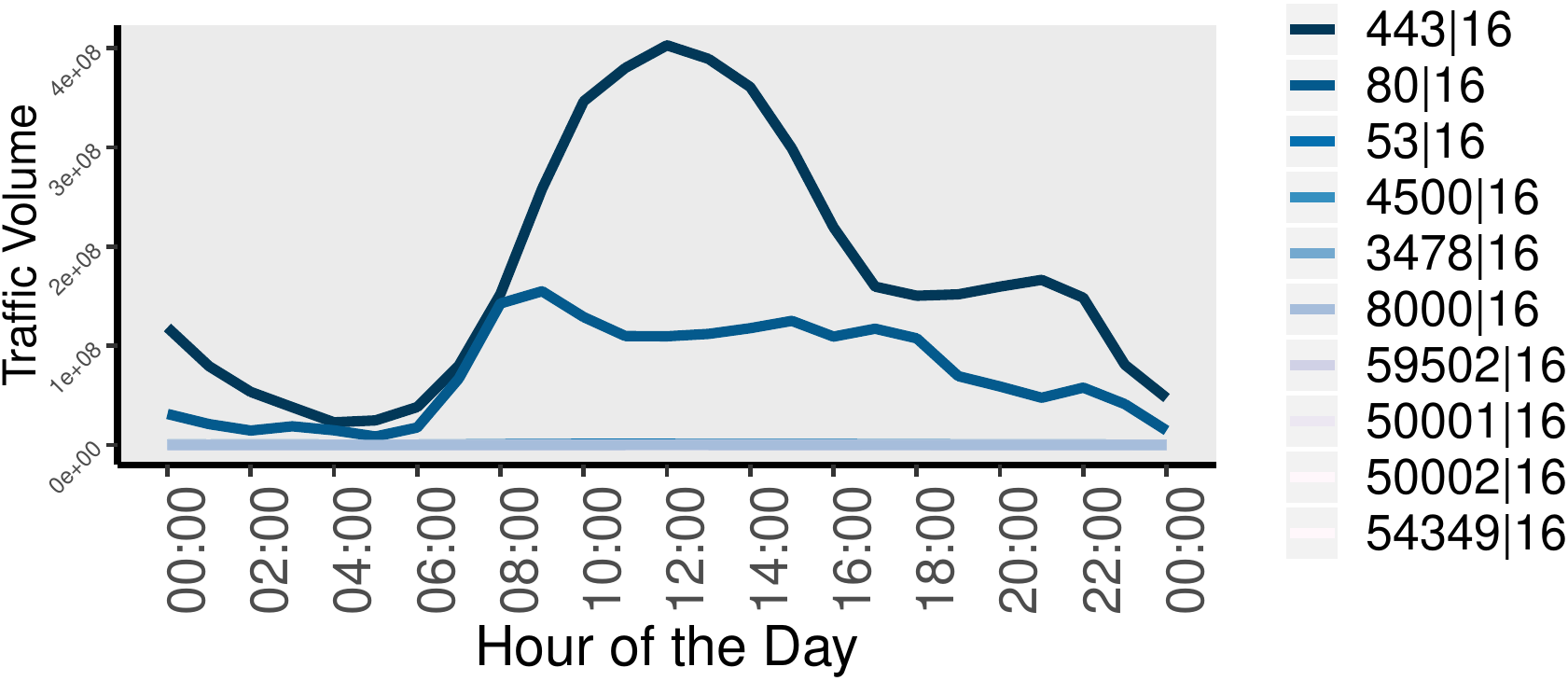}
\label{fig:timeseries-mawi-ports}
}
\hfill
\centering
\subfigure[Traffic matrix.]{
\includegraphics[width=.4\linewidth]{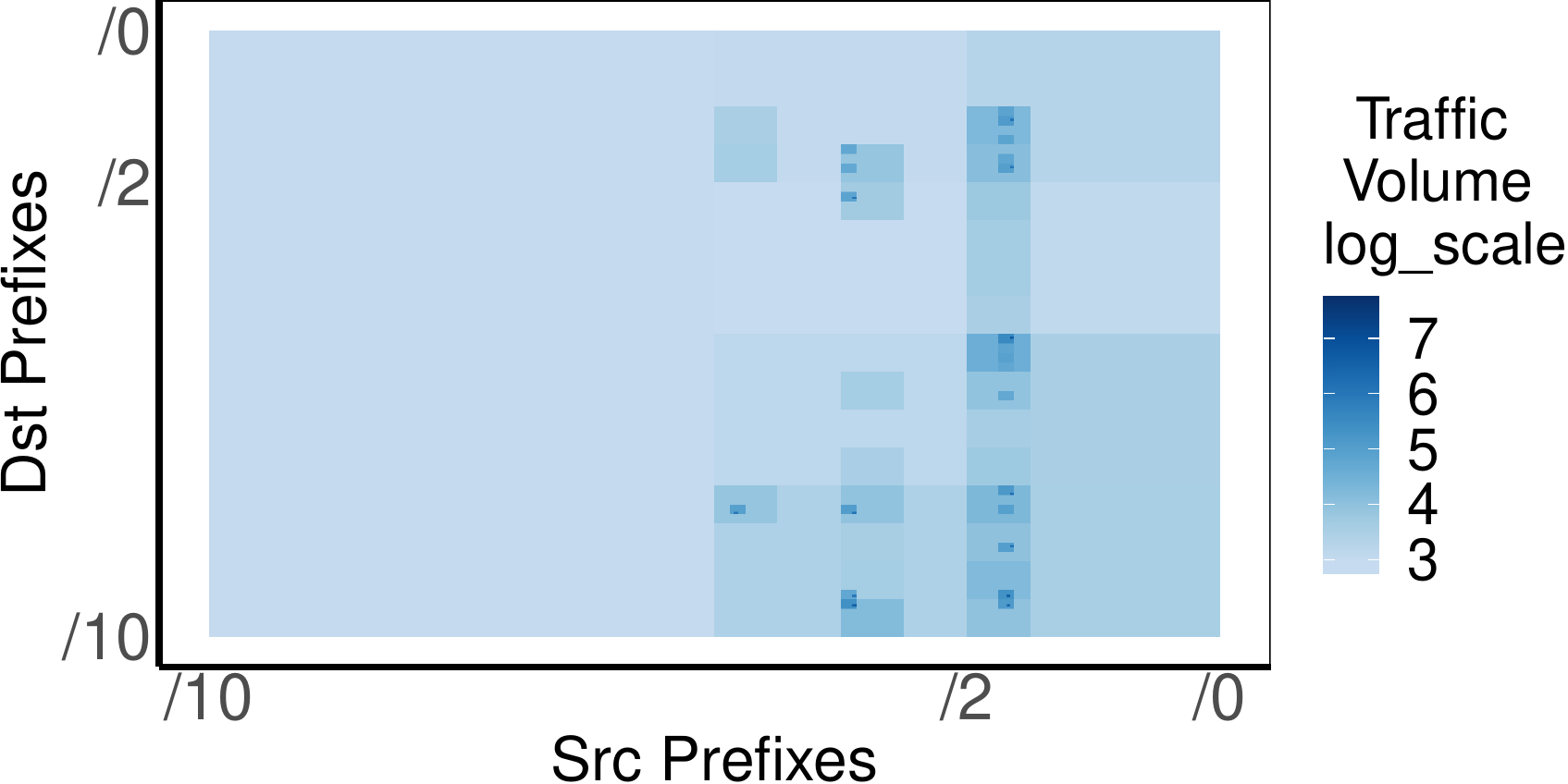}
\label{fig:trafficmatrix-mawi}
}
\caption{MAWI: Data exploration.}
\end{figure*}

\noindent{\bf Unveiling Application Trends:}\label{sec:apptrends}
With \flowstream we can easily infer the 10 most popular applications within a
time period using a top10 query with site\_id=ANY and src\-\_port=\-ANY:
\\
{\small
\noindent{\hspace*{.25cm}\texttt{SELECT top(10,any,byte) FROM (time 2018-05-09 00:00 to 2018-05-09 23:59) WHERE site\_id=ANY and src\_port=ANY}}
}\\
To then see how the popularity of each top~10 port changed over time we use
the query pop-bin60 for each port. Therefore, the query would be:\\
{\small
\noindent{\hspace*{.25cm}\texttt{SELECT pop(any,byte,bin60) FROM (time 2018-05-09 00:00 to 2018-05-09 23:59) WHERE site\_id=ANY and src\_port=X}}
}\\

See Figure~\ref{fig:timeseries-mawi-ports} for the results for the MAWI dataset.
We use the MAWI dataset for reproducibility as we will release the sample
  queries and their output along with the code.
The query takes less than 1.4~seconds. Web and DNS related ports 80, 443,
and 53 dominate. The same is true for the \ISP. Still, during
peak, other port numbers are prominent as well, e.g., port 3074. This port is
used by Xbox LIVE and Games for Windows--Live. The peak traffic
time also is the peak activity time for gaming, at least for residential
customers of this Tier-1 ISP.

\noindent{\bf Traffic matrix:}
Computing a traffic matrix involves determining all src/dst pairs with a
traffic volume larger than a value X. With \flowstream, one can use the above\_t,
for src\_i=ANYp and dst\_ip=ANY. Therefore, the following query can be used: 
\\{\small
\noindent{\hspace*{.25cm}\texttt{SELECT above(X,udp,byte) FROM (time 2018-05-09 00:00 to 2018-05-09 23:59) WHERE site\_id=ANY and src\_ip=ANY and dst\_ip=ANY}}
}\\
To highlight this capability we determine the
src/dst traffic matrix for the MAWI data, see
Figure~\ref{fig:trafficmatrix-mawi}. 
It shows the traffic matrix at different aggregation levels to detect which pairs of the source (src) and
destination
(dst) prefixes (at different
granularity levels) are responsible for a large fraction of traffic exchange.
For visualization, we use a two-dimensional heatmap where
the x-axis corresponds to src IPs, the y-axis to dst IPs, and the color to the
traffic volume normalized by the number of IPs within the area, i.e., traffic flowing
from a src prefix to a dst prefix. This query took less than 13~seconds.

\noindent{\bf Investigating DDoS attacks:}\label{sec:usecase-ddos}
Network attacks, and in particular, distributed denial-of-service (DDoS)
attacks are an ongoing nuisance for network operators as well as network
users. A large body of research papers has focused on techniques for detecting
DDoS attacks, see,
e.g.,~\cite{zargar2013survey,carl2006denial,douligeris2004ddos,lee2008ddos},
including references and citations. Indeed, the multitude and the impact of
DDoS attacks, see, e.g.,~\cite{NTP-DDoS,Jonker:2017:MTU:3131365.3131383}, have
given rise to a variety of different mitigation techniques, see
e.g.,~\cite{dietzel2018stellar,measuring-adoption-ddos}. Still, detecting
DDoS attacks reliably as well as diagnosing their root causes is critical for
starting countermeasures or taking preventive future actions. \Flowstream is
an ideal system for tackling this challenge.

One of the most common signatures of DDoS attacks is a sudden rise in traffic
for src/dst ports that are used within amplification
attacks~\cite{us-cert,memcached-Akamai,NTP-DDoS,rossow14amplification}. Among
such ports are 0, 123 (NTP), 11211 (memcached), 53 (DNS), and 1900 (SSDP), as
discussed above. Potential DDoS attacks can be found by using the heavy
changer query. It identifies time ranges during which they
occurred. We execute these queries for each hour:\\
{\small
\noindent{\hspace*{.25cm}\texttt{SELECT hc(100,any,byte) FROM (time 2019-04-01 00:00 to 2019-04-01 00:59)(time 2019-04-01 01:00 to 2019-04-01 01:59) WHERE site\_id=ITR and (dst\_port=ANY or src\_port=ANY)}}
}\\

Per hour this takes less than
0.3 seconds. Among the heavy changers are high volume ports related to Web
traffic, i.e., port 80, 443, as well as other ports where the volume can easily
vary. But, we also find some unusual ports, i.e., 123 (NTP) which are known to
be involved in DDoS attacks. Figure~\ref{fig:ddos-ISP} shows a DDoS
amplification attack in one of the sites of the ISP. This is a DDoS attack on
NTP (port 123). Here, a very large number of src IPs scattered across multiple
networks are involved but only a few dsts are targeted; namely two, whereby one of
them receives more than 95\% of the attack packets. It took us less than
5~minutes of human time and less than 1~minute computation time to find the attack
for port 123, the site, the src of the attacks, and identify the start and the
end of the attack. To illustrate the exploratory power of \flowstream, we
identified the hours where the attack took place, see
Figure~\ref{fig:ddos-ISP1}, within a second. Then, we drill-down to the 15
minutes granularity to infer the start and end of the attack, see
Figure~\ref{fig:ddos-ISP2}, with a second query that took two seconds of
execution time:
{\small
\noindent{\hspace*{.25cm}\texttt{SELECT pop(any,byte,bin15) FROM (time 2019-04-01 01:00 to 2019-04-01 01:59) WHERE site\_id=ITR and dst\_port=123|16}}
}
\\Note, detecting slowly increasing DDoS attacks needs a
different approach. Here, a diff query to an earlier time period can be used as
an indicator. 

\noindent{\bf Towards real-time DDoS Mitigation:}
Using insights from historical analysis of DDoS attacks it is possible to use
\flowstream also for near-live analysis if we keep recent \flowtrees at a shorter time
granularity, e.g., 1-minute bins: we can then either use the above queries to
monitor ports highly affected by DDoS attacks or we can use
heavy-changer queries to look for ports with unusual activity. If we see such
unusual activity, we can use the drill-down capabilities of \flowstream to
check if, e.g., the traffic is targeted at specific IPs, i.e., only involves a
small number of src or dst addresses, or involves spoofed addresses, i.e.,
a large number of IP addresses. If yes, \flowstream can be used to trigger an
alarm which may then blackhole the attack traffic, e.g., using a system such as 
Stellar~\cite{dietzel2018stellar} or traffic scrubbing systems~\cite{Jonker:2017:MTU:3131365.3131383}.
Recall that other techniques, e.g., telemetry, need to know a-priori the queries they have to execute. The power of
\flowstream is that is can answer arbitrary queries that are not known in advance and using the already available network
flow summaries supported by router vendors. Thus, \flowstream offers security capabilities that can help to identify
arbitrary security issues. It can also help in generating the appropriate queries to execute them in real-time when, e.g., telemetry is
used.

\noindent{\bf Lessons Learned:}
For our use cases neither the initial sampling in the flow captures nor the
\flowstream estimates were detrimental to achieving the goal.  However, we
noticed some implementation challenges, e.g., handling flows from routers
with unsynchronized clocks.
We decided to use the timestamp when the flow is arriving at \flowagg. Note
that this may lead to some small amount of misbinning if the router is distant
(in terms of network delay) from the aggregator. However, the impact is expected
to be limited and probably well within the typical uncertainty of flow captures.
Note that our approach even enables us to update \flowtrees of past time
bins, should a significant number of flows arrive delayed.

Another observation is that one can tune \flowstream according to the needs of
the users. Overall, we find that a query can be answered quickly if the
aggregation level of the available (cached) \flowtrees matches the query
granularity in terms of site sets and/or time granularity. The reason is that
this avoids merging \flowtrees on the fly. Thus, if many queries involve the
same subset of interfaces, e.g., per router, or all long-haul interfaces, it
may make sense to store additional \flowtrees, if only temporarily. For example,
keeping a \flowtree for all sites adds little overhead but speeds up queries
significantly.

\vspace{-2mm}
\section{Conclusion}\label{sec:conclusion}

Network flow captures are widely available and are essential for operators to
monitor the health of their networks and steer their evolution. Yet, due to
their ever-increasing size and complexity, their analysis is time-intensive and
challenging. In the past, this has substantially hindered ad-hoc queries across
multiple sites, for different time periods and over many network features. In this paper,
we design, develop, and evaluate \flowstream, a system that allows exploration of network-wide data and answering
ad-hoc a priori unknown queries within seconds. It achieves this using already existing network flow captures, without
the need for specialized hardware, and without the need to compile specific queries into telemetry programs that should
be known in advance and are slow to update.

\flowstream uses succinct summaries, \flowtrees, of raw flow captures and
provides an SQL-like interface, \flowql, that is easily usable by network
engineers. We showcase the performance and accuracy of \flowstream in two
operational settings: a large IXP and a tier-1 ISP.
Our results show that the query response time can be reduced by an order of
magnitude, and, thus, \Flowstream enables interactive network-wide queries and
offers unprecedented drill-down capabilities to identify the culprits, pinpoint
the involved sites, and determine the beginning and end of a network attack.

%
\balance
\bibliographystyle{IEEEtran}
\bibliography{paper}
%
%


\end{document}